\documentclass[fleqn,usenatbib,onecolumn]{mnras}

\usepackage{newtxtext,newtxmath}

\usepackage[T1]{fontenc}

\DeclareRobustCommand{\VAN}[3]{#2}
\let\VANthebibliography\thebibliography
\def\thebibliography{\DeclareRobustCommand{\VAN}[3]{##3}\VANthebibliography}

\usepackage{graphicx}	
\usepackage{amsmath}	
\usepackage{subcaption}   
\usepackage{tikz-feynman} 

\DeclareMathOperator*{\SumInt}{%
\mathchoice%
  {\ooalign{$\displaystyle\sum$\cr\hidewidth$\displaystyle\int$\hidewidth\cr}}
  {\ooalign{\raisebox{.14\height}{\scalebox{.7}{$\textstyle\sum$}}\cr\hidewidth$\textstyle\int$\hidewidth\cr}}
  {\ooalign{\raisebox{.2\height}{\scalebox{.6}{$\scriptstyle\sum$}}\cr$\scriptstyle\int$\cr}}
  {\ooalign{\raisebox{.2\height}{\scalebox{.6}{$\scriptstyle\sum$}}\cr$\scriptstyle\int$\cr}}
}

\title[Rayleigh and Raman scattering cross-sections and phase matrices of the ground-state hydrogen atom]{Rayleigh and Raman scattering cross-sections and phase matrices of the ground-state hydrogen atom, and their astrophysical implications}

\author[M. Kokubo]{
Mitsuru Kokubo$^{1}$\thanks{E-mail: mitsuru.kokubo@nao.ac.jp (MK)}\thanks{NAOJ fellow}
\\
$^{1}$National Astronomical Observatory of Japan, National Institutes of Natural Sciences, 2-21-1 Osawa, Mitaka, Tokyo 181-8588, Japan
}

\date{Accepted XXX. Received YYY; in original form ZZZ}

\pubyear{2023}

\begin{document}
\label{firstpage}
\pagerange{\pageref{firstpage}--\pageref{lastpage}}
\maketitle

\begin{abstract}

We present explicit expressions for Rayleigh and Raman scattering cross-sections and phase matrices of the ground $1s$ state hydrogen atom based on the Kramers-Heisenberg-Waller dispersion formula.
The Rayleigh scattering leaves the hydrogen atom in the ground-state while the Raman scattering leaves the hydrogen atom in either $ns$ ($n\geq2$; $s$-branch) or $nd$ ($n\geq3$; $d$-branch) excited state, and the Raman scattering converts incident ultraviolet (UV) photons around the Lyman resonance lines into optical-infrared (IR) photons.
We show that this Raman wavelength conversion of incident flat UV continuum in dense hydrogen gas with a column density of $N_{\text{H}} > 10^{21}~\text{cm}^{-2}$ can produce broad emission features centred at Balmer, Paschen, and higher-level lines, which would mimic Doppler-broadened hydrogen lines with the velocity width of $\gtrsim 1,000~\text{km}~\text{s}^{-1}$ that could be misinterpreted as signatures of Active Galactic Nuclei, supernovae, or fast stellar winds.
We show that the phase matrix of the Rayleigh and Raman $s$-branch scatterings is identical to that of the Thomson scattering while the Raman $d$-branch scattering is more isotropic, thus the Paschen and higher-level Raman features are depolarized compared to the Balmer features due to the flux contribution from the Raman $d$-branch.
We argue that observations of the line widths, line flux ratios, and linear polarization of multiple optical/IR hydrogen lines are crucial to discriminate between the Raman-scattered broad emission features and Doppler-broadened emission lines.

\end{abstract}




\section{Introduction}
\label{sec:intro}

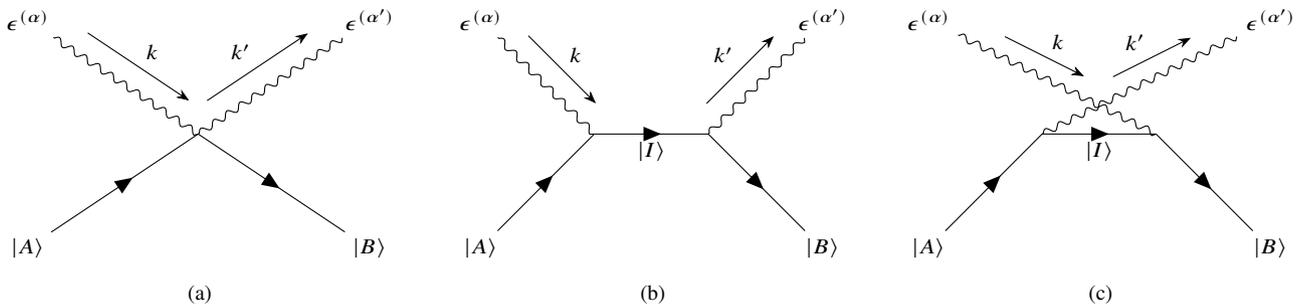
\begin{figure}
\begin{subfigure}{0.33\textwidth}
    \centering
    \begin{tikzpicture}
    \begin{feynman}
        \vertex at (0, 1.5) (i1) {\(\epsilon^{(\alpha)}\)};
        \vertex at (0,-1.5) (i2) {\(|A\rangle\)};
        \vertex at (2.25, 0) (a);
        \vertex at (4.5, 1.5) (f1) {\(\epsilon^{(\alpha')}\)};
        \vertex at (4.5,-1.5) (f2) {\(|B\rangle\)};
    \diagram*{
        (i1) -- [photon, momentum=\(k\)] (a) -- [photon, momentum=\(k'\)] (f1),
        (i2) -- [fermion] (a) -- [fermion] (f2),
    };
    \end{feynman}
    \end{tikzpicture}
    \caption{}
\end{subfigure}
\begin{subfigure}{0.33\textwidth}
    \centering
    \begin{tikzpicture}
    \begin{feynman}
        \vertex at (0, 1.5) (i1) {\(\epsilon^{(\alpha)}\)};
        \vertex at (0,-1.5) (i2) {\(|A\rangle\)};
        \vertex at (1.5, 0) (a);
        \vertex at (3.0, 0) (b);
        \vertex at (4.5, 1.5) (f1) {\(\epsilon^{(\alpha')}\)};
        \vertex at (4.5,-1.5) (f2) {\(|B\rangle\)};
        \diagram*{
            (i1) -- [photon, momentum=\(k\)]  (a),
            (i2) -- [fermion] (a),
            (a)  -- [fermion, edge label'=\(|I\rangle\)] (b),
            (b)  -- [photon, momentum=\(k'\)] (f1),
            (b)  -- [fermion] (f2),
        };
    \end{feynman}
    \end{tikzpicture}
    \caption{}
\end{subfigure}%
\begin{subfigure}{0.33\textwidth}
    \centering
    \begin{tikzpicture}
    \begin{feynman}
        \vertex at (0, 1.5) (i1) {\(\epsilon^{(\alpha)}\)};
        \vertex at (0,-1.5) (i2) {\(|A\rangle\)};
        \vertex at (1.5, 0) (a);
        \vertex at (3.0, 0) (b);
        \vertex at (4.5, 1.5) (f1) {\(\epsilon^{(\alpha')}\)};
        \vertex at (4.5,-1.5) (f2) {\(|B\rangle\)};
        \diagram*{
            (i1) -- [photon, momentum={[arrow shorten=0.3,xshift=-0.3cm,yshift=+0.1cm]\(k\)}]  (b),
            (i2) -- [fermion] (a),
            (a)  -- [fermion, edge label'=\(|I\rangle\)] (b),
            (a)  -- [photon, momentum={[arrow shorten=0.3,xshift=+0.3cm,yshift=+0.1cm]\(k'\)}] (f1),
            (b)  -- [fermion] (f2),
        };
    \end{feynman}
    \end{tikzpicture}
    \caption{}
\end{subfigure}%
\caption{Illustration of the three contributions to the scattering amplitude (Equation~\ref{eqn:kramersheisenberg_nonoriented}). Time flows from left to right. (a) The $\vec{A} \cdot \vec{A}$ `seagull' term. The incident photon is annihilated and the scattered photon is simultaneously created. (b) Resonant process of photon absorption followed by emission. (c) Non-resonant process of photon emission followed by absorption.}
\label{fig:feynmann_diagram}
\end{figure}

In the quantum theory of radiation, the interaction Hamiltonian $H_{\text{int}}$ between the atomic electron in a hydrogen atom and a radiation field is given in Gaussian units as \citep[e.g., Equation~2.94 of][]{sak67}:
\begin{eqnarray}
H_{\text{int}} = -\frac{e}{m_{e}c}\vec{A}\cdot\vec{p} + \frac{e^2}{2m_{e}c^2}\vec{A}\cdot\vec{A},
\end{eqnarray}
where $\vec{A}$ is the electromagnetic field operator and $\vec{p}$ is the momentum operator.
$H_{\text{int}}$ is made up of a linear ($\vec{A}\cdot\vec{p}$) and a quadratic ($\vec{A}\cdot\vec{A}$) term for $\vec{A}$.
According to the time-dependent perturbation theory, in the first order, the linear term gives rise to emission and absorption, and the quadratic term gives rise to the scattering of a single photon (Figure~\ref{fig:feynmann_diagram}a).
Moreover, in the second order, a double action of the linear term also contributes to the scattering of a single photon, which can be split into two processes: one is a photon absorption followed by emission, and the other is a photon emission followed by absorption (Figure~\ref{fig:feynmann_diagram}b and c).

The neutral hydrogen atom in most astrophysical environments is in the ground-state where the atomic electron is in the $1s$ orbital.
The photon scattering by the ground-state hydrogen under the electric dipole approximation happens in two ways (Figure~\ref{fig:energy_diagram}): first, the elastic Rayleigh scattering where the final atomic state is the same as the initial $1s$ state ($1s \rightarrow 1s$), and second, the inelastic Raman scattering where the final atomic state is either $ns$ orbital ($1s \rightarrow ns$; $n \geq 2$) or $nd$ orbital ($1s \rightarrow nd$; $n \geq 3$).
According to the energy conversion law between the initial and final hydrogen atom $\oplus$ photon states, the Raman scattering occurs when the incident photon has a frequency higher than Lyman-$\alpha$.
After the Raman scattering, the hydrogen atomic state is excited to $ns$ or $nd$ state and the frequency of the outgoing photon is reduced as $\hbar\omega' = \hbar\omega - (E_{n} - E_{1})$, where $\omega$ and $\omega'$ are the frequencies of the incident and outgoing photons, respectively, and $E_{1}$ and $E_{n}$ are the eigenenergies of the initial ($1s$) and final ($ns$ or $nd$) atomic states.
In this way, the Raman scattering of the ground-state hydrogen converts the incident UV photons with wavelengths shorter than Lyman-$\alpha$ into optical/infrared (IR) photons.

The first astrophysical example of the hydrogen Raman scattering feature was identified by \cite{sch89}, who ascribed broad ($\sim 20\text{\AA}$, or $\sim 1000~\text{km}~\text{s}^{-1}$) emission features at 6830\AA\ and 7088\AA\ commonly observed in symbiotic stars to Raman-converted \ion{O}{vi} resonance doublet UV lines $\lambda\lambda$1032, 1038.
In the symbiotic star system, the \ion{O}{vi} lines are generated around the mass-accreting white dwarf and the Rayleigh/Raman scattering of the \ion{O}{vi} photons occurs in a thick neutral region near the mass-losing giant.
The wavelength conversion and broadening of the \ion{O}{vi} UV lines (with the intrinsic velocity width of $\sim 150~\text{km}~\text{s}^{-1}$) are naturally explained by the $1s \rightarrow 2s$ Raman scattering, for which the wavelength conversion equations are $\lambda' = \lambda/(1-\lambda/\lambda_{\text{Ly}\alpha})\sim 6.7 \lambda$ and $\Delta \lambda'/\lambda' = (\lambda'/\lambda)\Delta \lambda/\lambda \sim 6.7\Delta\lambda/\lambda$.
Since there are many helium and metal emission lines in the UV wavelength range \citep[e.g.,][]{nus89}, several identifications of broad optical emission features as Raman-scattered UV lines have been reported not only in symbiotic stars \citep{gro93} but also in a variety of UV-bright and neutral material-rich objects, such as planetary nebulae \citep{peq97,cho20}, photodissociation regions \citep{hen21}, and Active Galactic Nuclei (AGN) \citep{nus89}.

Because the Raman scattering cross-section of the ground-state hydrogen exhibits multiple peaks at Lyman resonance wavelengths with Lorentzian-like extended wings, the Raman scattering of featureless UV continuum results in extinguished UV continuum with broad absorption features at Lyman resonances and broad optical/IR emission features at around Balmer, Paschen, and higher-level lines \citep{nus89}.
The broad optical hydrogen emission features with velocity widths of $\gtrsim 1,000~\text{km}~\text{s}^{-1}$ have been observed in symbiotic stars \citep{lee00,cha18} and planetary nebulae \citep{lee00b,arr03}, which can be identified as the signature of the Raman-scattering of the UV continuum \citep[though other possibilities such as electron scattering and fast stellar wind cannot be excluded; e.g.,][]{arr03,sko06}.
Interestingly, \cite{dop16} discover broad H$\alpha$ wing of $\sim 7600~\text{km}~\text{s}^{-1}$ in the Orion Nebula and in five \ion{H}{ii} regions in the Large and the Small Magellanic Clouds, which might suggest that the hydrogen Raman scattering features are common in star-forming regions \citep[see also][]{hen21,zag23}.
These broad optical hydrogen emission features could be so broad ($\gtrsim 10,000~\text{km}~\text{s}^{-1}$) that they might mimic Doppler-broadened emission lines from astrophysical objects with extremely high kinetic energy, such as supernovae (specifically type IIn supernovae), fast stellar winds, and AGN broad line region \citep[e.g.,][]{nus89,lee98,cha15,hat23}, which would lead to wrong physical interpretations of the nature of observing objects.
One of the purposes of this paper is to investigate the observational properties of the Raman features helpful in discriminating between the Raman features and Doppler-broadened emission lines.

As we shall see in Section~\ref{sec:analysis}, differential cross-sections for the Rayleigh and Raman scattering can be evaluated using the Kramers-Heisenberg-Waller dispersion formula \citep{kra25,dir27,wal28}\footnote{The Kramers-Heisenberg dispersion formula (Equation~\ref{eqn:kramersheisenberg}) was first derived by \cite{kra25} and afterward completed with its first term (`seagull' term) by \cite{wal28}.}.
The calculations of the scattering cross-sections of the ground-state hydrogen atom can be separated into radial and angular parts according to the Wigner-Eckart theorem.
The radial part is the overlap integral of the radial wavefunctions of the hydrogen atom, which have been discussed in many papers and textbooks but whose analytic solutions expressed in terms of elementary functions have only been explicitly derived for the first few low-level transitions.
As discussed below, such analytic solutions of the overlap integral can be obtained through algebraic manipulation of general formulae.
The angular part involves a sum over the magnetic substates of the hydrogen atom, with which the scattering phase function is determined for each scattering branch.
Here we stress the importance of the scattering phase function (or more generally, the scattering phase matrix) because it defines the polarization properties of the scattered photon, which leave unique observable consequences in spectropolarimetric data.
It is not widely recognized in the astronomy community that the scattering branches of the ground-state hydrogen atom each have different phase matrices; for example, the scattering phase matrix of ground-state hydrogen of any scattering branch is approximated to be the Rayleigh scattering phase matrix in the widely used hydrogen scattering radiative transfer code STaRS \citep[][]{cha15,cha20}, and with this approximation the calculated polarization degree can be overestimated due to the ignorance of depolarization effects arising from the contribution of a certain Raman scattering branch.
It is worth examining the polarization characteristics of the hydrogen scattering to provide a diagnostic to discriminate between the Raman-scattered broad emission features and Doppler-broadened emission lines.

In this paper, we will derive general, explicit formulae for the calculations of the Rayleigh and Raman scattering cross-sections of the ground-state hydrogen atom, and examine polarization properties induced by the scattering by introducing scattering phase functions and phase matrices.
These cross-sections have a wide variety of applications in astrophysical phenomena involving the hydrogen Rayleigh and Raman scattering, such as symbiotic stars, planetary nebulae, star-forming regions, Active Galactic Nuclei, and Damped Lyman-$\alpha$/Lyman limit absorbers \citep[e.g.,][and references therein]{nus89,cha15,mor16,dop16,hen21,zag21,mir22}. 
The formulae presented in this work will serve as a basis for future more detailed radiative transfer simulation studies.

In Section~\ref{sec:analysis}, the Kramers-Heisenberg-Waller formula for the ground-state hydrogen atom is described, and explicit expressions for the differential and total cross-sections of the Rayleigh and Raman scattering are derived along with their scattering phase functions and phase matrices.
In Section~\ref{sec:discussion}, the exact Rayleigh scattering cross-section based on the Kramers-Heisenberg\text{-Waller} formula is compared with various approximate formulae found in the literature and observational properties of the optical/IR emission features created by the Raman scattering of the UV photons are discussed.
Finally, we summarize the conclusions in Section~\ref{sec:conclusion}.
Throughout the paper, we use 
$\lambda_{\text{LyLimit}} = R_{\text{H}}^{-1} = 911.763$\AA\ as the inverse of the Rydberg constant for hydrogen\footnote{$R_{\text{H}} = 10967758.3$~m${}^{-1}$ is related to the Rydberg constant for heavy atoms $R_{\infty} = 10973731.6$~m${}^{-1}$ as $R_{\text{H}} = R_{\infty}\mu/m_{e} = R_{\infty}m_p/(m_p+m_e)$ where $m_{p}$, $m_{e}$, and $\mu$ are the proton mass, electron mass, and reduced mass of the hydrogen atom.}, $\alpha^{-1} = 137.036$ as the fine-structure constant, and
$c=299792~\text{km}~\text{s}^{-1}$ as the speed of light.
The classical radius of the electron $r_{0}$ and the Bohr radius $a_{0}$ are respectively given as $r_{0} = \sqrt{3\sigma_{T}/8\pi}$ and $a_{0} = \alpha^{-2}r_{0} = \alpha (4\pi)^{-1} \lambda_{\text{LyLimit}}$, where $\sigma_{T} = (8\pi/3)r_{0}^2 = (8\pi/3)\alpha^4a_{0}^2 = 6.65246\times 10^{-25}~\text{cm}^2$ is the Thomson scattering cross section \citep{roh22}.

\section{Cross-sections of the Rayleigh and Raman scattering by a ground-state hydrogen atom}
\label{sec:analysis}

\subsection{Kramers-Heisenberg-Waller formula}

\begin{figure}
\center{
\includegraphics[clip, width=3.4in]{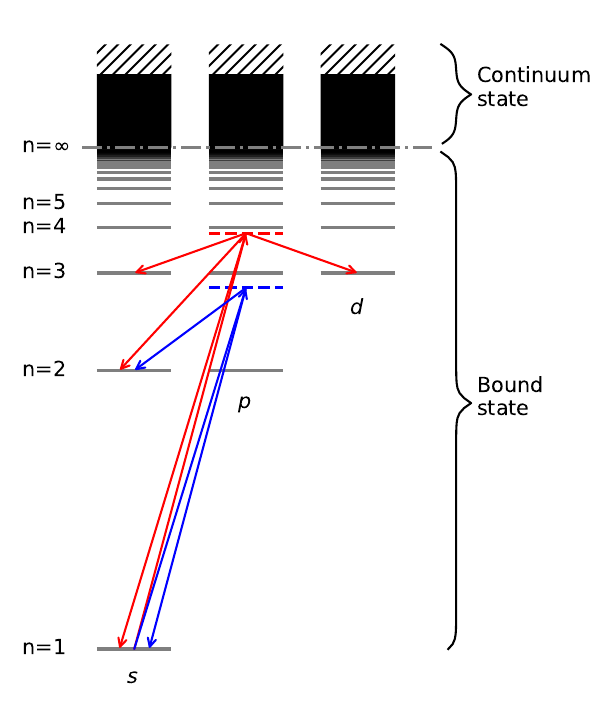}
}\vspace{0cm}
 \caption{Illustration of the energy levels of the bound-states and continuum-states of the hydrogen atom. The horizontal axis indicates $s$, $p$, and $d$ orbitals, and the vertical axis indicates the principal quantum number or the energy level; $E_{n}=-1/2n^2$ for the bound-state and $E_{n'}=1/2n'^2$ for the continuum-state ($n$: positive integer, $n'$: positive real number). The magnetic substates are assumed to be degenerated. The dashed horizontal lines denote the excited intermediate states $|I\rangle$. The arrows denote various scattering paths from the ground $1s$ state $|A\rangle$, including Rayleigh ($1s \rightarrow 1s$) and Raman ($1s \nrightarrow 1s$) scattering. The final state $|B\rangle$ is the $s$ orbital when the incident photon energy $\hbar\omega$ is less than $E_{\text{Ly}\beta} = E_{3}-E_{1} = 12.09~\text{eV}$ (scattering into wavelengths around Ly$\alpha$, Ly$\beta$, and H$\alpha$), while the final state can fall into either $s$ or $d$ orbitals when $\hbar\omega > E_{\text{Ly}\beta}$.}
 \label{fig:energy_diagram}
\end{figure}

The focus of this paper is the photon scattering process by an atomic electron in a hydrogen atom, which is the second-order perturbation term after the first-order (absorption and emission of a resonant photon) perturbation term (Figure~\ref{fig:feynmann_diagram}).
The scattering amplitude of this process under the electric dipole approximation is given by the Kramers-Heisenberg-Waller matrix element $M_{BA}$ and the Kramers-Heisenberg-Waller formula gives the differential scattering cross-section for a single atom as $d\sigma/d\Omega = r_{0}^2\left(\omega'/\omega\right)|M_{BA}|^2$ \citep[e.g.,][]{pla34,pen69,sak67,sas69,lee97,hub14}, where $A$ and $B$ stand for initial and final states of the hydrogen atom, and the angular frequencies of the incident and scattered photons are denoted by $\omega = 2\pi c/\lambda = c|\vec{k}|$ and $\omega' = 2\pi c/\lambda' = c|\vec{k}'|$, respectively.
Polarization basis vectors perpendicular to $\vec{k}$ and $\vec{k}'$ vectors are defined as real unit vectors $\vec{\epsilon}^{(\alpha)}$ ($\alpha=1,2$) and $\vec{\epsilon}^{(\alpha')}$ ($\alpha'=1,2$) \citep{sak67}.
Hereafter, $|A\rangle$, $|B\rangle$, and $|I\rangle$ indicate the initial, final, and intermediate atomic energy eigenstates whose energy eigenvalues are $E_{A}$, $E_{B}$, and $E_{I}$, respectively.
$\vec{x} = (x, y, z)$ and $\vec{p} = (p_x, p_y, p_z)$ denote the position and momentum operators acting on the atomic state, and the matrix element of an operator $O$ is represented as $(O)_{BA} \equiv \langle B|O|A \rangle$. 
Using these notations, the Kramers-Heisenberg-Waller matrix element $M_{BA}$ can be written as follows:
\begin{eqnarray}
M_{BA} &=& \delta_{AB}\vec{\epsilon}^{(\alpha)} \cdot \vec{\epsilon}^{(\alpha')} - \frac{1}{m_e} \SumInt_{I}\left[ \frac{(\vec{p}\cdot\vec{\epsilon}^{(\alpha')})_{BI}(\vec{p}\cdot\vec{\epsilon}^{(\alpha)})_{IA}}{E_{I}-E_{A}-\hbar\omega} + \frac{(\vec{p}\cdot\vec{\epsilon}^{(\alpha)})_{BI}(\vec{p}\cdot\vec{\epsilon}^{(\alpha')})_{IA}}{E_{I}-E_{A}+\hbar\omega'} \right] \nonumber\\
 &=&  \delta_{AB}\vec{\epsilon}^{(\alpha)} \cdot \vec{\epsilon}^{(\alpha')} - \frac{m_e}{\hbar^2} \SumInt_{I} (E_{I}-E_{B})(E_{I}-E_{A}) \left[ \frac{(\vec{x}\cdot\vec{\epsilon}^{(\alpha')})_{BI}(\vec{x}\cdot\vec{\epsilon}^{(\alpha)})_{IA}}{E_{I}-E_{A}-\hbar\omega} + \frac{(\vec{x}\cdot\vec{\epsilon}^{(\alpha)})_{BI}(\vec{x}\cdot\vec{\epsilon}^{(\alpha')})_{IA}}{E_{I}-E_{A}+\hbar\omega'} \right].
\label{eqn:kramersheisenberg}
\end{eqnarray}
An identity $\langle B|p_{i}|A\rangle$ = $\frac{m}{i \hbar}\langle B| [H_{0}, x_{i}]|A\rangle$ = $\frac{m}{i \hbar}(E_{B}-E_{A})\langle B|x_{i}|A\rangle$ is used to get the second line of Equation~\ref{eqn:kramersheisenberg}, where $H_{0}$ is the Hamiltonian of the unperturbed atomic system \citep[e.g.,][]{lee97}.
The three terms in Equation~\ref{eqn:kramersheisenberg} represent the second order `seagull' term and the double action of the first order absorption and emission of photons by the atom, and the later two terms involve intermediate virtual states of the atom $|I\rangle$ (Figure~\ref{fig:feynmann_diagram}).
Note that the intermediate state $|I\rangle$ represents not only bound-states but also free-states \citep{sas69,nus89}; the former are the discrete eigenstates of negative energy and the latter are the free electron states of positive energy in the proton's Coulomb potential (Figure~\ref{fig:energy_diagram}).
The symbol $\SumInt$ in Equation~\ref{eqn:kramersheisenberg} thus means to take a sum over the discrete bound-states and to perform an integral over the continuous free-states.

The hydrogen atomic state is represented by $|n,l,m\rangle$, where $n$, $l$, and $m$ are the principal, azimuthal, and magnetic quantum numbers. 
The corresponding wavefunction is $\psi_{nlm} = R_{nl}(r)Y_{lm}(\theta,\phi)$, where $R_{nl}(r)$ is the radial wavefunction of the hydrogen atom (Appendix~\ref{sec:overlap_integrals}) and $Y_{lm}(\theta,\phi)$ is the spherical harmonics.
For an electric dipole transition between $l,m$ and $l',m'$, the selection rule is $l'= l \pm 1$.
Also, because the perturbing Hamiltonian $H_{\text{int}}$ does not contain any spin operators, the spin quantum number $m_{s}$ cannot change during the transition ($m_{s}'=m_{s}$).
We restrict ourselves to the case where the initial hydrogen atomic state is in its ground-state $1s$ ($|A\rangle = |1,0,0\rangle$). 
Thus the intermediate state must have the $p$ orbital ($l=1$; $|I\rangle = |n_{I},1,m_{I}\rangle$) and the final state must have either $s$ or $d$ orbitals ($l=0$ or $2$; $|B\rangle = |n_{B}, 1\pm 1,m_{B}\rangle$) due to the electric dipole selection rule.
The energy levels of the initial and final bound-states are $E_{A} = -1/2$ and $E_{B} = -1/(2n_{B}^2)$, respectively, in Hartree atomic units (Figure~\ref{fig:energy_diagram}).
The energy level of the intermediate state $E_{I}$ is $E_{I} = -1/(2n_{I}^2) < 0$ for bound-states and $E_{I} = 1/(2n_{I}^2) > 0$ for free-states in Hartree atomic units.

The energy conservation is satisfied between the initial and final states (not the intermediate state): 
\begin{eqnarray}
E_{A} + \hbar \omega = E_{B} + \hbar \omega'.
\end{eqnarray}
From the energy conservation law we obtain 
\begin{eqnarray}
\hbar\omega' = \hbar\omega - (E_{{B}}-E_{A}) = \hbar\omega - \frac{1-n_{B}^{-2}}{2} = \frac{\lambda_{\text{LyLimit}}}{2\lambda} - \frac{1-n_{B}^{-2}}{2}
\end{eqnarray}
where $\hbar\omega_{\text{LyLimit}} = hc/\lambda_{\text{LyLimit}}=1/2$ in Hartree atomic units, and the Raman wavelength conversion and line-broadening formulae in the case of the Raman scattering ($n_{B} \geq 2$) follow:
\begin{eqnarray}
\frac{\lambda'}{\lambda} &=& \frac{\omega}{\omega'} =  \frac{1}{1-\frac{\lambda}{(1-n_{B}^{-2})^{-1}\lambda_{\text{LyLimit}}}} \label{eqn:wavelength_conversion}\\
\frac{d\lambda'}{\lambda'} &=& \left(\frac{\lambda'}{\lambda}\right)\frac{d\lambda}{\lambda}.  \label{eqn:differential_wavelength_conversion}
\end{eqnarray}
By definition, in the case of the Rayleigh scattering ($n_{B}=1$), the wavelength conversion does not happen ($\lambda' = \lambda$).
For the hydrogen atom, an excited final state left after a Raman scattering ($n_{B} \geq 2$) eventually decays into the ground-state via cascade or two-photon decays that would contribute to narrow emission lines of Rydberg series or continuum above the Balmer limit \citep[e.g.,][]{ost06}, which are out of the scope of this paper.

The differential cross-section for nonoriented atoms should be defined by taking the average over magnetic quantum numbers of the initial state $m_{A}$ (now $l_{A}=0$ thus $\frac{1}{2l_{A}+1}\sum_{m_{A}} = 1$) and the sum over that of the final state $m_{B}$ of the single-atom differential scattering cross-section \citep[e.g.,][]{pla34,pen69}.
Since we can assume that the energy levels of the hydrogen atom degenerate both on the azimuthal quantum number $l$ and the magnetic quantum number $m$, the summation over $I$ can be split into the summations over $m_{I}=\{-1,0,1\}$ and the principal quantum number $n_{I}$, $\SumInt_{I} = \SumInt_{n_{I}}\sum_{m_{I}}$.
Thus, for a final state specified as $|B\rangle = |n_{B},l_{B},m_{B}\rangle$, the differential cross-section for nonoriented hydrogen atoms ($Z=1$) can be explicitly evaluated as
\begin{footnotesize}
\begin{eqnarray}
\frac{d\sigma}{d\Omega} &=& r_{0}^2 \left(\frac{\omega'}{\omega}\right) \sum_{m_{B}} |M_{BA}|^2 \nonumber\\
&=& r_{0}^2 \left(\frac{\omega'}{\omega}\right) \sum_{m_{B}} \left| \delta_{AB}\vec{\epsilon}^{(\alpha)} \cdot \vec{\epsilon}^{(\alpha')} - \frac{m_e}{\hbar^2} \SumInt_{n_{I}} \left( (E_{I}-E_{B})(E_{I}-E_{A}) \left[ \frac{\sum_{m_{I}}(\vec{x}\cdot\vec{\epsilon}^{(\alpha')})_{BI}(\vec{x}\cdot\vec{\epsilon}^{(\alpha)})_{IA}}{E_{I}-E_{A}-\hbar\omega} + \frac{\sum_{m_{I}}(\vec{x}\cdot\vec{\epsilon}^{(\alpha)})_{BI}(\vec{x}\cdot\vec{\epsilon}^{(\alpha')})_{IA}}{E_{I}-E_{A}+\hbar\omega'} \right] \right) \right|^2.
\label{eqn:kramersheisenberg_nonoriented}
\end{eqnarray}
\end{footnotesize}
Equation~\ref{eqn:kramersheisenberg_nonoriented} represents the differential scattering cross-section between the photon modes with ($k, \vec{\epsilon}^{(\alpha)}$) and ($k', \vec{\epsilon}^{(\alpha')}$).
The elastic Rayleigh scattering is defined as the process where $|A\rangle = |B\rangle$ and $\omega=\omega'$, and the inelastic Raman scattering is the case when $|A\rangle \neq |B\rangle$ and $\omega \neq \omega'$ (\citealt{nus89,hub14}; Figure~\ref{fig:energy_diagram}).
In the case of the Rayleigh scattering, a slight manipulation of Equation~\ref{eqn:kramersheisenberg_nonoriented} gives us a more convenient form of the formula \citep[Equation~2.164 of][]{sak67}:
\begin{eqnarray}
\left. \frac{d\sigma}{d\Omega}\right\vert_{\text{Rayleigh}} &=& r_{0}^2 \sum_{m_{B}} \left| -\frac{m_e}{\hbar^2} \SumInt_{n_I}(E_{I}-E_{A})\left[ \frac{\hbar \omega \sum_{m_{I}} (\vec{x}\cdot\vec{\epsilon}^{(\alpha')})_{AI}(\vec{x}\cdot\vec{\epsilon}^{(\alpha)})_{IA}}{E_{I}-E_{A}-\hbar\omega} - \frac{\hbar\omega \sum_{m_{I}} (\vec{x}\cdot\vec{\epsilon}^{(\alpha)})_{AI}(\vec{x}\cdot\vec{\epsilon}^{(\alpha')})_{IA}}{E_{I}-E_{A}+\hbar\omega} \right]\right|^2 \label{eqn:rayleigh} 
\end{eqnarray}
where the three terms in Equation~\ref{eqn:kramersheisenberg_nonoriented} combine into the two terms in Equation~\ref{eqn:rayleigh} \citep{lee97,yoo02}.

Explicit calculations of the Kramers-Heisenberg-Waller matrix elements can be done with standard algebraic techniques that can be found in many literature \citep[e.g.,][]{pla34,kar61,bur65,pen69,ber82}.
Although there has been a vast amount of earlier works presenting the exact calculations of the Rayleigh/Raman cross sections using various methods \citep[e.g.,][and references therein]{mit62,zer64b,gav67,rap68,zon69,sas69,gon71,isl89,nus89,sad92,roh22}, I found there has been no literature fully discussing the scattering process (cross sections and phase matrices) by the gound-state hydrogen atom in an explicit manner, which is of particular interest for the astronomy community \citep[see the relevant discussion in][]{roh22}.
Although it is not very difficult to derive the scattering properties of the hydrogen atoms from general formulae given in the quantum mechanics textbooks/literature (as shown in this paper), it is worthwhile to present explicit expressions for the cross sections and phase matrices that are readily be used in, e.g., the radiative transfer simulations for future reference.
For this reason, here we outline the derivations of the explicit expressions for the differential and total Rayleigh/Raman cross sections of the hydrogen atom and consequently present full descriptions of scattering phase matrices of these scattering processes.
For completeness, some equations and formulae necessary to calculate the cross sections are given in Appendices~\ref{sec:wigner-eckart} (the Wigner-Eckart theorem) and \ref{sec:overlap_integrals} (overlap integrals).

\subsection{Explicit expressions for the scattering cross-sections}

\begin{figure}
\center{
\includegraphics[clip, width=6.4in]{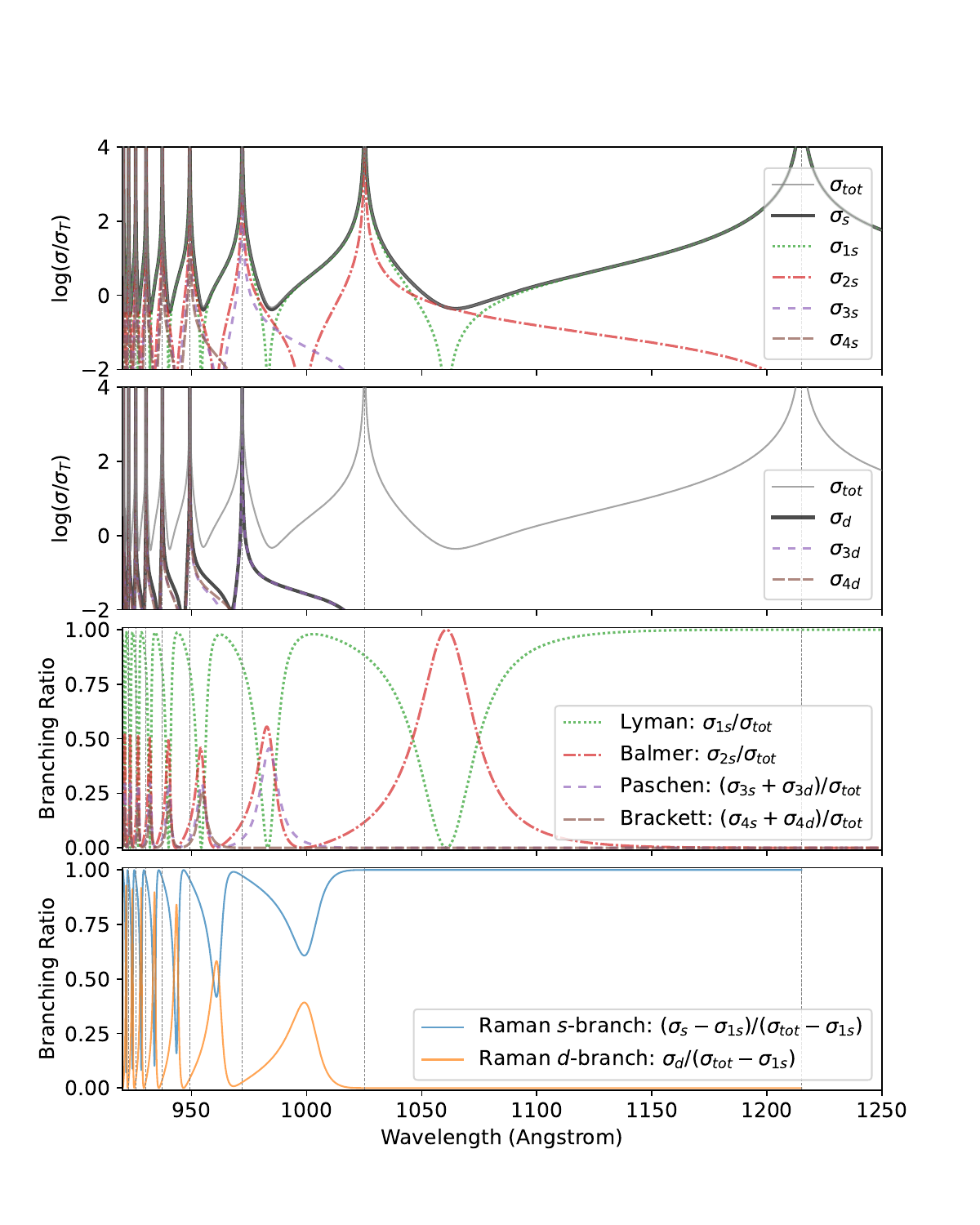}
}\vspace{0cm}
 \caption{ Top two panels: Rayleigh and Raman scattering cross-sections of the ground-state hydrogen atom up to $n_{B}=4$ as a function of the wavelength of the incident photon. The vertical lines indicate the Lyman resonance wavelengths. $\sigma_{n_{B}s}$ and $\sigma_{n_{B}d}$ respectively denote the cross-sections where the final state $|B\rangle$ is $s$ and $d$ orbitals with the principal quantum number $n_{B}$.  $\sigma_{\text{tot}}$, $\sigma_{s}$, and $\sigma_{d}$ are the sums of the cross-sections up to $n_{B}=4$: $\sigma_{\text{tot}} \equiv \sigma_{s} + \sigma_{d} = \sum_{n_{B}=1}^{4}\sigma_{n_{B}s} + \sum_{n_{B}=3}^{4}\sigma_{n_{B}d}$. 
 Bottom two panels: the scattering branching ratios. The branching ratios to $n_{B}=1$ (Lyman=Rayleigh), $2$ (Balmer), $3$ (Paschen), and $4$ (Brackett) are defined as $\sigma_{1s}/\sigma_{\text{tot}}$, $\sigma_{2s}/\sigma_{\text{tot}}$, $(\sigma_{3s}+\sigma_{3d})/\sigma_{\text{tot}}$, and $(\sigma_{4s}+\sigma_{4d})/\sigma_{\text{tot}}$, respectively. The total Raman scattering cross-section is $\sigma_{\text{tot}}-\sigma_{1s}$, and the Raman $s$-branch and $d$-branchs are defined as $(\sigma_{s}-\sigma_{1s})/(\sigma_{\text{tot}}-\sigma_{1s})$ and $\sigma_{d}/(\sigma_{\text{tot}}-\sigma_{1s})$, respectively.}
 \label{fig:crosssections}
\end{figure}

\subsubsection{$1s \rightarrow 1s$ Rayleigh scattering and $1s \rightarrow n_{B}s$ Raman scattering cross-sections}
\label{sec:1s_ns_transition}

When the final state is the $s$ orbital ($l_{B}=0$, $m_{B}=0$), from the Wigner-Eckart theorem (Appendix~\ref{sec:wigner-eckart}) we obtain \citep[Equation~\ref{eqn:sum_of_matrix}; e.g.,][]{lee97,lee13}: 
\small
\begin{eqnarray}
\sum_{m_{I}=-1,0,1}^{}(\vec{x} \cdot \vec{\epsilon}^{(\alpha')})_{BI}(\vec{x} \cdot \vec{\epsilon}^{(\alpha)})_{IA} = \sum_{m_{I}=-1,0,1}^{}(\vec{x} \cdot \vec{\epsilon}^{(\alpha)})_{BI}(\vec{x} \cdot \vec{\epsilon}^{(\alpha')})_{IA} = \frac{1}{3} R_{n_{I}1}^{10} R_{n_{B}0}^{n_{I}1} \left( \vec{{\epsilon}}^{(\alpha')} \cdot \vec{{\epsilon}}^{(\alpha)} \right)
\label{eqn:matrix_s1}
\end{eqnarray}
\normalsize
and the differential scattering cross-section (Equation~\ref{eqn:kramersheisenberg_nonoriented}) for transitions with $l_{B}=0$ can be shown to be proportional to the square of the inner product of the polarization basic vectors $(\vec{\epsilon}^{(\alpha)} \cdot \vec{\epsilon}^{(\alpha')})^2$:
\begin{align}
\left.\frac{d\sigma}{d\Omega}\right|_{|B\rangle = |n_{B},0\rangle} &=& r_{0}^2 \left(\frac{\omega'}{\omega}\right) \left(\vec{\epsilon}^{(\alpha)} \cdot \vec{\epsilon}^{(\alpha')}\right)^2 \left| \delta_{AB} - \frac{m_e}{\hbar^2} \SumInt_{n_{I}} \left( \frac{R_{n_{B}0}^{n_{I}1} R_{n_{I}1}^{10}(E_{I}-E_{B})(E_{I}-E_{A})}{3} \left[ \frac{1}{E_{I}-E_{A}-\hbar\omega} + \frac{1}{E_{I}-E_{A}+\hbar\omega'} \right] \right) \right|^2,
\label{eqn:differential_crosssection_nB1}
\end{align}
where $R_{n\:l}^{n'\:l'}$ represents the overlap integral of the radial wavefunctions of the hydrogen atom $R_{nl}$ and $R_{n'l'}$, $R_{n\:l}^{n'\:l'} \equiv \int_{0}^{\infty} r^3R_{nl}R_{n'l'}dr$.
The bound-bound and bound-free overlap integrals, or reduced matrix elements of the electric dipole moment operator of rank 1, can be analytically evaluated by using the well-known \cite{gor29}’s formulae (Equations~B.1 and C.1 in \citealt{kar61}; see also \citealt{hat46,bet57,gol68}).
The detailed descriptions of the overlap integrals are presented in Section~\ref{sec:overlap_integrals}, and the analytic expressions of the overlap integrals are summarized in Tables~\ref{tbl:radial_integrals_boundbound} and \ref{tbl:radial_integrals_boundfree}.

We can see that, in this particular case where both of the initial and final states have zero orbital angular momentum, the angular distribution of the scattering is of a {\it scalar} nature \citep[][]{pla34,pen69,ber82}, like the classical Thomson scattering.
As we will discuss in detail in Section~\ref{sec:phase_function}, the phase function of the {\it scalar} scattering $\propto (\vec{\epsilon}^{(\alpha)} \cdot \vec{\epsilon}^{(\alpha')})^2$ is the same as the Rayleigh's phase function \citep[e.g.,][]{yoo02,ali15}.

Recall that the energy levels of the bound-states are $E_{A} = -1/2$ and $E_{B} = -1/(2n_{B}^2)$, and $E_{I} = -1/(2n_{I}^2) < 0$ for bound-states and $E_{I} = 1/(2n_{I}^2) > 0$ for free-states in the Hartree atomic units ($\hbar=m_{e}=1$), and the energy conversion law relates the photon's wavelength and hydrogen's eigenenergy (Equation~\ref{eqn:wavelength_conversion}).
By the direct substitutions of the analytic formulae of the overlap integrals of the radial wavefunctions in Tables~\ref{tbl:radial_integrals_boundbound} and \ref{tbl:radial_integrals_boundfree} into Equation~\ref{eqn:differential_crosssection_nB1}, we obtain the explicit expression of the Rayleigh scattering cross-section ($n_{B} = 1$, i.e., $|A\rangle=|B\rangle = |1,0,0\rangle$) as:
\begin{eqnarray}
\left.\frac{d\sigma}{d\Omega}\right|_{|B\rangle = |1,0\rangle} &=& r_{0}^2 \left(\vec{\epsilon}^{(\alpha)} \cdot \vec{\epsilon}^{(\alpha')}\right)^2\nonumber\\
&\times& \left| 1 - \sum_{n_{I} \geq 2} \left( 2^{7}3^{-1}n_{I}^{3}(n_{I}^2-1)^{-3} \left( \frac{n_{I}-1}{n_{I}+1} \right)^{2n_{I}} \left[ \frac{1}{1-\frac{1}{n_{I}^2}-\frac{\lambda_{\text{LyLimit}}}{\lambda}} + \frac{1}{1-\frac{1}{n_{I}^2}+\frac{\lambda_{\text{LyLimit}}}{\lambda}} \right] \right) \right.\nonumber\\
&-&\left. \int_{0}^{\infty}dn_{I}' \left( \frac{2^{7}3^{-1}n_{I}'^{3}(n_{I}'^2+1)^{-3}  e^{-4n_{I}'\text{arccot}(n_{I}')} }{(1-\exp(-2\pi n_{I}'))} \left[ \frac{1}{1+\frac{1}{n_{I}'^2}-\frac{\lambda_{\text{LyLimit}}}{\lambda}} + \frac{1}{1+\frac{1}{n_{I}'^2}+\frac{\lambda_{\text{LyLimit}}}{\lambda}} \right] \right) \right|^2, \nonumber\\
&=& r_{0}^2 \left(\vec{\epsilon}^{(\alpha)} \cdot \vec{\epsilon}^{(\alpha')}\right)^2\nonumber\\
&\times& \left| \sum_{n_{I} \geq 2} \left( 2^{7}3^{-1}n_{I}^{5}(n_{I}^2-1)^{-4} \left( \frac{n_{I}-1}{n_{I}+1} \right)^{2n_{I}} \left[ \frac{\frac{\lambda_{\text{LyLimit}}}{\lambda}}{1-\frac{1}{n_{I}^2}-\frac{\lambda_{\text{LyLimit}}}{\lambda}} - \frac{\frac{\lambda_{\text{LyLimit}}}{\lambda}}{1-\frac{1}{n_{I}^2}+\frac{\lambda_{\text{LyLimit}}}{\lambda}} \right] \right) \right.\nonumber\\
&+&\left. \int_{0}^{\infty}dn_{I}' \left( \frac{2^{7}3^{-1}n_{I}'^{5}(n_{I}'^2+1)^{-4}  e^{-4n_{I}'\text{arccot}(n_{I}')} }{(1-\exp(-2\pi n_{I}'))} \left[ \frac{\frac{\lambda_{\text{LyLimit}}}{\lambda}}{1+\frac{1}{n_{I}'^2}-\frac{\lambda_{\text{LyLimit}}}{\lambda}} - \frac{\frac{\lambda_{\text{LyLimit}}}{\lambda}}{1+\frac{1}{n_{I}'^2}+\frac{\lambda_{\text{LyLimit}}}{\lambda}} \right] \right) \right|^2,
\label{eqn:crosssection_rayleigh}
\end{eqnarray}
where the second equality follows from Equation~\ref{eqn:rayleigh}.
Here $\SumInt$ in Equation~\ref{eqn:differential_crosssection_nB1} is explicitly split into the sum over discrete $n_{I}$ and the integral over continuous $n_{I}'$.
The integrand of the bound-free part has a singularity at $n_{I} = (\lambda_{\text{LyLimit}}/\lambda-1)^{-1/2}$ when $\lambda < \lambda_{\text{LyLimit}}$, thus in general the integral in Equation~\ref{eqn:crosssection_rayleigh} should be regarded as the principal value integral \citep[e.g.,][]{sas69,gru16,mcn18}.
Also, the bound-bound part involves unphysical resonances at $\lambda = \lambda_{\text{LyLimit}}/(1-n_{I}^{-2})$ due to the neglect of the finite lifetimes of the intermediate states, i.e., radiation damping \citep[][]{sak67,mcn18}.
Note that detailed treatments of the resonances become important only at the very centres of the resonant lines \footnote{ For example, the radiation damping term for Lyman-$\alpha$ in Equation~\ref{eqn:peebles} becomes important at a wavelength range of $|\lambda - \lambda_{\text{Ly}\alpha}| = (\Gamma_{12}\lambda_{\text{Ly}\alpha}^2)/(4 \pi c) \sim 2.5 \times 10^{-5}\text{\AA}$, which is smaller than the Lamb shift and fine-structure splitting \citep[e.g.,][]{roh22} that are beyond the scope of this paper. }.
Since we are interested in the wavelength range at $\lambda > \lambda_{\text{LyLimit}}$ and off-resonant scattering, we do not give special treatment to these singularities thereafter.

In the same way, the $1s \rightarrow n_{B}s$ Raman scattering cross-sections ($n_{B} \geq 2$) can explicitly be expressed as:
\small
\begin{eqnarray}
\left.\frac{d\sigma}{d\Omega}\right|_{|B\rangle = |n_{B},0\rangle} &=& r_{0}^2 \left[ 1-\left(1-n_{B}^{-2}\right)\frac{\lambda}{\lambda_{\text{LyLimit}}} \right]\left(\vec{\epsilon}^{(\alpha)} \cdot \vec{\epsilon}^{(\alpha')}\right)^2 \nonumber\\
&\times& \left| \sum_{n_{I}} \left( 2^{3}3^{-1}n_{B}^{-2}n_{I}^{-1/2} (n_{I}^2-1)^{-3/2} (n_{I}^2-n_{B}^2) \left( \frac{n_{I}-1}{n_{I}+1} \right)^{n_{I}} R_{n_{B}0}^{n_{I}1} \left[ \frac{1}{1-\frac{1}{n_{I}^2}-\frac{\lambda_{\text{LyLimit}}}{\lambda}} + \frac{1}{n_{B}^{-2}-\frac{1}{n_{I}^2}+\frac{\lambda_{\text{LyLimit}}}{\lambda}} \right] \right) \right. \nonumber\\
&+& \left. \int_{0}^{\infty} dn_{I}' \left( \frac{2^{3}3^{-1}n_{B}^{-2}n_{I}'^{-1/2} (n_{I}'^2+1)^{-3/2} (n_{I}'^2+n_{B}^2) e^{-2n_{I}'\text{arccot}(n_{I}')} R_{n_{B}0}^{n_{I}'1} }{\sqrt{1-\exp(-2\pi n_{I}')}} \left[ \frac{1}{1+\frac{1}{n_{I}'^2}-\frac{\lambda_{\text{LyLimit}}}{\lambda}} + \frac{1}{n_{B}^{-2}+\frac{1}{n_{I}'^2}+\frac{\lambda_{\text{LyLimit}}}{\lambda}} \right] \right) \right|^2
\label{eqn:crosssection_raman_s0}
\end{eqnarray}
\normalsize
that becomes 0 at $\lambda > \lambda_{\text{LyLimit}}/(1-n_{B}^{-2})$ due to the energy conservation law between the initial and final states.
From Equation~\ref{eqn:crosssection_raman_s0}, we obtain the Raman scattering cross-sections for $n_{B} = 2$, $3$, and $4$ as:
\small
\begin{eqnarray}
\left.\frac{d\sigma}{d\Omega}\right|_{|B\rangle = |2,0\rangle} &=& r_{0}^2 \left(1-\frac{3}{4}\frac{\lambda}{\lambda_{\text{LyLimit}}}\right) \left(\vec{\epsilon}^{(\alpha)} \cdot \vec{\epsilon}^{(\alpha')}\right)^2 \nonumber\\
&\times& \left| \sum_{n_{I} \geq 3} \left( 2^{19/2}3^{-1}n_{I}^{3}(n_{I}^2-1)^{-1} (n_{I}^2-4)^{-2} \left(\frac{n_{I}-1}{n_{I}+1}\right)^{n_{I}}\left(\frac{n_{I}-2}{n_{I}+2}\right)^{n_{I}} \left[ \frac{1}{1-\frac{1}{n_{I}^2}-\frac{\lambda_{\text{LyLimit}}}{\lambda}} + \frac{1}{\frac{1}{4}-\frac{1}{n_{I}^2}+\frac{\lambda_{\text{LyLimit}}}{\lambda}} \right] \right) \right. \nonumber \\
&+& \left. \int_{0}^{\infty} dn_{I}' \left( \frac{ 2^{19/2}3^{-1}n_{I}'^{3}(n_{I}'^2+1)^{-1} (n_{I}'^2+4)^{-2} e^{-2n_{I}'\left(\text{arccot}(n_{I}')+\text{arccot}(n_{I}'/2)\right)}}{\left(1-\exp(-2\pi n_{I}') \right)} \left[ \frac{1}{1+\frac{1}{n_{I}'^2}-\frac{\lambda_{\text{LyLimit}}}{\lambda}} + \frac{1}{\frac{1}{4}+\frac{1}{n_{I}'^2}+\frac{\lambda_{\text{LyLimit}}}{\lambda}} \right] \right) \right|^2 \nonumber.\\
\left.\frac{d\sigma}{d\Omega}\right|_{|B\rangle = |3,0\rangle} &=& r_{0}^2 \left(1-\frac{8}{9}\frac{\lambda}{\lambda_{\text{LyLimit}}}\right) \left(\vec{\epsilon}^{(\alpha)} \cdot \vec{\epsilon}^{(\alpha')}\right)^2 \nonumber\\
&\times&\left| \sum_{n_{I}=2; n_{I}\geq 4} \left( 2^{7}3^{1/2}n_{I}^{3}(n_{I}^2-1)^{-1}(n_{I}^2-9)^{-3}(7n_{I}^2-27) \left(\frac{n_{I}-1}{n_{I}+1}\right)^{n_{I}} \left(\frac{n_{I}-3}{n_{I}+3}\right)^{n_{I}} \left[ \frac{1}{1-\frac{1}{n_{I}^2}-\frac{\lambda_{\text{LyLimit}}}{\lambda}} + \frac{1}{\frac{1}{9}-\frac{1}{n_{I}^2}+\frac{\lambda_{\text{LyLimit}}}{\lambda}} \right] \right) \right. \nonumber\\
&+& \left. \int_{0}^{\infty} dn_{I}' \left( \frac{2^{7}3^{1/2}n_{I}'^{3}(n_{I}'^2+1)^{-1} (n_{I}'^2+9)^{-3} (7n_{I}'^2+27) e^{-2n_{I}'\left(\text{arccot}(n_{I}')+\text{arccot}(n_{I}'/3)\right)}}{\left(1-\exp(-2\pi n_{I}') \right)} \left[ \frac{1}{1+\frac{1}{n_{I}'^2}-\frac{\lambda_{\text{LyLimit}}}{\lambda}} + \frac{1}{\frac{1}{9}+\frac{1}{n_{I}'^2}+\frac{\lambda_{\text{LyLimit}}}{\lambda}} \right] \right) \right|^2 \nonumber\\
\left.\frac{d\sigma}{d\Omega}\right|_{|B\rangle = |4,0\rangle} &=& r_{0}^2 \left(1-\frac{15}{16}\frac{\lambda}{\lambda_{\text{LyLimit}}}\right) \left(\vec{\epsilon}^{(\alpha)} \cdot \vec{\epsilon}^{(\alpha')}\right)^2 \nonumber\\
&\times&\left| \sum_{n_{I}=2, 3; n_{I}\geq 5} \left( 2^{12}3^{-2}n_{I}^{3}(n_{I}^2-1)^{-1}(n_{I}^2-16)^{-4}(23n_{I}^4-288n_{I}^2+768) \left(\frac{n_{I}-1}{n_{I}+1}\right)^{n_{I}} \left(\frac{n_{I}-4}{n_{I}+4}\right)^{n_{I}} \left[ \frac{1}{1-\frac{1}{n_{I}^2}-\frac{\lambda_{\text{LyLimit}}}{\lambda}} + \frac{1}{\frac{1}{16}-\frac{1}{n_{I}^2}+\frac{\lambda_{\text{LyLimit}}}{\lambda}} \right] \right) \right. \nonumber\\
&+& \left. \int_{0}^{\infty} dn_{I}' \left( \frac{2^{12}3^{-2}n_{I}'^{3}(n_{I}'^2+1)^{-1}(n_{I}'^2+16)^{-4}(23n_{I}'^4+288n_{I}'^2+768) e^{-2n_{I}'\left(\text{arccot}(n_{I}')+\text{arccot}(n'/4)\right)}}{\left(1-\exp(-2\pi n_{I}') \right)} \left[ \frac{1}{1+\frac{1}{n_{I}'^2}-\frac{\lambda_{\text{LyLimit}}}{\lambda}} + \frac{1}{\frac{1}{16}+\frac{1}{n_{I}'^2}+\frac{\lambda_{\text{LyLimit}}}{\lambda}} \right] \right) \right|^2.\nonumber
\end{eqnarray}
\normalsize
The differential cross-sections for any $n_{B}$ can readily be obtained by using the radial integrals in Tables~\ref{tbl:radial_integrals_boundbound}--\ref{tbl:radial_integrals_boundfree} or using directly Equations~\ref{eqn:R_minus_bb}--\ref{eqn:R_plus_bf}.

To derive the total cross-section from the differential cross-section (Equation~\ref{eqn:differential_crosssection_nB1}), consider a situation where the incident photon of the polarization mode $\vec{\epsilon}^{(1)}$ propagates toward $z$ direction ($\vec{k}/k = (0, 0, 1)$ and $\vec{\epsilon}^{(1)} = (1, 0, 0)$), and the outgoing photon propagates toward a direction $\vec{k'}/k' = (\sin\theta \cos\phi, \sin\theta \sin\phi, \cos\theta)$.
The two orthogonal polarization basic vectors of the outgoing photon can be defined as $\vec{\epsilon}^{(1')} = (-\sin\phi, \cos\phi, 0)$ and $\vec{\epsilon}^{(2')} = (\cos\theta\cos\phi, \cos\theta\sin\phi, -\sin\theta)$.
The differential scattering cross-section summed over the two polarization modes of the outgoing photon is proportional to $(\vec{\epsilon}^{(1)} \cdot \vec{\epsilon}^{(1')})^2 + (\vec{\epsilon}^{(1)} \cdot \vec{\epsilon}^{(2')})^2 = \sin^2\phi + \cos^2\theta\cos^2\phi$, and the total scattering cross-section is proportional to its integral over the full solid angle\footnote{The differential cross-section for unpolarized photon beam is proportional to $\frac{1}{2\pi}\int_{0}^{2\pi} d\phi (\sin^2\phi + \cos^2\theta\cos^2\phi) = \frac{1}{2}(1+\cos^2\theta$), and its integral over the full solid angle gives the same factor of $8\pi/3$. Thus, the total cross-section for an unpolarized photon beam is the same as that for a photon of a particular polarization mode given in Equation~\ref{eqn:total_crosssection_rayleigh} \citep[][\S 3.4]{ryb79}.}: $\int_{\Omega}d\Omega(\sin^2\phi + \cos^2\theta\cos^2\phi) = 8\pi/3$.
Therefore, the total scattering cross-section of the $1s \rightarrow n_{B}s$ Rayleigh and Raman scattering can be obtained by replacing $r_{0}^2 \left(\vec{\epsilon}^{(\alpha)} \cdot \vec{\epsilon}^{(\alpha')}\right)^2$ in the differential cross-section (Equation~\ref{eqn:differential_crosssection_nB1}) by $8\pi r_{0}^2/3 = \sigma_{T}$:
\begin{eqnarray}
\sigma_{n_{B}s} \equiv \left.\sigma \right|_{|B\rangle = |n_{B},0\rangle} &=& \sigma_{T} \left(\frac{\omega'}{\omega}\right) \left| \delta_{AB} - \frac{m_e}{\hbar^2} \SumInt_{n_{I}} \left( \frac{R_{n_{B}0}^{n_{I}1} R_{n_{I}1}^{10}(E_{I}-E_{B})(E_{I}-E_{A})}{3} \left[ \frac{1}{E_{I}-E_{A}-\hbar\omega} + \frac{1}{E_{I}-E_{A}+\hbar\omega'} \right] \right) \right|^2.
\label{eqn:total_crosssection_rayleigh}
\end{eqnarray}


From the differential Rayleigh cross section (Equation~\ref{eqn:crosssection_rayleigh}), the total Rayleigh scattering cross section is obtained as:
\small
\begin{eqnarray}
\sigma_{1s} &=& \sigma_{T} \nonumber\\
&\times& \left| 1 - \sum_{n_{I} \geq 2} \left( 2^{7}3^{-1}n_{I}^{3}(n_{I}^2-1)^{-3} \left( \frac{n_{I}-1}{n_{I}+1} \right)^{2n_{I}} \left[ \frac{1}{1-\frac{1}{n_{I}^2}-\frac{\lambda_{\text{LyLimit}}}{\lambda}} + \frac{1}{1-\frac{1}{n_{I}^2}+\frac{\lambda_{\text{LyLimit}}}{\lambda}} \right] \right) \right.\nonumber\\
&-&\left. \int_{0}^{\infty}dn_{I}' \left( \frac{2^{7}3^{-1}n_{I}'^{3}(n_{I}'^2+1)^{-3}  e^{-4n_{I}'\text{arccot}(n_{I}')} }{(1-\exp(-2\pi n_{I}'))} \left[ \frac{1}{1+\frac{1}{n_{I}'^2}-\frac{\lambda_{\text{LyLimit}}}{\lambda}} + \frac{1}{1+\frac{1}{n_{I}'^2}+\frac{\lambda_{\text{LyLimit}}}{\lambda}} \right] \right) \right|^2,\nonumber\\
&=& \sigma_{T} \nonumber\\
&\times& \left| \sum_{n_{I} \geq 2} \left( 2^{7}3^{-1}n_{I}^{5}(n_{I}^2-1)^{-4} \left( \frac{n_{I}-1}{n_{I}+1} \right)^{2n_{I}} \left[ \frac{\frac{\lambda_{\text{LyLimit}}}{\lambda}}{1-\frac{1}{n_{I}^2}-\frac{\lambda_{\text{LyLimit}}}{\lambda}} - \frac{\frac{\lambda_{\text{LyLimit}}}{\lambda}}{1-\frac{1}{n_{I}^2}+\frac{\lambda_{\text{LyLimit}}}{\lambda}} \right] \right) \right.\nonumber\\
&+&\left. \int_{0}^{\infty}dn_{I}' \left( \frac{2^{7}3^{-1}n_{I}'^{5}(n_{I}'^2+1)^{-4}  e^{-4n_{I}'\text{arccot}(n_{I}')} }{(1-\exp(-2\pi n_{I}'))} \left[ \frac{\frac{\lambda_{\text{LyLimit}}}{\lambda}}{1+\frac{1}{n_{I}'^2}-\frac{\lambda_{\text{LyLimit}}}{\lambda}} - \frac{\frac{\lambda_{\text{LyLimit}}}{\lambda}}{1+\frac{1}{n_{I}'^2}+\frac{\lambda_{\text{LyLimit}}}{\lambda}} \right] \right) \right|^2.
\label{eqn:rayleigh_explicit}
\end{eqnarray}
\normalsize
This is the exact formula for the $1s \rightarrow 1s$ Rayleigh scattering cross section of the ground-state hydrogen, which has been used in some literature (see Section~\ref{sec:approximate_formulae}) to calculate, e.g., the Ly$\alpha$ damping wing cross section relevant to studies of cosmic reionization \citep{bac15,mor16} and cosmic microwave background anisotropies and the distribution of matter in the Universe \citep{ali15}.

The $n_{B} = 2, 3$, and $4$ Raman scattering total cross sections can be expressed as:
\small
\begin{eqnarray}
\sigma_{2s} &=& \sigma_{T} \left(1-\frac{3}{4}\frac{\lambda}{\lambda_{\text{LyLimit}}}\right) \nonumber\\
&\times& \left| \sum_{n_{I} \geq 3} \left( 2^{19/2}3^{-1}n_{I}^{3}(n_{I}^2-1)^{-1} (n_{I}^2-4)^{-2} \left(\frac{n_{I}-1}{n_{I}+1}\right)^{n_{I}}\left(\frac{n_{I}-2}{n_{I}+2}\right)^{n_{I}} \left[ \frac{1}{1-\frac{1}{n_{I}^2}-\frac{\lambda_{\text{LyLimit}}}{\lambda}} + \frac{1}{\frac{1}{4}-\frac{1}{n_{I}^2}+\frac{\lambda_{\text{LyLimit}}}{\lambda}} \right] \right) \right. \nonumber \\
&+& \left. \int_{0}^{\infty} dn_{I}' \left( \frac{ 2^{19/2}3^{-1}n_{I}'^{3}(n_{I}'^2+1)^{-1} (n_{I}'^2+4)^{-2} e^{-2n_{I}'\left(\text{arccot}(n_{I}')+\text{arccot}(n_{I}'/2)\right)}}{\left(1-\exp(-2\pi n_{I}') \right)} \left[ \frac{1}{1+\frac{1}{n_{I}'^2}-\frac{\lambda_{\text{LyLimit}}}{\lambda}} + \frac{1}{\frac{1}{4}+\frac{1}{n_{I}'^2}+\frac{\lambda_{\text{LyLimit}}}{\lambda}} \right] \right) \right|^2 \nonumber.\\
\sigma_{3s} &=& \sigma_{T} \left(1-\frac{8}{9}\frac{\lambda}{\lambda_{\text{LyLimit}}}\right) \nonumber\\
&\times&\left| \sum_{n_{I}=2; n_{I}\geq 4} \left( 2^{7}3^{1/2}n_{I}^{3}(n_{I}^2-1)^{-1}(n_{I}^2-9)^{-3}(7n_{I}^2-27) \left(\frac{n_{I}-1}{n_{I}+1}\right)^{n_{I}} \left(\frac{n_{I}-3}{n_{I}+3}\right)^{n_{I}} \left[ \frac{1}{1-\frac{1}{n_{I}^2}-\frac{\lambda_{\text{LyLimit}}}{\lambda}} + \frac{1}{\frac{1}{9}-\frac{1}{n_{I}^2}+\frac{\lambda_{\text{LyLimit}}}{\lambda}} \right] \right) \right. \nonumber\\
&+& \left. \int_{0}^{\infty} dn_{I}' \left( \frac{2^{7}3^{1/2}n_{I}'^{3}(n_{I}'^2+1)^{-1} (n_{I}'^2+9)^{-3} (7n_{I}'^2+27) e^{-2n_{I}'\left(\text{arccot}(n_{I}')+\text{arccot}(n_{I}'/3)\right)}}{\left(1-\exp(-2\pi n_{I}') \right)} \left[ \frac{1}{1+\frac{1}{n_{I}'^2}-\frac{\lambda_{\text{LyLimit}}}{\lambda}} + \frac{1}{\frac{1}{9}+\frac{1}{n_{I}'^2}+\frac{\lambda_{\text{LyLimit}}}{\lambda}} \right] \right) \right|^2 \nonumber\\
\sigma_{4s} &=& \sigma_{T} \left(1-\frac{15}{16}\frac{\lambda}{\lambda_{\text{LyLimit}}}\right) \nonumber\\
&\times&\left| \sum_{n_{I}=2, 3; n_{I}\geq 5} \left( 2^{12}3^{-2}n_{I}^{3}(n_{I}^2-1)^{-1}(n_{I}^2-16)^{-4}(23n_{I}^4-288n_{I}^2+768) \left(\frac{n_{I}-1}{n_{I}+1}\right)^{n_{I}} \left(\frac{n_{I}-4}{n_{I}+4}\right)^{n_{I}} \left[ \frac{1}{1-\frac{1}{n_{I}^2}-\frac{\lambda_{\text{LyLimit}}}{\lambda}} + \frac{1}{\frac{1}{16}-\frac{1}{n_{I}^2}+\frac{\lambda_{\text{LyLimit}}}{\lambda}} \right] \right) \right. \nonumber\\
&+& \left. \int_{0}^{\infty} dn_{I}' \left( \frac{2^{12}3^{-2}n_{I}'^{3}(n_{I}'^2+1)^{-1}(n_{I}'^2+16)^{-4}(23n_{I}'^4+288n_{I}'^2+768) e^{-2n_{I}'\left(\text{arccot}(n_{I}')+\text{arccot}(n'/4)\right)}}{\left(1-\exp(-2\pi n_{I}') \right)} \left[ \frac{1}{1+\frac{1}{n_{I}'^2}-\frac{\lambda_{\text{LyLimit}}}{\lambda}} + \frac{1}{\frac{1}{16}+\frac{1}{n_{I}'^2}+\frac{\lambda_{\text{LyLimit}}}{\lambda}} \right] \right) \right|^2\nonumber,
\end{eqnarray}
\normalsize
and it is straightforward to obtain the Raman scattering cross sections of higher $n_{B}$ by using Equation~\ref{eqn:total_crosssection_rayleigh} using the overlap integrals given in Tables~\ref{tbl:radial_integrals_boundbound}--\ref{tbl:radial_integrals_boundfree} or using Equations~\ref{eqn:R_minus_bb}--\ref{eqn:R_plus_bf}

The wavelength-dependent cross-sections $\sigma_{n_{B}s}$ up to $n_{B}=4$ are shown in the first panel of Figure~\ref{fig:crosssections}.
Hereinafter, the sum over $n_{I}$ is calculated by truncating at $n_{I} = 1000$, and the integral is performed by using {\it scipy.integrate.quad} \citep{scipy20} over $n_{I}' = 0 - 1000$.
As noted above, the Raman scattering cross-section $\sigma_{n_{B}s}$ ($n_{B} \geq 2$) is non-zero only at $\lambda < \lambda_{\text{LyLimit}}/(1-n_{B}^{-2})$.
The dominant contribution to the gross cross-section $\sigma_{\text{tot}} \equiv \sigma_{1s}+\sigma_{2s}+\sigma_{3s}+\sigma_{4s}+\sigma_{3d}+\sigma_{4d}$ comes from the $1s \rightarrow 1s$ Rayleigh branch $\sigma_{1s}$, but $\sigma_{1s}$ vanishes at wavelength ranges in between the resonant wavelengths \citep[][]{zon69,roh22} where the Raman branch becomes predominant.
At $\lambda > \lambda_{\text{Ly}\alpha}$, the cross-section $\sigma_{1s}$ asymptotically approaches $\propto \lambda^{-4}$, i.e., the $\lambda^{-4}$-law of the classical Rayleigh scattering (see Section~\ref{sec:approximate_formulae}).

\subsubsection{$1s \rightarrow n_{B}d$ Raman scattering cross-sections}

When the final state is the $d$ orbital ($l_{B}=2$, $m_{B}=\{-2,-1,0,1,2\})$, by using the Wigner-Eckart theorem (Appendix~\ref{sec:wigner-eckart}), $\sum_{m_{I}=-1,0,1}^{}(\vec{x} \cdot \vec{\epsilon}^{(\alpha')})_{BI}(\vec{x} \cdot \vec{\epsilon}^{(\alpha)})_{IA}$ for each of the magnetic quantum number $m_{B}$ becomes (Equation~\ref{eqn:sum_of_matrix}):
\small
\begin{eqnarray}
m_{B}=2: &&\sum_{m_{I}=-1,0,1}^{}(\vec{x} \cdot \vec{\epsilon}^{(\alpha')})_{BI}(\vec{x} \cdot \vec{\epsilon}^{(\alpha)})_{IA} = \sum_{m_{I}=-1,0,1}^{}(\vec{x} \cdot \vec{\epsilon}^{(\alpha)})_{BI}(\vec{x} \cdot \vec{\epsilon}^{(\alpha')})_{IA} \nonumber\\
&=& R_{n_{I}1}^{10} R_{n_{B}2}^{n_{I}1} \sqrt{\frac{1}{30}} \left( {\epsilon}_{x}^{(\alpha')}{\epsilon}_{x}^{(\alpha)} - {\epsilon}_{y}^{(\alpha')}{\epsilon}_{y}^{(\alpha)} - i{\epsilon}_{x}^{(\alpha')}{\epsilon}_{y}^{(\alpha)} - i{\epsilon}_{y}^{(\alpha')}{\epsilon}_{x}^{(\alpha)} \right) \nonumber\\
m_{B}=1: &&\sum_{m_{I}=-1,0,1}^{}(\vec{x} \cdot \vec{\epsilon}^{(\alpha')})_{BI}(\vec{x} \cdot \vec{\epsilon}^{(\alpha)})_{IA} = \sum_{m_{I}=-1,0,1}^{}(\vec{x} \cdot \vec{\epsilon}^{(\alpha)})_{BI}(\vec{x} \cdot \vec{\epsilon}^{(\alpha')})_{IA} \nonumber\\
&=& \sqrt{\frac{1}{30}} R_{n_{I}1}^{10} R_{n_{B}2}^{n_{I}1}        \left(-{\epsilon}_{x}^{(\alpha')}{\epsilon}_{z}^{(\alpha)}+i{\epsilon}_{y}^{(\alpha')}{\epsilon}_{z}^{(\alpha)}\right) \nonumber\\
&+& \sqrt{\frac{1}{30}} R_{n_{I}1}^{10} R_{n_{B}2}^{n_{I}1} \left(-{\epsilon}_{z}^{(\alpha')}{\epsilon}_{x}^{(\alpha)}+i{\epsilon}_{z}^{(\alpha')}{\epsilon}_{y}^{(\alpha)}\right)\nonumber\\
&=& R_{n_{I}1}^{10} R_{n_{B}2}^{n_{I}1} \sqrt{\frac{1}{30}} \left(-{\epsilon}_{x}^{(\alpha')}{\epsilon}_{z}^{(\alpha)}-{\epsilon}_{z}^{(\alpha')}{\epsilon}_{x}^{(\alpha)}+i{\epsilon}_{y}^{(\alpha')}{\epsilon}_{z}^{(\alpha)} +i{\epsilon}_{z}^{(\alpha')}{\epsilon}_{y}^{(\alpha)} \right)\nonumber\\
m_{B}=0: &&\sum_{m_{I}=-1,0,1}^{}(\vec{x} \cdot \vec{\epsilon}^{(\alpha')})_{BI}(\vec{x} \cdot \vec{\epsilon}^{(\alpha)})_{IA} = \sum_{m_{I}=-1,0,1}^{}(\vec{x} \cdot \vec{\epsilon}^{(\alpha)})_{BI}(\vec{x} \cdot \vec{\epsilon}^{(\alpha')})_{IA} \nonumber\\
&=& \sqrt{\frac{1}{180}} R_{n_{I}1}^{10} R_{n_{B}2}^{n_{I}1}               \left( -{\epsilon}_{x}^{(\alpha')}{\epsilon}_{x}^{(\alpha)} -{\epsilon}_{y}^{(\alpha')}{\epsilon}_{y}^{(\alpha)} - i{\epsilon}_{x}^{(\alpha')}{\epsilon}_{y}^{(\alpha)} + i{\epsilon}_{y}^{(\alpha')}{\epsilon}_{x}^{(\alpha)} \right) \nonumber\\
&+& \sqrt{\frac{1}{180}} R_{n_{I}1}^{10} R_{n_{B}2}^{n_{I}1}               \left( -{\epsilon}_{x}^{(\alpha')}{\epsilon}_{x}^{(\alpha)} -{\epsilon}_{y}^{(\alpha')}{\epsilon}_{y}^{(\alpha)} + i{\epsilon}_{x}^{(\alpha')}{\epsilon}_{y}^{(\alpha)} - i{\epsilon}_{y}^{(\alpha')}{\epsilon}_{x}^{(\alpha)} \right) \nonumber\\
&+& 4\sqrt{\frac{1}{180}}  R_{n_{I}1}^{10} R_{n_{B}2}^{n_{I}1}               {\epsilon}_{z}^{(\alpha')} {\epsilon}_{z}^{(\alpha)}\nonumber\\
&=& R_{n_{I}1}^{10} R_{n_{B}2}^{n_{I}1} \sqrt{\frac{1}{45}} \left( -{\epsilon}_{x}^{(\alpha')}{\epsilon}_{x}^{(\alpha)} -{\epsilon}_{y}^{(\alpha')}{\epsilon}_{y}^{(\alpha)} + 2{\epsilon}_{z}^{(\alpha')} {\epsilon}_{z}^{(\alpha)} \right) \nonumber\\
m_{B}=-1: &&\sum_{m_{I}=-1,0,1}^{}(\vec{x} \cdot \vec{\epsilon}^{(\alpha')})_{BI}(\vec{x} \cdot \vec{\epsilon}^{(\alpha)})_{IA} = \sum_{m_{I}=-1,0,1}^{}(\vec{x} \cdot \vec{\epsilon}^{(\alpha)})_{BI}(\vec{x} \cdot \vec{\epsilon}^{(\alpha')})_{IA} \nonumber\\
&=& \sqrt{\frac{1}{30}} R_{n_{I}1}^{10} R_{n_{B}2}^{n_{I}1}         \left({\epsilon}_{x}^{(\alpha')} {\epsilon}_{z}^{(\alpha)} +i{\epsilon}_{y}^{(\alpha')} {\epsilon}_{z}^{(\alpha)} \right) \nonumber\\
&+& \sqrt{\frac{1}{30}} R_{n_{I}1}^{10} R_{n_{B}2}^{n_{I}1}        \left({\epsilon}_{z}^{(\alpha')}{\epsilon}_{x}^{(\alpha)}+i{\epsilon}_{z}^{(\alpha')}{\epsilon}_{y}^{(\alpha)}\right)\nonumber\\
&=& R_{n_{I}1}^{10} R_{n_{B}2}^{n_{I}1} \sqrt{\frac{1}{30}} \left({\epsilon}_{x}^{(\alpha')} {\epsilon}_{z}^{(\alpha)} + {\epsilon}_{z}^{(\alpha')}{\epsilon}_{x}^{(\alpha)} + i{\epsilon}_{y}^{(\alpha')} {\epsilon}_{z}^{(\alpha)} + i{\epsilon}_{z}^{(\alpha')}{\epsilon}_{y}^{(\alpha)} \right) \nonumber\\
m_{B}=-2: &&\sum_{m_{I}=-1,0,1}^{}(\vec{x} \cdot \vec{\epsilon}^{(\alpha')})_{BI}(\vec{x} \cdot \vec{\epsilon}^{(\alpha)})_{IA} = \sum_{m_{I}=-1,0,1}^{}(\vec{x} \cdot \vec{\epsilon}^{(\alpha)})_{BI}(\vec{x} \cdot \vec{\epsilon}^{(\alpha')})_{IA} \nonumber\\
&=& R_{n_{I}1}^{10} R_{n_{B}2}^{n_{I}1} \sqrt{\frac{1}{30}} \left( {\epsilon}_{x}^{(\alpha')}{\epsilon}_{x}^{(\alpha)} - {\epsilon}_{y}^{(\alpha')}{\epsilon}_{y}^{(\alpha)} + i{\epsilon}_{x}^{(\alpha')}{\epsilon}_{y}^{(\alpha)} + i{\epsilon}_{y}^{(\alpha')}{\epsilon}_{x}^{(\alpha)} \right) \nonumber,
\end{eqnarray}
\normalsize
and it can be shown by straightforward algebraic manipulations of Equation~\ref{eqn:kramersheisenberg_nonoriented} that the summation of the squares of the Kramers-Heisenberg-Waller matrix element over $m_{B}$ yield the differential cross-section for the $1s \rightarrow n_{B}d$ Raman scattering proportional to a factor $(3+(\vec{\epsilon}^{(\alpha)} \cdot \vec{\epsilon}^{(\alpha')})^2)$:
\small
\begin{eqnarray}
\left.\frac{d\sigma}{d\Omega}\right|_{|B\rangle = |n_{B}, 2\rangle} &=& r_{0}^2 \left(\frac{\omega'}{\omega}\right) \frac{m_{e}^2}{\hbar^4} \frac{2}{90}\left[ 3+\left(\vec{\epsilon}^{(\alpha')} \cdot \vec{\epsilon}^{(\alpha)}\right)^2 \right] \left| \SumInt_{n_{I}} \left( R_{n_{B}2}^{n_{I}1}R_{n_{I}1}^{10} (E_{I}-E_{B})(E_{I}-E_{A}) \left[ \frac{1}{E_{I}-E_{A}-\hbar\omega} + \frac{1}{E_{I}-E_{A}+\hbar\omega'} \right] \right) \right|^2 \label{eqn:differential_crosssection_nB2}
\end{eqnarray}
\normalsize
The cross-section becomes 0 at $\lambda > \lambda_{\text{LyLimit}}/(1-n_{B}^{-2})$ due to the energy conservation law.
In this case of the $1s \rightarrow n_{B}d$ Raman scattering, the scattering is of a {\it symmetric} nature \citep[][]{pla34,pen69,ber82}, and the scattering phase function $\propto (3+(\vec{\epsilon}^{(\alpha)} \cdot \vec{\epsilon}^{(\alpha')})^2)$ is closer to isotropic than the pure Rayleigh's phase function for the $1s \rightarrow n_{B}s$ scattering (see Section~\ref{sec:phase_function}).

Following the prescription in Section~\ref{sec:1s_ns_transition}, the total cross-section can be obtained by taking the solid angle integral of the differential cross-section summed over the two independent polarization modes of the outgoing photon. 
The sum of $(3+(\vec{\epsilon}^{(\alpha)} \cdot \vec{\epsilon}^{(\alpha')})^2)$ over $\alpha'$ can be parameterized as\footnote{The differential cross-section for unpolarized photon beam is proportional to $\frac{1}{2\pi}\int_{0}^{2\pi} d\phi (6 + \sin^2\phi + \cos^2\theta\cos^2\phi) = \frac{1}{2}(13+\cos^2\theta$), and whose integral over the full solid angle is $80\pi/3$.}
$3+(\vec{\epsilon}^{(1)} \cdot \vec{\epsilon}^{(1')})^2 + 3+(\vec{\epsilon}^{(1)} \cdot \vec{\epsilon}^{(2')})^2 = 6 + \sin\phi^2 + \cos^2\theta \cos^2\phi$, then we obtain: $\int_{\Omega}d\Omega ( 6+\sin\phi^2 + \cos^2\theta \cos^2\phi ) = 80\pi/3$. 
Thus, the total scattering cross-section of the $1s \rightarrow n_{B}d$ Raman scattering (denoted as $\sigma_{n_{B}d} \equiv \left.\sigma \right|_{|B\rangle = |n_{B},2\rangle}$ hereafter) can be obtained by replacing $r_{0}^2 [3 + (\vec{\epsilon}^{(\alpha)} \cdot \vec{\epsilon}^{(\alpha')})^2 ]$ in the differential cross-section (Equation~\ref{eqn:differential_crosssection_nB2}) by $80\pi r_{0}^2/3 = 10\sigma_{T}$:
\small
\begin{eqnarray}
\sigma_{n_{B}d} \equiv \left.\sigma\right|_{|B\rangle = |n_{B}, 2\rangle} &=& \sigma_{T} \left(\frac{\omega'}{\omega}\right) \frac{m_{e}^2}{\hbar^4} \frac{2}{9} \left| \SumInt_{n_{I}} \left( R_{n_{B}2}^{n_{I}1}R_{n_{I}1}^{10} (E_{I}-E_{B})(E_{I}-E_{A}) \left[ \frac{1}{E_{I}-E_{A}-\hbar\omega} + \frac{1}{E_{I}-E_{A}+\hbar\omega'} \right] \right) \right|^2.
\end{eqnarray}
\normalsize
The $n_{B} = 3$, and $4$ Raman scattering total cross sections can be expressed as:
\small
\begin{eqnarray}
\sigma_{3d} &=& \sigma_T \left(1-\frac{8}{9}\frac{\lambda}{\lambda_{\text{LyLimit}}}\right) \nonumber\\
&\times& \left| \sum_{n_{I}=2; n_{I}\geq 4} \left( 2^{9}3^{1/2}5^{-1/2}n_{I}^{5} (n_{I}^2-1)^{-1} (n_{I}^2-9)^{-3} \left(\frac{n_{I}-1}{n_{I}+1}\right)^{n_{I}} \left(\frac{n_{I}-3}{n_{I}+3}\right)^{n_{I}} \left[ \frac{1}{1-\frac{1}{n_{I}^2}-\frac{\lambda_{\text{LyLimit}}}{\lambda}} + \frac{1}{\frac{1}{9}-\frac{1}{n_{I}^2}+\frac{\lambda_{\text{LyLimit}}}{\lambda}} \right] \right) \right. \nonumber\\
&+& \left. \int_{0}^{\infty} dn_{I}' \left( \frac{2^{9}3^{1/2}5^{-1/2}n_{I}'^{5} (n_{I}'^2+1)^{-1} (n_{I}'^2+9)^{-3} e^{-2n_{I}'(\text{arccot}(n_{I}')+\text{arccot}(n_{I}'/3))}}{1-\exp(-2\pi n_{I}')} \left[ \frac{1}{1+\frac{1}{n_{I}'^2}-\frac{\lambda_{\text{LyLimit}}}{\lambda}} + \frac{1}{\frac{1}{9}+\frac{1}{n_{I}'^2}+\frac{\lambda_{\text{LyLimit}}}{\lambda}} \right] \right) \right|^2 \nonumber\\
\sigma_{4d} &=& \sigma_T \left(1-\frac{15}{16}\frac{\lambda}{\lambda_{\text{LyLimit}}}\right) \nonumber\\
&\times& \left| \sum_{n_{I}=2, 3; n_{I}\geq 5} \left( 2^{27/2}3^{-2}5^{-1/2}n_{I}^{5} (n_{I}^2-1)^{-1} (n_{I}^2-16)^{-4} (7n_{I}^2-48) \left(\frac{n_{I}-1}{n_{I}+1}\right)^{n_{I}} \left(\frac{n_{I}-4}{n_{I}+4}\right)^{n_{I}} \left[ \frac{1}{1-\frac{1}{n_{I}^2}-\frac{\lambda_{\text{LyLimit}}}{\lambda}} + \frac{1}{\frac{1}{16}-\frac{1}{n_{I}^2}+\frac{\lambda_{\text{LyLimit}}}{\lambda}} \right] \right) \right. \nonumber\\
&+& \left. \int_{0}^{\infty} dn_{I}' \left( \frac{2^{27/2}3^{-2}5^{-1/2}n_{I}'^{5} (n_{I}'^2+1)^{-1} (n_{I}'^2+16)^{-4} (7n_{I}^2+48) e^{-2n_{I}'(\text{arccot}(n_{I}')+\text{arccot}(n_{I}'/4))}}{1-\exp(-2\pi n_{I}')} \left[ \frac{1}{1+\frac{1}{n_{I}'^2}-\frac{\lambda_{\text{LyLimit}}}{\lambda}} + \frac{1}{\frac{1}{16}+\frac{1}{n_{I}'^2}+\frac{\lambda_{\text{LyLimit}}}{\lambda}} \right] \right) \right|^2\nonumber.
\end{eqnarray}
\normalsize

The wavelength-dependent cross-sections $\sigma_{3d}$ and $\sigma_{4d}$ are shown in the second panel of Figure~\ref{fig:crosssections}.
Similar to the $1s \rightarrow n_{B}s$ Raman scattering, the $1s \rightarrow n_{B}d$ Raman scattering branch becomes important around the $\sigma_{1s}$ minima.

\subsection{Scattering phase function and phase matrix}
\label{sec:phase_function_and_phase_matrix}

\begin{figure}
\center{
\includegraphics[clip, width=4.4in]{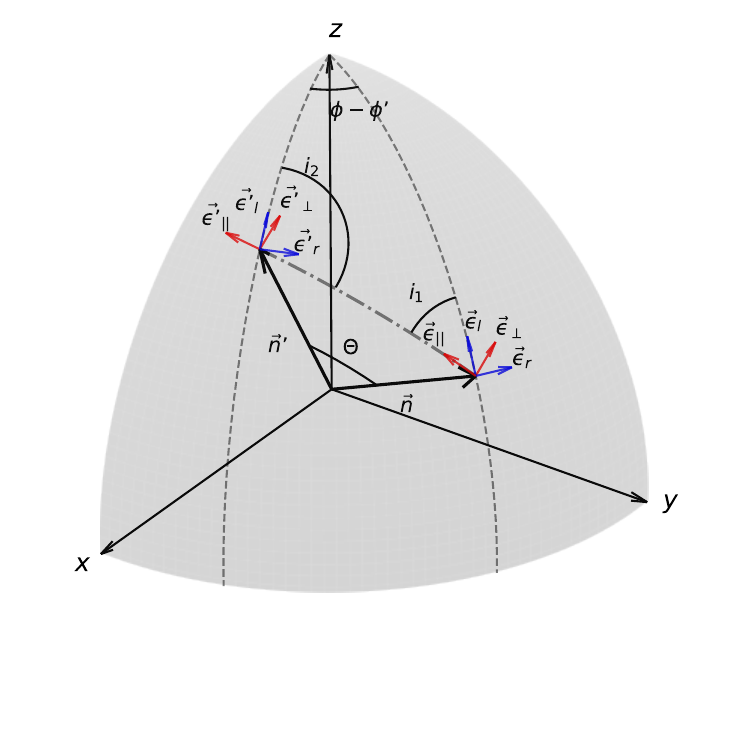}
}\vspace{-2cm}
 \caption{Geometry of the plane of scattering defined by the unit propagation vectors for the incident and scattered lights ($\vec{n}$ and $\vec{n}'$, respectively), and the meridian planes containing the unit propagation vectors. Two polarization basis sets perpendicular to $\vec{n}$ and $\vec{n}'$ are denoted by $\vec{\epsilon}$ and $\vec{\epsilon}'$, respectively, where the $\parallel-\perp$ basis is defined with respect to the scattering plane and the $l$-$r$ basis is defined with respect to the meridian plane. The three unit vectors $\vec{e}_{z}=(0,0,1)$, $\vec{n}$, and $\vec{n}'$ define a spherical triangle, whose angles at $\vec{n}$ and $\vec{n}'$ are denoted as $i_{1}$ and $i_{2}$, respectively.}
 \label{fig:scattering_plane}
\end{figure}

As we have seen in the previous section, the $1s \rightarrow n_{B}s$ (Rayleigh and Raman $s$-branch) and $1s \rightarrow n_{B}d$ (Raman $d$ branch) scattering cross-sections exhibit different angular distributions, referred to as the scattering of the {\it scalar} and {\it symmetric} nature, respectively.
This leads to differences in the polarization characteristics of the scattered photons.
Here we discuss the polarization characteristics of the {\it scalar} and {\it symmetric} scattering, deriving phase functions and phase matrices which will be helpful for future studies of more detailed radiative transfer simulations \citep[see][for more general discussion about the angular dependence, depolarization, and phase matrices of the Rayleigh and Raman scattering]{pla34,pen69,ste76,ste80,ste94}.

\subsubsection{Phase function for the scattering of unpolarized incident light}
\label{sec:phase_function}

Let us consider partial linear polarization of scattered light for unpolarized incident light.
This can be analyzed by first considering linearly polarized incident light and then averaging over directions of the polarization.

We denote the propagation directions of the incident and scattered lights as $\vec{n} = (\sin\theta\cos\phi, \sin\theta\sin\phi, \cos\theta)$ and $\vec{n}' = (\sin\theta'\cos\phi', \sin\theta'\sin\phi', \cos\theta')$, respectively.
The {\it plane of scattering} is defined as the plane spanned by $\vec{n}$ and $\vec{n}'$, and the angle between $\vec{n}$ and $\vec{n}'$ is denoted as $\Theta$, whose cosine is $\cos\Theta = \vec{n} \cdot \vec{n}' = \sin\theta\sin\theta'\cos(\phi'-\phi) + \cos\theta\cos\theta'$.
Following \cite{cha60}, orthogonal unit basis vectors perpendicular to $\vec{n}$ (denoted as $\vec{\epsilon}_{\parallel}, \vec{\epsilon}_{\perp}$) and $\vec{n}'$ (denoted as $\vec{\epsilon}_{\parallel}', \vec{\epsilon}_{\perp}'$) are defined such that one is parallel ($\parallel$) to the plane of scattering and the other is perpendicular ($\perp$) to it (Figure~\ref{fig:scattering_plane}):
\begin{eqnarray}
(\vec{\epsilon}_{\parallel}, \vec{\epsilon}_{\perp}) &=& \left(-\frac{\cos\Theta}{\sin\Theta}\vec{n} + \frac{1}{\sin\Theta}\vec{n}', \vec{\epsilon}_{\parallel} \times \vec{n}\right)\\
(\vec{\epsilon}_{\parallel}', \vec{\epsilon}_{\perp}') &=& \left(-\frac{1}{\sin\Theta}\vec{n} + \frac{\cos\Theta}{\sin\Theta}\vec{n}', \vec{\epsilon}_{\parallel}' \times \vec{n}'\right) = \left(-\frac{1}{\sin\Theta}\vec{n} + \frac{\cos\Theta}{\sin\Theta}\vec{n}', \vec{\epsilon}_{\perp}\right).
\end{eqnarray}
Consider that the incident light is linearly polarized in a direction given by a unit vector $\epsilon = \cos\beta\vec{\epsilon}_{\parallel} + \sin\beta\vec{\epsilon}_{\perp}$, then the squared inner products of $\epsilon$ and $(\vec{\epsilon}_{\parallel}', \vec{\epsilon}_{\perp}')$ are:
\begin{eqnarray}
(\epsilon \cdot \vec{\epsilon}_{\perp}')^2 &=& \sin^2\beta. \label{eqn:squared_basic_perp}\\
(\epsilon \cdot \vec{\epsilon}_{\parallel}')^2 &=& \cos^2\beta\cos^2\Theta \label{eqn:squared_basic_parallel}
\end{eqnarray}

In the case of the {\it scalar} scattering ($d\sigma/d\Omega \propto (\epsilon^{(\alpha)} \cdot \epsilon^{(\alpha')})^2$), by taking a sum of the two modes of the scattered light in Equations~\ref{eqn:squared_basic_perp}-\ref{eqn:squared_basic_parallel} and an average over the incident polarization direction $\beta$, we can see that the intensity of the scattered light for the unpolarized incident light is found to be proportional to
\begin{eqnarray}
\frac{1}{2\pi} \int_{0}^{2\pi}d\beta \left[ (\sin^2\beta) + (\cos^2\beta\cos^2\Theta)\right] = \left[ \left(\frac{1}{2}\right) + \left(\frac{1}{2}\cos^2\Theta\right) \right] = \frac{1}{2}\left( 1+\cos^2\Theta \right).
\label{eqn:scattered_intensity_scalar}
\end{eqnarray}
From the first and second terms in the square brackets in Equation~\ref{eqn:scattered_intensity_scalar}, we respectively define $\perp$ and $\parallel$ components of normalized scattered photon beam intensities as:
\begin{eqnarray}
I'^{0}_{\perp} &=& \frac{3}{4}\\
I'^{0}_{\parallel} &=& \frac{3}{4}\cos^2\Theta\\
I'^{0}&=&I'^{0}_{\perp}+I'^{0}_{\parallel} = \frac{3}{4}(1+\cos^2\Theta),\label{eqn:normed_intensity_scalar}
\end{eqnarray}
where the normalization condition is $\frac{1}{4\pi}\int_{\Omega'} I'^{0} d\Omega' = 1$ (\S 6 of \citealt{pla34}; \S 60 of \citealt{ber82}).
The superscript $0$ indicates the {\it scalar} scattering.
Equation~\ref{eqn:normed_intensity_scalar} indicates that {\it the phase function}\footnote{The phase function $p(\cos\Theta)$ is defined as $I'(\theta', \phi') \propto \int_{\Omega}p(\cos\Theta)I(\theta, \phi)d\Omega$ normalized as $\frac{1}{4\pi}\int_{\Omega'}p(\cos\Theta)d\Omega' = 1$, where $I$ and $I'$ are incident and scattered intensities.} of this scattering process is the Rayleigh's phase function: $p(\cos\Theta) = \frac{3}{4}(1+\cos^2\Theta) = P_{0}(\cos\Theta) + \frac{1}{2}P_{2}(\cos\Theta)$, where $P_{l}(x)$ is the Legendre polynomials\footnote{The Legendre polynomials of order 0, 1, and 2 are respectively $P_{0}(x)=1$, $P_{1}(x)=x$, and $P_{2}(x)=\frac{1}{2}(3x^2-1)$} of order $l$ \citep[][\S 3]{cha60}.
The ratio of the  $\parallel$ to $\perp$ components of the scattered photon intensities, or the depolarization factor, is $I'^{0}_{\parallel}/I'^{0}_{\perp} = \cos^2\Theta$, and the degree of the linear polarization $\Pi^{0}$ is
\begin{eqnarray}
{\Pi}^{0}(\Theta) = \frac{I'^{0}_{\perp}-I'^{0}_{\parallel}}{I'^{0}_{\perp}+I'^{0}_{\parallel}} = \frac{1-\cos^2\Theta}{1+\cos^2\Theta}. 
\end{eqnarray}
As is the case for the classical Thomson scattering \citep[e.g.,][]{ryb79}, the scattered light is 100\% polarized at $\Theta = 90^{\circ}$.

In the case of the {\it \it symmetric} scattering ($d\sigma/d\Omega \propto  3+(\epsilon^{(\alpha)} \cdot \epsilon^{(\alpha')})^2$), the intensity of the scattered light for the unpolarized incident light is proportional to
\begin{eqnarray}
\frac{1}{2\pi} \int_{0}^{2\pi}d\beta \left[ (3+\sin^2\beta) + (3+\cos^2\beta\cos^2\Theta)\right] = \left[ \left(\frac{7}{2}\right) + \left(\frac{6+\cos^2\Theta}{2}\right) \right] = \frac{1}{2}\left( 13+\cos^2\Theta \right).
\end{eqnarray}
The $\perp$, $\parallel$, and total normalized scattered intensities can be defined as 
\begin{eqnarray}
I'^{s}_{\perp} &=& \frac{21}{40}\\
I'^{s}_{\parallel} &=& \frac{3}{40}(6+\cos^2\Theta)\\
I'^{s}&=&I'^{s}_{\perp}+I'^{s}_{\parallel} = \frac{3}{40}(13+\cos^2\Theta), \label{eqn:normed_intensity_symmetric}
\end{eqnarray}
where $\frac{1}{4\pi}\int_{\Omega} I'^{s} d\Omega = 1$, and the superscript $s$ indicates the {\it symmetric} scattering (\S 6 of \citealt{pla34}; \S 60 of \citealt{ber82}).
Equation~\ref{eqn:normed_intensity_symmetric} defines the scattering phase function of this process as $p(\cos\Theta) = \frac{3}{40}(13+\cos^2\Theta) = P_{0}(\cos\Theta) + \frac{1}{20}P_{2}(\cos\Theta)$.
The depolarization factor is $I'^{s}_{\parallel}/I'^{s}_{\perp} = (6+\cos^2\Theta)/7$, and the polarization degree is
\begin{eqnarray}
{\Pi}^{s}(\theta) &=& \frac{I'^{s}_{\perp}-I'^{s}_{\parallel}}{I'^{s}_{\perp}+I'^{s}_{\parallel}} = \frac{1-\cos^2\Theta}{13+\cos^2\Theta}.
\end{eqnarray}
We can see that the {\it symmetric} scattering is more isotropic than the {\it scalar} scattering, which may be traced back to the fact that atomic dipole moments not only in the orthogonal directions but also the parallel direction of the propagation of the incident radiation contribute to the scattering process \citep[e.g.,][]{ham47}.
As a result, the {\it symmetric} scattering exhibits much smaller degrees of polarization of $1/13 = 7.7$\% at most (at $\Theta = 90^{\circ}$).

\subsubsection{Phase matrix}
\label{sec:phase_matrix}

The phase function discussed in Section~\ref{sec:phase_function} can be regarded as a special case of a more general concept, {\it the phase matrix}, with which we can consider the scattering of an arbitrarily polarized incident light \citep[][\S 16]{cha60}.
The Rayleigh ({\it scalar}) scattering phase matrix is discussed in detail in \cite{cha60}, and below we extend it to the case of the {\it symmetric} scattering.

A one-parameter family of a matrix $R = R(\gamma)$ may be introduced as a transformation matrix between the incident Stokes vector $\vec{I}=(I_{\parallel}, I_{\perp}, U, V)^{T}$ and scattered Stokes vector $\vec{I'}=(I_{\parallel}', I_{\perp}', U', V')^{T}$, $\vec{I'}\propto R{\vec{I}}$ as \citep[][]{ham47,cha60,ste94}:
\begin{eqnarray}
\setlength{\arraycolsep}{0.2cm}
R(\gamma) = \frac{3}{2(1+2\gamma)} \begin{pmatrix}
\cos^2\Theta + \gamma\sin^2\Theta & \gamma & 0 & 0 \\
\gamma & 1 & 0 & 0 \\
0 & 0 & (1-\gamma)\cos\Theta & 0 \\
0 & 0 & 0 & (1-3\gamma)\cos\Theta
\end{pmatrix},
\label{eqn:phase_matrix}
\end{eqnarray}
where $I=I_{\parallel}+I_{\perp}$, $U$, and $V$ are (Chandrasekhar's notation of) the Stokes parameters \citep{cha46,cha47}\footnote{The polarization angle is measured clockwise when looking at the source, which is opposite from the IAU standard \citep[e.g.,][]{seo22}.}.
This matrix $R$, referred to as the {\it scattering matrix}, describes the evolution of the Stokes vector by a single scattering when the plane of scattering serves as the plane of reference for the Stokes parameters.
The expression given in Equation~\ref{eqn:phase_matrix} is from the classical theory of scattering by anisotropic particles \citep{cha60}, but it provides a convenient parametrization to our specific quantum mechanical problems \citep{ham47,ste94}\footnote{\cite{ham47}, \cite{cha60}, and \cite{ste80,ste94} give more general parametrizations of the scattering matrix for the quantum mechanical resonant scattering problems. For reference, the relationships between $\gamma$ and their parametrizations, after appropriate conversions of the polarization basis \citep{seo22}, are $E_{1} = W_{2} = (1-\gamma)/(1+2\gamma)$, $E_{2}=1-E_{1}=(3\gamma)/(1+2\gamma)$, $E_{3} = W_{1} =(1-3\gamma)/(1+2\gamma)$. $\gamma=0$ equals $(E_{1}, E_{2}, E_{3}, W_{1}, W_{2}) = (1, 0, 1, 1, 1)$, and $\gamma=3/4$ equals $(E_{1}, E_{2}, E_{3}, W_{1}, W_{2}) = (1/10, 9/10, -1/2, -1/2, 1/10)$.}.
The scattering matrix $R$ can be split into diagonal Rayleigh scattering matrix and residual (=isotropic scattering) matrix as \citep{cha60}:
\begin{eqnarray}
\setlength{\arraycolsep}{0.2cm}
R(\gamma) &=& \frac{3(1-\gamma)}{2(1+2\gamma)}\left(\begin{array}{c@{\quad}c@{\quad}c@{\quad}c}
\cos^2\Theta & 0 & 0 & 0 \\
0 & 1 & 0 & 0 \\
0 & 0 & \cos\Theta & 0 \\
0 & 0 & 0 & \cos\Theta
\end{array}\right) + \frac{3\gamma}{2(1+2\gamma)}\left(\begin{array}{c@{\quad}c@{\quad}c@{\quad}c}
1 & 1 & 0 & 0 \\
1 & 1 & 0 & 0 \\
0 & 0 & 0 & 0 \\
0 & 0 & 0 & -2\cos\Theta
\end{array}\right) \nonumber\\
&=& \frac{1-\gamma}{1+2\gamma}R(\gamma=0) + \frac{3\gamma}{1+2\gamma}R(\gamma=1).
\label{eqn:phase_matrix_reduced}
\end{eqnarray}
We find that the scattering matrices for the {\it scalar} and {\it symmetric} scattering respectively correspond to $R(\gamma=0)$ and $R(\gamma=3/4)$, explicitly:
\begin{eqnarray}
\setlength{\arraycolsep}{0.2cm}
R (\gamma=0) = \frac{3}{2}\begin{pmatrix}
\cos^2\Theta & 0 & 0 & 0 \\
0 & 1 & 0 & 0 \\
0 & 0 & \cos\Theta & 0\\
0 & 0 & 0 & \cos\Theta 
\end{pmatrix}
\end{eqnarray}
and
\begin{eqnarray}
R(\gamma=3/4) &=& \frac{1}{10}R(\gamma=0)+\frac{9}{10}R(\gamma=1) \nonumber \\
&=& \frac{3}{20}\left(\begin{array}{c@{\quad}c@{\quad}c@{\quad}c}
\begin{array}{c}
\cos^2\Theta
\end{array}
& \begin{array}{c}
0
\end{array}
& \begin{array}{c}
0
\end{array}
& \begin{array}{c}
0
\end{array} \\
\begin{array}{c}
0
\end{array}
& \begin{array}{c}
1
\end{array}
& \begin{array}{c}
0
\end{array}
& \begin{array}{c}
0
\end{array} \\
\begin{array}{c}
0
\end{array}
& \begin{array}{c}
0
\end{array}
& \begin{array}{c}
\cos\Theta
\end{array}
& \begin{array}{c}
0
\end{array} \\
\begin{array}{c}
0
\end{array} 
& \begin{array}{c}
0
\end{array}
& \begin{array}{c}
0
\end{array}
& \begin{array}{c}
\cos\Theta
\end{array}
\end{array}\right) + \frac{9}{20}\left(\begin{array}{c@{\quad}c@{\quad}c@{\quad}c}
\begin{array}{c}
1
\end{array}
& \begin{array}{c}
1
\end{array}
& \begin{array}{c}
0
\end{array}
& \begin{array}{c}
0
\end{array} \\
\begin{array}{c}
1
\end{array}
& \begin{array}{c}
1
\end{array}
& \begin{array}{c}
0
\end{array}
& \begin{array}{c}
0
\end{array} \\
\begin{array}{c}
0
\end{array}
& \begin{array}{c}
0
\end{array}
& \begin{array}{c}
0
\end{array}
& \begin{array}{c}
0
\end{array} \\
\begin{array}{c}
0
\end{array} 
& \begin{array}{c}
0
\end{array}
& \begin{array}{c}
0
\end{array}
& \begin{array}{c}
-2\cos\Theta
\end{array}
\end{array}\right) \nonumber\\
&=& \frac{3}{20}\left(\begin{array}{c@{\quad}c@{\quad}c@{\quad}c}
\begin{array}{c}
\cos^2\Theta + 3
\end{array}
& \begin{array}{c}
3
\end{array}
& \begin{array}{c}
0
\end{array}
& \begin{array}{c}
0
\end{array} \\
\begin{array}{c}
3
\end{array}
& \begin{array}{c}
4
\end{array}
& \begin{array}{c}
0
\end{array}
& \begin{array}{c}
0
\end{array} \\
\begin{array}{c}
0
\end{array}
& \begin{array}{c}
0
\end{array}
& \begin{array}{c}
\cos\Theta
\end{array}
& \begin{array}{c}
0
\end{array} \\
\begin{array}{c}
0
\end{array} 
& \begin{array}{c}
0
\end{array}
& \begin{array}{c}
0
\end{array}
& \begin{array}{c}
-5\cos\Theta
\end{array}
\end{array}\right).
\end{eqnarray}
For the unpolarized incident light \citep[$I_{\parallel}=I_{\perp}=I/2, U=0, V=0$;][\S 16]{cha60}, the scattered intensity vector is $\vec{I}' = (I_{\parallel}', I_{\perp}', U', V') \propto R\vec{I} = \frac{3}{4}I\left(\cos^2\Theta, 1, 0, 0\right)$ for $\gamma=0$, and $R\vec{I} = \frac{3}{40}I\left(6+\cos^2\Theta, 7, 0, 0\right)$ for $\gamma=3/4$, reducing to the transformations of $I_{\parallel}$ and $I_{\perp}$ due to the {\it scalar} and {\it symmetric} scattering, respectively, as analyzed in Section~\ref{sec:phase_function}.
Equation~\ref{eqn:phase_matrix} indicates that the circular polarization $V$ is scattered independently of other Stokes parameters.
Since the $V$ component in $R(\gamma=3/4)$ has a negative sign, the left-handed circular polarization can be converted into right-handed circular polarization in the case of forward {\it symmetric} scattering.
If $V=0$ in the incident light, the scattered light is free of circularity \citep[e.g.,][Chapter~9.5]{ste94}.


A more convenient expression of the matrix directly related to observable quantities is obtained by using the local meridian plane as a plane of reference for the Stokes parameters \citep[][]{cha60}.
The {\it phase matrix} $P$ is defined such that the basis vectors of the reference frame for the incident (scattered) Stokes parameters are parallel and perpendicular to the local meridian planes of the incident (scattered) light (see Figure~\ref{fig:scattering_plane}).
The local meridian plane basis vectors are:
\begin{eqnarray}
(\vec{\epsilon}_{l}, \vec{\epsilon}_{r}) &=& \left( (-\cos\theta\cos\phi, \cos\theta\sin\phi, \sin\theta)^{T}, (-\sin\phi, \cos\phi, 0)^{T} \right)\\
(\vec{\epsilon}_{l}', \vec{\epsilon}_{r}') &=& \left( (-\cos\theta'\cos\phi', \cos\theta'\sin\phi', \sin\theta')^{T}, (-\sin\phi', \cos\phi', 0)^{T} \right).
\end{eqnarray}
The Stokes vector $\vec{I}=(I_{\parallel}, I_{\perp}, U, V)^{T}$ transforms into $\vec{I}_{\varphi}=(I_{\varphi}, I_{\varphi+\pi/2}, U_{\varphi}, V_{\varphi})^{T}$ by a clockwise rotation of the basis by $\varphi$ as $\vec{I}_{\varphi} = L(\varphi)\vec{I}$, where the $4 \times 4$ rotation matrix $L(\varphi)$ is explicitly given as \citep[][\S 15.5]{cha60}
\begin{eqnarray}
\setlength{\arraycolsep}{0.2cm}
L(\varphi) = \begin{pmatrix}
 \cos^2\varphi & \sin^2\varphi &  \frac{1}{2}\sin 2\varphi & 0 \\
 \sin^2\varphi & \cos^2\varphi & -\frac{1}{2}\sin 2\varphi & 0 \\
-\sin 2\varphi & \sin 2\varphi &             \cos 2\varphi & 0 \\
0 & 0 & 0 & 1 \\
\end{pmatrix}.
\label{eqn:phase_matrix_L}
\end{eqnarray}
We denote the Stokes vectors of the incident and scattered lights expressed with respect to the $\epsilon_{l}-\epsilon_{r}$ basis and $\epsilon_{l}'-\epsilon_{r}'$ basis as $\vec{\tilde{I}} = (I_{l}, I_{r}, \tilde{U}, \tilde{V})^{T}$ and $\vec{\tilde{I}}' = (I_{l}', I_{r}', \tilde{U}', \tilde{V}')^{T}$, respectively.
According to Figure~\ref{fig:scattering_plane}, we can see $\vec{\tilde{I}} = L(i_{1})\vec{I}$ and $\vec{\tilde{I}}' = L(\pi-i_{2})\vec{I}'$, thus $\vec{\tilde{I}}' = L(\pi-i_{2})\vec{I}' \propto L(\pi-i_{2})R\vec{I} = L(\pi-i_{2})RL(-i_{1})\vec{\tilde{I}}$, with which the phase matrix $P$ can be expressed by using $R$ (Equation~\ref{eqn:phase_matrix_reduced}) and $L$ (Equation~\ref{eqn:phase_matrix_L}) as:
\begin{eqnarray}
P(\gamma) = L(\pi - i_{2})R(\gamma)L(-i_{1}) = \frac{1-\gamma}{1+2\gamma}L(\pi - i_{2})R(\gamma=0)L(-i_{1}) + \frac{3\gamma}{1+2\gamma}R(\gamma=1) = \frac{1-\gamma}{1+2\gamma}P(\gamma=0) + \frac{3\gamma}{1+2\gamma}R(\gamma=1),
\end{eqnarray}
where an identity $L(\pi - i_{2})R(\gamma=1)L(-i_{1}) = R(\gamma=1)$ is used.

With explicit calculations, $P(\gamma=0) = L(\pi - i_{2})R(\gamma=0)L(-i_{1})$ is shown to be:
\begin{eqnarray}
P(\gamma=0) &=& \frac{3}{2}\left(\begin{array}{c@{\quad}c@{\quad}c@{}c}
\begin{array}{c}
(l,l)^2
\end{array}
& \begin{array}{c}
(r,l)^2
\end{array}
& \begin{array}{c}
(l,l)(r,l)
\end{array}
& \begin{array}{c}
0
\end{array} \\
\begin{array}{c}
(l,r)^2
\end{array}
& \begin{array}{c}
(r,r)^2
\end{array}
& \begin{array}{c}
(l,r)(r,r)
\end{array}
& \begin{array}{c}
0
\end{array} \\
\begin{array}{c}
2(l,l)(l,r)
\end{array}
& \begin{array}{c}
2(r,r)(r,l)
\end{array}
& \begin{array}{c}
(l,l)(r,r)+(r,l)(l,r)
\end{array}
& \begin{array}{c}
0
\end{array} \\
\begin{array}{c}
0
\end{array} 
& \begin{array}{c}
0
\end{array}
& \begin{array}{c}
0
\end{array}
& \begin{array}{c}
(l,l)(r,r)-(r,l)(l,r)
\end{array}
\end{array}\right)
\end{eqnarray}
where
\begin{subequations}
\begin{eqnarray}
(r,l) &\equiv& \sin i_{1} \cos i_{2} \cos\Theta + \cos i_{1} \sin i_{2}\\
(l,r) &\equiv& -\sin i_{2} \cos i_{1} \cos\Theta - \cos i_{2} \sin i_{1}\\
(r,r) &\equiv& \sin i_{1} \sin i_{2} \cos\Theta - \cos i_{1} \cos i_{2}\\
(l,l) &\equiv& \sin i_{1} \sin i_{2} - \cos i_{1} \cos i_{2} \cos\Theta.
\end{eqnarray}
\end{subequations}
Note that $(l,l)(r,r)-(r,l)(l,r) = \cos\Theta \left(\sin^2(i_{1}-i_{2})+\cos^2(i_{1}-i_{2})\right) = \cos\Theta = \cos\theta\cos\theta' + \sin\theta\sin\theta'\cos(\phi-\phi')$, which means that the Stokes parameter $V$ is unaffected by the rotation of the reference frame \citep{cha47}.
From the the formulas of spherical trigonometry\footnote{The five-part formulas give $\sin(\phi-\phi')\cos\theta = \cos i_{2}\sin i_{1} + \sin i_{2}\cos i_{1}\cos\Theta$ and $\sin(\phi-\phi')\cos\theta' = \cos i_{1}\sin i_{2} + \sin i_{1}\cos i_{2}\cos\Theta$, proving Equations~\ref{eqn:rl} and \ref{eqn:lr}. The supplemental cosine rules give $\cos(\phi-\phi') = -\cos i_{1}\cos i_{2} +\sin i_{1} \sin i_{2} \cos\Theta$, proving Equation~\ref{eqn:rr}. Finally, Cagnoli's Equation gives $\sin\theta\sin\theta' + \cos\theta\cos\theta'\cos(\phi-\phi') = \sin i_{2}\sin i_{1} - \cos i_{2}\cos i_{1}\cos \Theta$, proving Equation~\ref{eqn:ll} \citep[][Part~II, Equations~5, 6, 78]{chauvenet1852treatise}.}, the matrix elements can be explicitly expressed in terms of the spherical coordinates of $\vec{n}$ and $\vec{n}'$ as \citep[Equation~218 of][\S17]{cha60}
\begin{subequations}
\begin{eqnarray}
(r,l) &=& \cos\theta'\sin\Delta\phi \label{eqn:rl}\\
(l,r) &=& -\cos\theta\sin\Delta\phi \label{eqn:lr}\\
(r,r) &=& \cos\Delta\phi \label{eqn:rr}\\
(l,l) &=& \sin\theta\sin\theta' + \cos\theta\cos\theta'\cos\Delta\phi \label{eqn:ll},
\end{eqnarray}
\end{subequations}
where $\Delta\phi \equiv \phi-\phi'$.

From these equations, we obtain the phase matrix for the {\it scalar} scattering as:
\begin{eqnarray}
P(\gamma=0) &=& \frac{3}{2}\left(\begin{array}{c@{\quad}c@{\quad}c@{\quad}c}
\left(\begin{array}{c}
\cos\theta\cos\theta'\cos\Delta\phi\\
+\sin\theta\sin\theta' 
\end{array}\right)^2
& \begin{array}{c}
\cos^2\theta'\sin^2\Delta\phi
\end{array}
& \begin{array}{c}
\frac{1}{2}\cos\theta\cos^2\theta'\sin2\Delta\phi\\
+\sin\theta\sin\theta'\cos\theta'\sin\Delta\phi
\end{array}
& \begin{array}{c}
0
\end{array} \\
\begin{array}{c}
\cos^2\theta\sin^2\Delta\phi
\end{array}
& \begin{array}{c}
\cos^2\Delta\phi
\end{array}
& \begin{array}{c}
-\frac{1}{2}\cos\theta\sin2\Delta\phi
\end{array}
& \begin{array}{c}
0
\end{array} \\
\begin{array}{c}
-\cos^2\theta\cos\theta'\sin2\Delta\phi \\
-2\sin\theta\cos\theta\sin\theta'\sin\Delta\phi
\end{array}
& \begin{array}{c}
\cos\theta'\sin2\Delta\phi
\end{array}
& \begin{array}{c}
\sin\theta\sin\theta'\cos\Delta\phi\\
+\cos\theta\cos\theta'\cos2\Delta\phi
\end{array}
& \begin{array}{c}
0
\end{array} \\
\begin{array}{c}
0
\end{array} 
& \begin{array}{c}
0
\end{array}
& \begin{array}{c}
0
\end{array}
& \begin{array}{c}
\cos\theta\cos\theta' \\
+\sin\theta\sin\theta'\cos\Delta\phi
\end{array}
\end{array}\right)
\label{eqn:phase_matrix_p_explicit_scalar}\\
&=& \frac{3}{4}Q\left(\begin{array}{c@{\quad}c@{\quad}c@{\quad}c}
\begin{array}{c}
2\sin^2\theta\sin^2\theta'+\cos^2\theta\cos^2\theta'
\end{array}
& \begin{array}{c}
\cos^2\theta'
\end{array}
& \begin{array}{c}
0
\end{array}
& \begin{array}{c}
0
\end{array} \\
\begin{array}{c}
\cos^2\theta
\end{array}
& \begin{array}{c}
1
\end{array}
& \begin{array}{c}
0
\end{array}
& \begin{array}{c}
0
\end{array} \\
\begin{array}{c}
0
\end{array}
& \begin{array}{c}
0
\end{array}
& \begin{array}{c}
0
\end{array}
& \begin{array}{c}
0
\end{array} \\
\begin{array}{c}
0
\end{array} 
& \begin{array}{c}
0
\end{array}
& \begin{array}{c}
0
\end{array}
& \begin{array}{c}
\cos\theta\cos\theta'
\end{array}
\end{array}\right) \nonumber\\
&+& \frac{3}{4}\sin\theta\sin\theta' Q\left(\begin{array}{c@{\quad}c@{\quad}c@{\quad}c}
\begin{array}{c}
4\cos\theta\cos\theta' \cos\Delta\phi
\end{array}
& \begin{array}{c}
0
\end{array}
& \begin{array}{c}
2\cos\theta' \sin\Delta\phi
\end{array}
& \begin{array}{c}
0
\end{array} \\
\begin{array}{c}
0
\end{array}
& \begin{array}{c}
0
\end{array}
& \begin{array}{c}
0
\end{array}
& \begin{array}{c}
0
\end{array} \\
\begin{array}{c}
-2\cos\theta\sin\Delta\phi
\end{array}
& \begin{array}{c}
0
\end{array}
& \begin{array}{c}
\cos\Delta\phi
\end{array}
& \begin{array}{c}
0
\end{array} \\
\begin{array}{c}
0
\end{array} 
& \begin{array}{c}
0
\end{array}
& \begin{array}{c}
0
\end{array}
& \begin{array}{c}
\cos\Delta\phi
\end{array}
\end{array}\right) \nonumber\\
&+& \frac{3}{4}Q\left(\begin{array}{c@{\quad}c@{\quad}c@{\quad}c}
\begin{array}{c}
\cos^2\theta\cos^2\theta'\cos 2\Delta\phi
\end{array}
& \begin{array}{c}
-\cos^2\theta'\cos 2\Delta\phi
\end{array}
& \begin{array}{c}
\cos\theta\cos^2\theta'\sin 2\Delta\phi
\end{array}
& \begin{array}{c}
0
\end{array} \\
\begin{array}{c}
-\cos^2\theta\cos 2\Delta\phi
\end{array}
& \begin{array}{c}
\cos 2\Delta\phi
\end{array}
& \begin{array}{c}
-\cos\theta\sin 2\Delta\phi
\end{array}
& \begin{array}{c}
0
\end{array} \\
\begin{array}{c}
-\cos^2\theta\cos\theta'\sin 2\Delta\phi
\end{array}
& \begin{array}{c}
\cos\theta'\sin 2\Delta\phi
\end{array}
& \begin{array}{c}
\cos\theta\cos\theta'\cos 2\Delta\phi
\end{array}
& \begin{array}{c}
0
\end{array} \\
\begin{array}{c}
0
\end{array} 
& \begin{array}{c}
0
\end{array}
& \begin{array}{c}
0
\end{array}
& \begin{array}{c}
0
\end{array}
\end{array}\right),
\label{eqn:phase_matrix_p_explicit_scalar_split}
\end{eqnarray}
where
\begin{eqnarray}
Q &=& \left(\begin{array}{c@{\quad}c@{\quad}c@{\quad}c}
\begin{array}{c}
1
\end{array}
& \begin{array}{c}
0
\end{array}
& \begin{array}{c}
0
\end{array}
& \begin{array}{c}
0
\end{array} \\
\begin{array}{c}
0
\end{array}
& \begin{array}{c}
1
\end{array}
& \begin{array}{c}
0
\end{array}
& \begin{array}{c}
0
\end{array} \\
\begin{array}{c}
0
\end{array}
& \begin{array}{c}
0
\end{array}
& \begin{array}{c}
2
\end{array}
& \begin{array}{c}
0
\end{array} \\
\begin{array}{c}
0
\end{array} 
& \begin{array}{c}
0
\end{array}
& \begin{array}{c}
0
\end{array}
& \begin{array}{c}
2
\end{array}
\end{array}\right). \nonumber
\end{eqnarray}
This phase matrix is identical to that given in \cite{cha60} (Equation~219, \S 17) for the electron Thomson scattering.
The first term in Equation~\ref{eqn:phase_matrix_p_explicit_scalar_split} is independent of $\Delta\phi$, and only this term is relevant to the transfer equation for radiative transfer problems with axially symmetric radiation field \citep[][\S 17, \S 69]{cha60}.

We can derive the phase matrix for the {\it symmetric} scattering as:
\begin{eqnarray}
\setlength{\arraycolsep}{0.2cm}
& &P(\gamma=3/4) = \frac{1}{10}P(\gamma=0) + \frac{9}{10}R(\gamma=1)\nonumber\\
&=& \frac{3}{20}\left(\begin{array}{c@{\quad}c@{\quad}c@{\quad}c}
\left(\begin{array}{c}
\cos\theta\cos\theta'\cos\Delta\phi\\
+\sin\theta\sin\theta' 
\end{array}\right)^2 + 3
& \begin{array}{c}
\cos^2\theta'\sin^2\Delta\phi + 3
\end{array}
& \begin{array}{c}
\frac{1}{2}\cos\theta\cos^2\theta'\sin2\Delta\phi\\
+\sin\theta\sin\theta'\cos\theta'\sin\Delta\phi
\end{array}
& \begin{array}{c}
0
\end{array} \\
\begin{array}{c}
\cos^2\theta\sin^2\Delta\phi + 3
\end{array}
& \begin{array}{c}
\cos^2\Delta\phi + 3
\end{array}
& \begin{array}{c}
-\frac{1}{2}\cos\theta\sin2\Delta\phi
\end{array}
& \begin{array}{c}
0
\end{array} \\
\begin{array}{c}
-\cos^2\theta\cos\theta'\sin2\Delta\phi \\
-2\sin\theta\cos\theta\sin\theta'\sin\Delta\phi
\end{array}
& \begin{array}{c}
\cos\theta'\sin2\Delta\phi
\end{array}
& \begin{array}{c}
\sin\theta\sin\theta'\cos\Delta\phi\\
+\cos\theta\cos\theta'\cos2\Delta\phi
\end{array}
& \begin{array}{c}
0
\end{array} \\
\begin{array}{c}
0
\end{array} 
& \begin{array}{c}
0
\end{array}
& \begin{array}{c}
0
\end{array}
& \begin{array}{c}
-5\cos\theta\cos\theta' \\
-5\sin\theta\sin\theta'\cos\Delta\phi
\end{array}
\end{array}\right) \label{eqn:phase_matrix_p_explicit_symmetric} \\
&=& \frac{3}{40}Q\left(\begin{array}{c@{\quad}c@{\quad}c@{\quad}c}
\begin{array}{c}
2\sin^2\theta\sin^2\theta'+\cos^2\theta\cos^2\theta' + 6
\end{array}
& \begin{array}{c}
\cos^2\theta' + 6
\end{array}
& \begin{array}{c}
0
\end{array}
& \begin{array}{c}
0
\end{array} \\
\begin{array}{c}
\cos^2\theta + 6
\end{array}
& \begin{array}{c}
6
\end{array}
& \begin{array}{c}
0
\end{array}
& \begin{array}{c}
0
\end{array} \\
\begin{array}{c}
0
\end{array}
& \begin{array}{c}
0
\end{array}
& \begin{array}{c}
0
\end{array}
& \begin{array}{c}
0
\end{array} \\
\begin{array}{c}
0
\end{array} 
& \begin{array}{c}
0
\end{array}
& \begin{array}{c}
0
\end{array}
& \begin{array}{c}
-5\cos\theta\cos\theta'
\end{array}
\end{array}\right) \nonumber\\
&+& \frac{3}{40}\sin\theta\sin\theta' Q\left(\begin{array}{c@{\quad}c@{\quad}c@{\quad}c}
\begin{array}{c}
4\cos\theta\cos\theta' \cos\Delta\phi
\end{array}
& \begin{array}{c}
0
\end{array}
& \begin{array}{c}
2\cos\theta' \sin\Delta\phi
\end{array}
& \begin{array}{c}
0
\end{array} \\
\begin{array}{c}
0
\end{array}
& \begin{array}{c}
0
\end{array}
& \begin{array}{c}
0
\end{array}
& \begin{array}{c}
0
\end{array} \\
\begin{array}{c}
-2\cos\theta\sin\Delta\phi
\end{array}
& \begin{array}{c}
0
\end{array}
& \begin{array}{c}
\cos\Delta\phi
\end{array}
& \begin{array}{c}
0
\end{array} \\
\begin{array}{c}
0
\end{array} 
& \begin{array}{c}
0
\end{array}
& \begin{array}{c}
0
\end{array}
& \begin{array}{c}
-5\cos\Delta\phi
\end{array}
\end{array}\right) \nonumber\\
&+& \frac{3}{40}Q\left(\begin{array}{c@{\quad}c@{\quad}c@{\quad}c}
\begin{array}{c}
\cos^2\theta\cos^2\theta'\cos 2\Delta\phi
\end{array}
& \begin{array}{c}
-\cos^2\theta'\cos 2\Delta\phi
\end{array}
& \begin{array}{c}
\cos\theta\cos^2\theta'\sin 2\Delta\phi
\end{array}
& \begin{array}{c}
0
\end{array} \\
\begin{array}{c}
-\cos^2\theta\cos 2\Delta\phi
\end{array}
& \begin{array}{c}
\cos 2\Delta\phi
\end{array}
& \begin{array}{c}
-\cos\theta\sin 2\Delta\phi
\end{array}
& \begin{array}{c}
0
\end{array} \\
\begin{array}{c}
-\cos^2\theta\cos\theta'\sin 2\Delta\phi
\end{array}
& \begin{array}{c}
\cos\theta'\sin 2\Delta\phi
\end{array}
& \begin{array}{c}
\cos\theta\cos\theta'\cos 2\Delta\phi
\end{array}
& \begin{array}{c}
0
\end{array} \\
\begin{array}{c}
0
\end{array} 
& \begin{array}{c}
0
\end{array}
& \begin{array}{c}
0
\end{array}
& \begin{array}{c}
0
\end{array}
\end{array}\right).
\label{eqn:phase_matrix_p_explicit_symmetric_split}
\end{eqnarray}
Similar to the {\it scalar} case, only the first term in Equation~\ref{eqn:phase_matrix_p_explicit_symmetric_split} is important for radiative transfer problems with the axially symmetric radiation field.

The scattering matrix $R$ and phase matrix $P$ given above are defined as the transformation matrices for the Stokes vectors $\vec{I} = (I_{\parallel}, I_{\perp}, U, V)^{T}$ and $\vec{\tilde{I}} = (I_{l}, I_{r}, \tilde{U}, \tilde{V})^{T}$, respectively.
These vectors can be converted into a more commonly used form of the Stokes vectors $(I, Q, U, V)^{T}$ and $(\tilde{I}, \tilde{Q}, \tilde{U}, \tilde{V})^{T}$ as $T\vec{I}$ and $T\vec{\tilde{I}}$, respectively, where $I = \tilde{I} = I_{\parallel} + I_{\perp} = I_{l}+I_{r}$, $Q = I_{\parallel}-I_{\perp}$, $\tilde{Q} = I_{l}-I_{r}$, and
\begin{eqnarray}
T&=& \left(\begin{array}{c@{\quad}c@{\quad}c@{\quad}c}
\begin{array}{c}
1
\end{array}
& \begin{array}{c}
1
\end{array}
& \begin{array}{c}
0
\end{array}
& \begin{array}{c}
0
\end{array} \\
\begin{array}{c}
1
\end{array}
& \begin{array}{c}
-1
\end{array}
& \begin{array}{c}
0
\end{array}
& \begin{array}{c}
0
\end{array} \\
\begin{array}{c}
0
\end{array}
& \begin{array}{c}
0
\end{array}
& \begin{array}{c}
1
\end{array}
& \begin{array}{c}
0
\end{array} \\
\begin{array}{c}
0
\end{array} 
& \begin{array}{c}
0
\end{array}
& \begin{array}{c}
0
\end{array}
& \begin{array}{c}
1
\end{array}
\end{array}\right). \nonumber
\end{eqnarray}
Thus it is straightforward to obtain corresponding scattering and phase matrices for this definition of the Stokes vector as $TRT^{-1}$ and $TPT^{-1}$ \citep[as adopted by, e.g.,][]{ste76,ste94,seo22}.

Similarly, with the condition $V=0$ (no circular polarization), we may identify the Stokes vector components with the $2 \times 2$ density matrix of the photons $\rho$ often used in Monte-Carlo radiative transfer studies as $\vec{\rho} \equiv (\rho_{11}, \rho_{22}, \rho_{12}=\rho_{21})^{T} \propto (I_{r}, I_{l}, U/2)^{T} = T_{44}(I_{l}, I_{r}, U)^{T}$ \citep[e.g.,][]{ahn02,cha20,seo22}, and the single scattering of $\vec{\rho}$ may be described by a $3 \times 3$ phase matrix $T_{44}P_{44}T_{44}^{-1}$, where $P_{44}$ denotes the upper left $3 \times 3$ submatrix of $P$, and 
\small
\begin{eqnarray}
T_{44} &=& \left(\begin{array}{c@{\quad}c@{\quad}c}
\begin{array}{c}
0
\end{array}
& \begin{array}{c}
1
\end{array}
& \begin{array}{c}
0
\end{array} \\
\begin{array}{c}
1
\end{array}
& \begin{array}{c}
0
\end{array}
& \begin{array}{c}
0
\end{array} \\
\begin{array}{c}
0
\end{array}
& \begin{array}{c}
0
\end{array}
& \begin{array}{c}
\frac{1}{2}
\end{array}
\end{array}\right). \nonumber
\end{eqnarray}
For the {\it scalar} ($\gamma=0$) and {\it symmetric} ($\gamma=3/4$) scattering, we respectively obtain
\begin{eqnarray}
\left(\begin{array}{c}
\begin{array}{c}
\rho_{11}'
\end{array}\\
\begin{array}{c}
\rho_{22}'
\end{array}\\
\begin{array}{c}
\rho_{12}'
\end{array}
\end{array}\right) &=& \frac{3}{2}\left(\begin{array}{c@{\quad}c@{\quad}c}
\begin{array}{c}
\cos^2\Delta\phi
\end{array}
& \begin{array}{c}
\cos^2\theta\sin^2\Delta\phi
\end{array}
& \begin{array}{c}
-\cos\theta\sin2\Delta\phi
\end{array} \\
\begin{array}{c}
\cos^2\theta'\sin^2\Delta\phi
\end{array}
& \left(\begin{array}{c}
\cos\theta\cos\theta'\cos\Delta\phi\\
+ \sin\theta\sin\theta'
\end{array}\right)^2
& \begin{array}{c}
\cos\theta\cos^2\theta'\sin2\Delta\phi\\
+ 2\sin\theta\sin\theta'\cos\theta'\sin\Delta\phi
\end{array} \\
\begin{array}{c}
\frac{1}{2}\cos\theta'\sin2\Delta\phi
\end{array}
& \begin{array}{c}
-\frac{1}{2}\cos^2\theta\cos\theta'\sin2\Delta\phi\\
-\sin\theta\cos\theta\sin\theta'\sin\Delta\phi
\end{array}
& \begin{array}{c}
\sin\theta\sin\theta'\cos\Delta\phi\\
+\cos\theta\cos\theta'\cos2\Delta\phi
\end{array}
\end{array}\right) \left(\begin{array}{c}
\begin{array}{c}
\rho_{11}
\end{array}\\
\begin{array}{c}
\rho_{22}
\end{array}\\
\begin{array}{c}
\rho_{12}
\end{array}
\end{array}\right) \quad [\gamma=0] \label{eqn:density_matrix_scalar}\\
\left(\begin{array}{c}
\begin{array}{c}
\rho_{11}'
\end{array}\\
\begin{array}{c}
\rho_{22}'
\end{array}\\
\begin{array}{c}
\rho_{12}'
\end{array}
\end{array}\right) &=& \frac{3}{20}\left(\begin{array}{c@{\quad}c@{\quad}c}
\begin{array}{c}
\cos^2\Delta\phi + 3
\end{array}
& \begin{array}{c}
\cos^2\theta\sin^2\Delta\phi + 3
\end{array}
& \begin{array}{c}
-\cos\theta\sin2\Delta\phi
\end{array} \\
\begin{array}{c}
\cos^2\theta'\sin^2\Delta\phi + 3
\end{array}
& \left(\begin{array}{c}
\cos\theta\cos\theta'\cos\Delta\phi\\
+ \sin\theta\sin\theta'
\end{array}\right)^2  + 3
& \begin{array}{c}
\cos\theta\cos^2\theta'\sin2\Delta\phi\\
+ 2\sin\theta\sin\theta'\cos\theta'\sin\Delta\phi
\end{array} \\
\begin{array}{c}
\frac{1}{2}\cos\theta'\sin2\Delta\phi
\end{array}
& \begin{array}{c}
-\frac{1}{2}\cos^2\theta\cos\theta'\sin2\Delta\phi\\
-\sin\theta\cos\theta\sin\theta'\sin\Delta\phi
\end{array}
& \begin{array}{c}
\sin\theta\sin\theta'\cos\Delta\phi\\
+\cos\theta\cos\theta'\cos2\Delta\phi
\end{array}
\end{array}\right) \left(\begin{array}{c}
\begin{array}{c}
\rho_{11}
\end{array}\\
\begin{array}{c}
\rho_{22}
\end{array}\\
\begin{array}{c}
\rho_{12}
\end{array}
\end{array}\right) \quad [\gamma=3/4]. \label{eqn:density_matrix_symmetric}
\end{eqnarray}
\normalsize
Equation~\ref{eqn:density_matrix_scalar} is identical to the transformation rules for the density matrix elements for the Rayleigh scattering given in \cite{cha20} except for the normalization, and Equation~\ref{eqn:density_matrix_symmetric} is newly derived here \citep[see e.g.,][and references therein for different conventions used in expressing the Stokes vectors, scattering matrix, phase matrix, and density matrix]{seo22}.

As we will discuss in Section~\ref{sec:discussion}, the Raman scattering features are produced by multiple Rayleigh scatterings followed by a Raman scattering to transform UV photons to optical photons, which immediately escape from the neutral hydrogen scattering region \citep{nus89,lee98,cha15}.
The phase matrices derived here can readily be applied to Monte-Carlo radiative transfer simulations to evaluate emergent polarization of photons escaped from scattering regions of complex geometry after the Rayleigh and Raman scatterings \citep[e.g.,][]{sch92,cha20,seo22}, which we will leave for the forthcoming paper.

\section{Discussion: Broad optical emission features due to the Raman wavelength conversion}
\label{sec:discussion}

\subsection{Definitions of optical thickness and branching ratios}

\begin{figure}
\center{
\includegraphics[clip, width=4.4in]{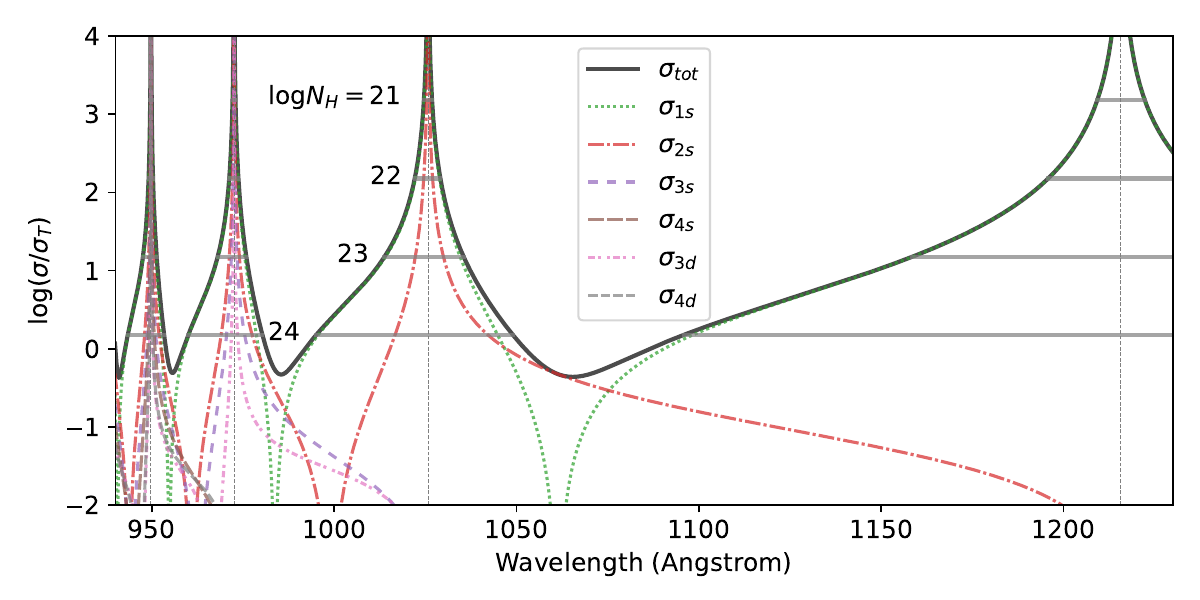}
\includegraphics[clip, width=4.4in]{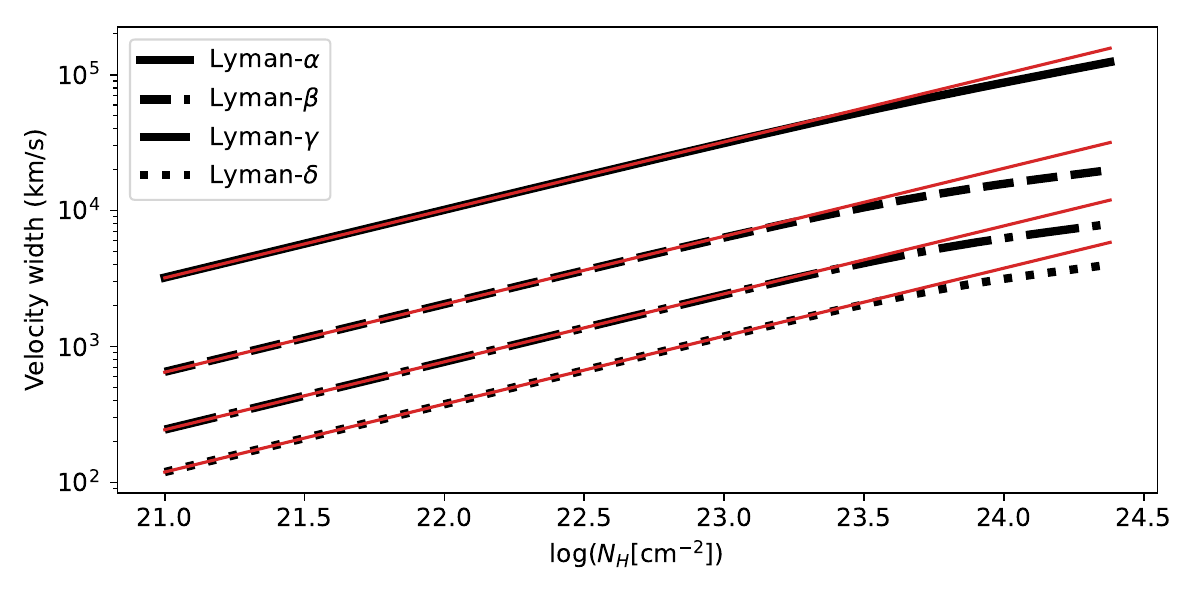}
}\vspace{0cm}
 \caption{Top: Same as the top panel of Figure~\ref{fig:crosssections}. The horizontal lines denote the wavelength intervals $\Delta\lambda$ in which the scattering optical depth $\tau_{\lambda} = \sigma_{\text{tot}}N_{\text{H}}$ gets greater than 1 for pure hydrogen scattering gas with $\log N_{\text{H}}(\text{cm}^{-2}) = 21, 22, 23$, and $24$. 
 Bottom: the velocity widths corresponding to $\Delta\lambda$ of each of the Lyman series: $c(\Delta\lambda/\lambda)_{\text{Ly}}$ for $\text{Ly} = \text{Ly}\alpha$, $\text{Ly}\beta$, $\text{Ly}\gamma$, and $\text{Ly}\delta$. The overlayed thin solid lines denote the power-law model $c(\Delta\lambda/\lambda)_{\text{Ly}} \propto N_{\text{H}}^{0.5}$ fitted over $N_{\text{H}} = 10^{21-23}~\text{cm}^{-2}$ (Equation~\ref{eqn:optically_thick}).}
 \label{fig:opticaldepth}
\end{figure}

Consider a pure neutral hydrogen gas with a column density of $N_{\text{H}}$ along the travel path of a single incident photon, in which all the hydrogen atoms are in the ground-state.
The scattering optical depth is
\begin{eqnarray}
\tau_{\lambda} = \sigma_{\text{tot}}N_{\text{H}} = \frac{\sigma_{\text{tot}}}{\sigma_{T}} \left( \frac{N_{\text{H}}}{10^{24.177}~\text{cm}^{-2}} \right)
\end{eqnarray}
where $\sigma_{\text{tot}}$ is the sum of all the Rayleigh and Raman scattering cross-sections: 
\begin{eqnarray}
\sigma_{\text{tot}} &=& \sigma_{s} + \sigma_{d},
\end{eqnarray}
where $\sigma_{s}$ and $\sigma_{d}$ are the sums of the $1s\rightarrow n_{B}s$ and $1s\rightarrow n_{B}d$ scattering cross-sections:
\begin{eqnarray}
\sigma_{s} &=& \sum_{n_{B}=1}^{n_{\text{max}}} \sigma_{n_{B}s} \label{eqn:sigma_tot_s}\\
\sigma_{d} &=& \sum_{n_{B}=3}^{n_{\text{max}}} \sigma_{n_{B}d}.\label{eqn:sigma_tot_d}
\end{eqnarray}

Since $\sigma_{\text{tot}}$ has sharp peaks with $\sigma_{\text{tot}} \gg \sigma_{T}$ at around the resonance wavelengths, the hydrogen gas with $N_{H} \gtrsim 10^{21}~\text{cm}^{-2}$ is optically-thick to the incident photons with the wavelengths close to the resonance wavelengths, as indicated in Figure~\ref{fig:opticaldepth}.
The wavelength interval of each resonance where $\tau_{\lambda} \geq 1$ is denoted as $\Delta \lambda$.
$\Delta\lambda$ normalized by the resonance wavelength, $(\Delta\lambda/\lambda)_{\text{Ly}}$, is approximately in proportion to $N_{\text{H}}^{1/2}$ as the cross-section at the resonance peak $\omega \approx\omega_{\text{Ly}}$ follows $\propto (\omega - \omega_{\text{Ly}})^{-2}$ \citep[e.g.,][]{cha15}.
As shown in Figure~\ref{fig:opticaldepth}, we obtain the fitting functions for the velocity widths $\Delta V_{\text{Ly}} = c(\Delta\lambda/\lambda)_{\text{Ly}}$ of each of the Lyman series, fitted over $N_{\text{H}} = 10^{21-23}~\text{cm}^{-2}$, as:
\begin{subequations}
\begin{eqnarray}
\Delta V_{\text{Ly}\alpha} &=& 10124~\text{km}~\text{s}^{-1} \left( \frac{N_{\text{H}}}{10^{22}~\text{cm}^{-2}} \right)^{1/2}\\
\Delta V_{\text{Ly}\beta}  &=&  2048~\text{km}~\text{s}^{-1} \left( \frac{N_{\text{H}}}{10^{22}~\text{cm}^{-2}} \right)^{1/2}\\
\Delta V_{\text{Ly}\gamma} &=&   772~\text{km}~\text{s}^{-1} \left( \frac{N_{\text{H}}}{10^{22}~\text{cm}^{-2}} \right)^{1/2}\\
\Delta V_{\text{Ly}\delta} &=&   377~\text{km}~\text{s}^{-1} \left( \frac{N_{\text{H}}}{10^{22}~\text{cm}^{-2}} \right)^{1/2}.
\end{eqnarray}
\label{eqn:optically_thick}
\end{subequations}
As is clear in Figure~\ref{fig:opticaldepth}, the neutral hydrogen gas of $N_{\text{H}} \gtrsim 10^{24.5}~\text{cm}^{-2}$ becomes optically-thick to all the UV photons at $\lambda \lesssim 1600$\AA.

Wavelength-dependent relative probabilities of various scattering paths for each scattering event, branching ratios ($BR$), are defined as the ratio of the cross-section of each scattering path to $\sigma_{\text{tot}}$ \citep[e.g.,][]{cha20}.
The $BR$s of the $n_{B}s$-branch and $n_{B}d$-branch are $BR(n_{B}s) = \sigma_{n_{B}s}/\sigma_{\text{tot}}$ and $BR(n_{B}d) = \sigma_{n_{B}d}/\sigma_{\text{tot}}$, and $s$-branch and $d$-branch are $BR(s) = \sigma_{s}/\sigma_{\text{tot}}$ and $BR(d) = \sigma_{d}/\sigma_{\text{tot}}$, respectively.
The Rayleigh branch and Raman branch can be defined as $BR(\text{Rayleigh}) = \sigma_{1s}/\sigma_{\text{tot}}$ and $BR(\text{Raman}) = (\sigma_{\text{tot}}-\sigma_{1s})/\sigma_{\text{tot}}$, respectively, and the Raman $s$-branch and Raman $d$-branch are $BR(\text{Raman}~s) = (\sigma_{s}-\sigma_{1s})/(\sigma_{\text{tot}}-\sigma_{1s})$ and $BR(\text{Raman}~d) = \sigma_{d}/(\sigma_{\text{tot}}-\sigma_{1s})$.
Figure~\ref{fig:crosssections} shows the $BR$s to $n_{B}=1$, $2$, $3$, $4$, and the $BR$s of the Raman $s$-branch and Raman $d$-branch, as a function of the wavelength of the incident photon, where $n_{\text{max}}=4$.

\subsection{ Broad Balmer, Paschen, Brackett, and higher-level hydrogen emission features due to the Raman conversion of the UV continuum photons }
\label{sec:broad_hydrogen_feature}

\begin{figure}
\center{
\includegraphics[clip, width=6.8in]{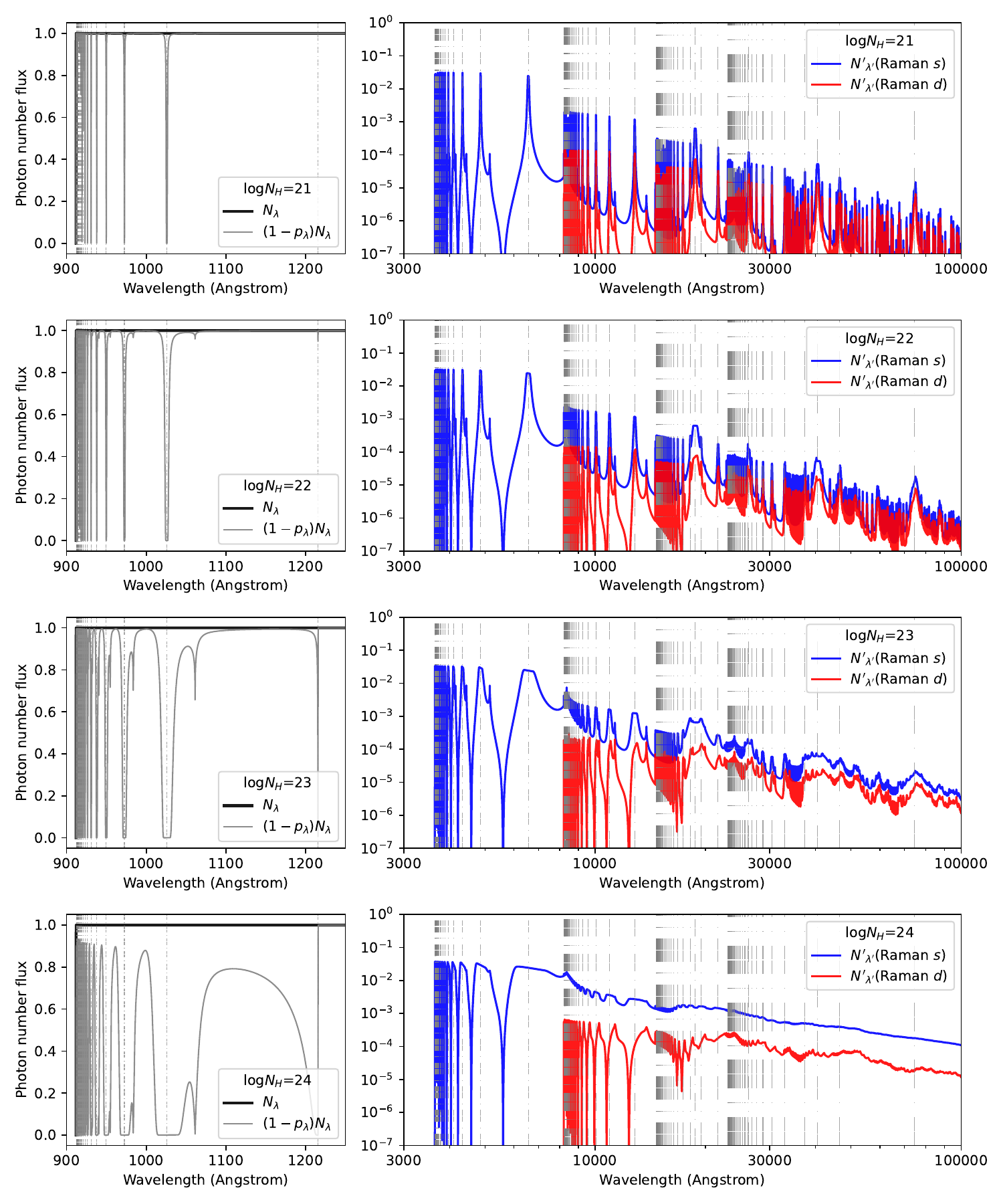}
}\vspace{0cm}
 \caption{ Left column: the incident UV photon number flux $N_{\lambda}$ ($\alpha_{\nu}=-1$) and Raman-extinguished photon number flux $(1-p_{\lambda})N_{\lambda}$ for $\log N_{\text{H}}(\text{cm}^{-2})=21$, $22$, $23$, and $24$. $N_{\lambda}$ is assumed to be a step function: $1$ at $\lambda > \lambda_{\text{LyLimit}}$ and 0 at $\lambda \leq \lambda_{\text{LyLimit}}$. The Raman-scattered photons $p_{\lambda}N_{\lambda}$ are converted into optical-IR photons (Equation~\ref{eqn:extinguished_amount}). The calculations of the Rayleigh and Raman scattering cross-sections are performed up to $n_{B}=12$ ($n_{\text{max}}=12$ in Equations~\ref{eqn:sigma_tot_s} and \ref{eqn:sigma_tot_d}). Right column: the Raman-converted optical-IR photon number flux $N'_{\lambda'}$ (Equations~\ref{eqn:raman_converted_flux}) from each of the Raman $s$-branch and Raman $s$-branch. The vertical lines indicate the Lyman, Balmer, Paschen, Brackett, and Pfund resonance wavelengths.}
 \label{fig:broadband_sed}
\end{figure}

\begin{figure}
\center{
\includegraphics[clip, width=6.8in]{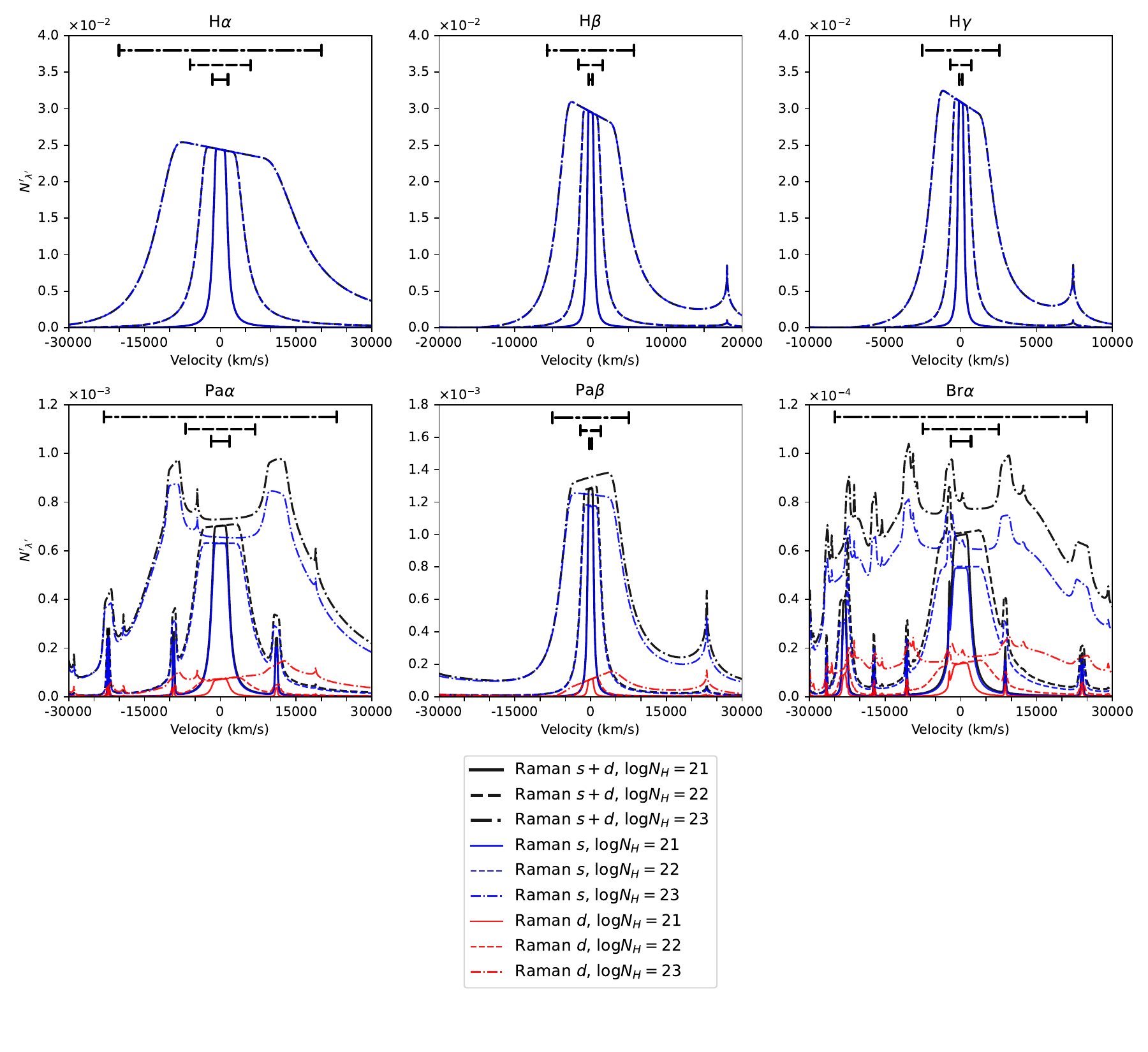}
}\vspace{-0.5cm}
 \caption{ Same as the right column of Figure~\ref{fig:broadband_sed}, but is plotted as a function of velocity defined as $c(\lambda - \lambda_{0})/\lambda_{0}$ where $\lambda_{0} = \lambda_{\text{H}\alpha}$, $\lambda_{\text{H}\beta}$, $\lambda_{\text{H}\gamma}$, $\lambda_{\text{Pa}\alpha}$, $\lambda_{\text{Pa}\beta}$, and $\lambda_{\text{Br}\alpha}$. The solid, dashed, and dashed-dotted curves denote $N'_{\lambda'}$ for $\log N_{\text{H}}(\text{cm}^{-2}) = 21$, $22$, and $23$, respectively. The horizontal lines denote the analytic models of the velocity width $\Delta V$ given by Equation~\ref{eqn:broadened_lines} for $\log N_{\text{H}}(\text{cm}^{-2}) = 21$, $22$, and $23$.}
 \label{fig:broad_emission_feature}
\end{figure}

Consider the Rayleigh/Raman scattering of UV photons by a neutral hydrogen gas of $10^{20}~\text{cm}^{-2} \lesssim N_{\text{H}} \lesssim 10^{24.5}~\text{cm}^{-2}$ in a wavelength range of $\lambda < \lambda_{\text{Ly}\alpha}$ where the Raman branch ratio $BR(\text{Raman})$ is non-zero (see the bottom panel of Figure~\ref{fig:crosssections}).
In this $N_{\text{H}}$ range, the neutral hydrogen gas is optically-thick to the UV photons around the Lyman resonances (Figure~\ref{fig:opticaldepth}).
Since $\sigma_{1s}$ dominates the contribution to $\sigma_{\text{tot}}$ over a wide range of the wavelength around the Lyman resonances, the incident UV photons travelling into the optically-thick gas experience multiple $1s \rightarrow 1s$ Rayleigh scattering until they escape (the number of scattering is approximately $n_{\text{sca}} \approx \tau_{\lambda}^{2} + \tau_{\lambda}$; \citealt{ryb79}).
In the course of the photon's journey inside the scattering region, a fraction of the UV photons are Raman-scattered and converted into optical/IR photons, then the Raman-scattered optical/IR photons immediately escape from the scattering region because $\sigma_{\text{tot}} \ll \sigma_{T}$ in the optical wavelengths \citep[e.g.,][]{nus89,lee98,cha15}.

The input UV-IR spectrum is assumed to be a featureless continuum $f_{\nu} \propto \nu^{\alpha_{\nu}}$ or $f_{\lambda} \propto \lambda^{-\alpha_{\nu}-2}$.
The corresponding photon number flux is $N_{\lambda}d\lambda = \lambda f_{\lambda}d\lambda \propto \lambda^{-\alpha_{\nu}-1}d\lambda$ \citep{lee98}.
The number of the scattering for an incident photon with wavelength $\lambda$ is approximated as $n_{\text{sca}} = \tau_{\lambda}^2+\tau_{\lambda}$, and the probability of being Raman-scattered during the $n_{\text{sca}}$ scatterings is $p_{\lambda} = 1 - (\sigma_{1s}/\sigma_{\text{tot}})^{n_{\text{sca}}}$.
Assuming that the Raman scattering extinguishes the input photon number flux while the Rayleigh scattering does not alter the input, the extinguished amount of the UV photon number flux is 
\begin{eqnarray}
p_{\lambda}N_{\lambda}d\lambda, 
\label{eqn:extinguished_amount}
\end{eqnarray}
which is Raman-converted into optical-IR photons.
From Equations~\ref{eqn:wavelength_conversion} and \ref{eqn:differential_wavelength_conversion}, the Raman-scattered photon number flux for a particular Raman branch $1s\rightarrow n_{B}s/n_{B}d$ as a function of $\lambda' = \lambda/[1-\lambda/((1-n_{B}^{-2})^{-1}\lambda_{\text{LyLimit}})]$ is given as
\begin{eqnarray}
N_{\lambda'}'(\text{Raman}~n_{B}s) &=& p_{\lambda}N_{\lambda}\frac{\sigma_{n_{B}s}}{\sigma_{\text{tot}}-\sigma_{1s}}\frac{d\lambda}{d\lambda'} = p_{\lambda}N_{\lambda}\frac{\sigma_{n_{B}s}}{\sigma_{\text{tot}}-\sigma_{1s}}\left(\frac{\lambda}{\lambda'}\right)^2 \quad [n_{B}\geq 2]\\
N_{\lambda'}'(\text{Raman}~n_{B}d) &=& p_{\lambda}N_{\lambda}\frac{\sigma_{n_{B}d}}{\sigma_{\text{tot}}-\sigma_{1s}}\frac{d\lambda}{d\lambda'} = p_{\lambda}N_{\lambda}\frac{\sigma_{n_{B}d}}{\sigma_{\text{tot}}-\sigma_{1s}}\left(\frac{\lambda}{\lambda'}\right)^2 \quad [n_{B}\geq 3],
\end{eqnarray}
and the total spectra from the Raman $s$-branch and $d$-branch are
\begin{eqnarray}
N_{\lambda'}'(\text{Raman}~s+d) &=& N_{\lambda'}'(\text{Raman}~s) + N_{\lambda'}'(\text{Raman}~d)
\label{eqn:raman_converted_flux}
\end{eqnarray}
where
\begin{eqnarray}
N_{\lambda'}'(\text{Raman}~s) &=& \sum_{n_{B}=2}^{n_{max}} N_{\lambda'}'(\text{Raman}~n_{B}s)\\
N_{\lambda'}'(\text{Raman}~d) &=& \sum_{n_{B}=3}^{n_{max}} N_{\lambda'}'(\text{Raman}~n_{B}d).
\end{eqnarray}

Figure~\ref{fig:broadband_sed} shows the incident UV continuum ($\alpha_{\nu}=-1$) and Raman-converted optical-IR photon number flux spectra calculated for $\log N_{\text{H}}(\text{cm}^{-2}) = 21$, $22$, $23$, and $24$, where $N_{\lambda}$ is assumed to be $1$ at $\lambda > \lambda_{\text{LyLimit}}$ and 0 otherwise.\footnote{If photons with $\lambda < \lambda_{\text{LyLimit}}$ are incident, photo-ionization and recombination dominate the resultant UV-optical spectrum (though the Raman-scattering could have non-negligible contribution to the diffuse continuum), which is beyond the scope of this paper.}
Here the calculations of the Rayleigh and Raman scattering cross-sections are performed up to $n_{B} = 12$ ($n_{\text{max}} = 12$ in Equations~\ref{eqn:sigma_tot_s} and \ref{eqn:sigma_tot_d})
As shown in the left column of Figure~\ref{fig:broadband_sed}, the larger is $N_{\text{H}}$, the wider the wavelength ranges around the Lyman resonances where the photons get optically-thick to the Raman-scattering cross-section.
The Raman-scattered photons $p_{\lambda}N_{\lambda}d\lambda$ are converted into optical-IR photons $N'_{\lambda'}(\text{Raman}~s)d\lambda'$ and $N'_{\lambda'}(\text{Raman}~d)d\lambda'$ following Equation~\ref{eqn:raman_converted_flux}. 
At around the Lyman resonances, all the incident UV photons are Raman-converted, thus the resultant Raman-scattered broad optical emission features generally exhibit flat-topped profile \citep[as pointed out by, e.g.,][]{cha15}.
Since the high-level Raman-scattered emission features overlap with each other at IR wavelengths, the unresolved Raman-scattered photons form a pseudo-continuum.

The Balmer broad features are solely produced via the Raman $2s$-branch and the scattering phase function is that of the {\it scalar} scattering, i.e., the Rayleigh phase function.
On the other hand, the Paschen ($n_{B}=3$), Brackett ($n_{B}=4$), and higher-level transitions ($n_{B} \geq 5$) are produced via the sum of $n_{B}s$-branch and $n_{B}d$-branch, thus the total scattering phase function is the weighted sum of the {\it scalar} and {\it symmetric} scattering phase functions (Section~\ref{sec:phase_function_and_phase_matrix}).
The more the $d$-branch contributes to the total Raman scattering, the less linear polarization of the resulting Raman-scattered emission feature exhibits.
We can see in Figure~\ref{fig:broadband_sed} that the Raman $d$-branch contribution relative to the total Raman $s+d$-branch emission can become $>10$\%, especially at off-resonance wavelengths, the depolarization effect due to the Raman $d$-branch {\it symmetric} scattering would be at a detectable level (Section~\ref{sec:phase_function_and_phase_matrix}).

Figure~\ref{fig:broad_emission_feature} shows the Raman-scattered optical-IR photon number flux $N'_{\lambda'}$ around several hydrogen transitions as a function of the velocity defined as defined as $c(\lambda - \lambda_{0})/\lambda_{0}$ where $\lambda_{0} = \lambda_{\text{H}\alpha}$, $\lambda_{\text{H}\beta}$, $\lambda_{\text{H}\gamma}$, $\lambda_{\text{Pa}\alpha}$, $\lambda_{\text{Pa}\beta}$, and $\lambda_{\text{Br}\alpha}$.
As seen in Figure~\ref{fig:broadband_sed}, the high-level Raman features are contaminated by other Raman features of higher-level transitions, and the isolated Raman broad emission features are the most clearly visible around the Balmer series and Paschen series other than Paschen-$\alpha$.
The narrow features at the red wind of the isolated $\lambda_{\text{H}\beta}$, $\lambda_{\text{H}\gamma}$, and $\lambda_{\text{Pa}\beta}$ Raman features originate from the minima of the Rayleigh branch $\sigma_{1s}/\sigma_{\text{tot}}$ (Figure~\ref{fig:crosssections}).
In Figure~\ref{fig:broad_emission_feature}, especially in the emission profile of the isolated Balmer and Paschen-$\beta$ features, we can see that the flat-topped profile due to the saturated scattering is tilted, which is because of the difference in the wavelength dependence between the Raman scattering cross-section and the Rayleigh scattering cross-section at around the Lyman resonances.
In the Paschen-$\beta$ profile, we can see that the Raman $d$-branch contribution shows a redward-asymmetric wavelength dependence, which means that the red wing of the Paschen-$\beta$ feature is depolarized relative to the blue wing.

The line width of the broad emission features due to the Raman scattering might be evaluated analytically as follows.
If we approximate the situation such that the number of scattering for the photons satisfying $\tau_{\lambda} \geq 1$ is $n_{\text{sca}} \rightarrow \infty$ and that for the other photons is $n_{\text{sca}} \rightarrow 0$, all the $\tau_{\lambda} \geq 1$ photons are eventually converted into Raman-scattered optical photons, and the Raman scattered photons form broad emission feature around the Balmer, Paschen, and other higher-level lines whose line widths can be evaluated with Equation~\ref{eqn:optically_thick} broadened by the wavelength ratio (Equations~\ref{eqn:wavelength_conversion} and \ref{eqn:differential_wavelength_conversion}):
\begin{subequations}
\begin{eqnarray}
\Delta V_{\text{H}\alpha}  &\approx& \left( \frac{\lambda_{\text{H}\alpha }}{\lambda_{\text{Ly}\beta}} \right)   \Delta V_{\text{Ly}\beta}  = 13107~\text{km}~\text{s}^{-1} \left( \frac{N_{\text{H}}}{10^{22}~\text{cm}^{-2}} \right)^{1/2}\\
\Delta V_{\text{H}\beta}   &\approx& \left( \frac{\lambda_{\text{H}\beta  }}{\lambda_{\text{Ly}\gamma}} \right)  \Delta V_{\text{Ly}\gamma} =  3860~\text{km}~\text{s}^{-1} \left( \frac{N_{\text{H}}}{10^{22}~\text{cm}^{-2}} \right)^{1/2}\\
\Delta V_{\text{H}\gamma}  &\approx& \left( \frac{\lambda_{\text{H}\gamma }}{\lambda_{\text{Ly}\delta}} \right)  \Delta V_{\text{Ly}\delta} =  1723~\text{km}~\text{s}^{-1} \left( \frac{N_{\text{H}}}{10^{22}~\text{cm}^{-2}} \right)^{1/2}\\
\Delta V_{\text{Pa}\alpha} &\approx& \left( \frac{\lambda_{\text{Pa}\alpha}}{\lambda_{\text{Ly}\gamma}} \right)  \Delta V_{\text{Ly}\gamma} = 14889~\text{km}~\text{s}^{-1} \left( \frac{N_{\text{H}}}{10^{22}~\text{cm}^{-2}} \right)^{1/2}\\
\Delta V_{\text{Pa}\beta}  &\approx& \left( \frac{\lambda_{\text{Pa}\beta }}{\lambda_{\text{Ly}\delta}} \right)  \Delta V_{\text{Ly}\delta} =  5090~\text{km}~\text{s}^{-1} \left( \frac{N_{\text{H}}}{10^{22}~\text{cm}^{-2}} \right)^{1/2}\\
\Delta V_{\text{Br}\alpha} &\approx& \left( \frac{\lambda_{\text{Br}\alpha}}{\lambda_{\text{Ly}\delta}} \right)  \Delta V_{\text{Ly}\delta} =  16085~\text{km}~\text{s}^{-1} \left( \frac{N_{\text{H}}}{10^{22}~\text{cm}^{-2}} \right)^{1/2}.
\end{eqnarray}
\label{eqn:broadened_lines}
\end{subequations}
As shown in Figure~\ref{fig:broad_emission_feature}, when the incident UV spectrum is flat, the resulting profile of the emission features is flat-topped because of the complete Raman conversion \citep[e.g.,][]{cha15}.
When the Lyman series emission lines are incident and the gas is optically-thick to these incident emission lines, the line widths of the resulting Raman-scattered optical emission lines are broadened by the wavelength ratios as the prefactors in Equations~\ref{eqn:broadened_lines} (Equation~\ref{eqn:wavelength_conversion}).
This is the simplified picture of the formation mechanism of the broad optical emission features due to the Raman scattering, and the rough estimates of $\Delta V$ in Equations~\ref{eqn:broadened_lines} provide reasonable order of magnitude for the line widths \citep[Figure~\ref{fig:broad_emission_feature}; see also][]{cha15}.
Note that under real astrophysical circumstances, the scattering region has complex geometry and the scattered light is the sum of the scattered fluxes from scattering clouds with various $N_{\text{H}}$ values, toward which we would observe the Raman scattering feature as a peaky profile with a smooth, extended Lorentzian-like wing \citep[e.g.,][]{dop16}.
Monte-Carlo radiative transfer studies are needed to examine the (viewing angle-dependent) emergent Raman-scattered spectra and their linear polarization degree from scattering regions with realistic geometry, which we will leave for the forthcoming paper.

It is interesting to consider observational properties to distinguish the Raman-scattered broad hydrogen emission features from the Doppler-broadened hydrogen emission lines due to the high-velocity dynamical motions of the line-emitting gas.
First, Equation~\ref{eqn:broadened_lines} suggests that the Raman-scattered broad emission features exhibit different line widths for different transitions.
The line widths of the commonly-observed optical H$\alpha$ and H$\beta$ lines are different by a factor of $\sim 3.4$, which is a useful observable to discriminate between the Raman-broadened and Doppler-broadened emission features; in the latter case, the line widths are common among different transitions.
Second, the profile of the Raman-scattered broad emission features exhibits red asymmetry directly reflecting the asymmetry in the Raman scattering cross-section \citep[Figure~\ref{fig:crosssections};][]{lee98}. 
In contrast, the Doppler-broadened emission lines tend to show a symmetric or blue-asymmetric profile.
Third, because the ratios of the Raman cross-sections determine the line flux ratios of the Raman-scattered features while the Einstein coefficients determine the line flux ratios between various transitions in normal Case~B photoionization regions, significant deviations from the line flux ratios expected in Case~B would suggest the Raman scattering nature of the broad emission features.
Then, the linear polarization of the H$\alpha$ and H$\beta$ features might serve as a diagnostic of the Raman scattering nature \citep[e.g.,][]{sch90,esp95}, as the Doppler-broadened emission lines are not expected to be polarized.
If we get both of the optical and IR spectropolarimetry, observations of the depolarization effect in the Raman-scattered broad features around the Paschen, Brackett, and higher-level transitions compared to the Balmer transitions due to the $d$-branch contribution would be a smoking gun evidence for the Raman scattering nature.

\subsection{Raman-scattering of UV emission lines from heavy elements}

When UV emission lines from heavy elements (accidentally close to the Lyman resonances) are incident on the neutral hydrogen atoms, we would observe Raman-scattered optical emission lines broadened by the wavelength ratios \citep[Equation~\ref{eqn:differential_wavelength_conversion}; e.g.,][]{nus89,dop16}.
The most famous example of the Raman wavelength conversion of the UV emission lines in the real astrophysical circumstances is the \ion{O}{vi} $\lambda\lambda$ 1031.912, 1037.613 resonance doublet lines whose Raman-scattered broadened lines are observed at 6826.264\AA\ and 7083.729\AA\ in symbiotic stars with the line-broadening factors $(\Delta \lambda'/\lambda')/(\Delta \lambda/\lambda)=\lambda'/\lambda$ are $6.62$ and $6.83$, respectively \citep[Equations~\ref{eqn:wavelength_conversion} and \ref{eqn:differential_wavelength_conversion};][]{nus89,sch89,lee97,hub14,lee16}.
The \ion{O}{vi} doublet lines are located at the red wing of the $\text{Ly}\beta$ ($\lambda_{\text{Ly}\beta}=1025.734$\AA), which means that only $\sigma_{2s}$ contributes to the Raman scattering of the lines.
Thus, the scattering angular distribution is purely that of the {\it scalar} (the Rayleigh phase function) and the scattered photons tend to be linearly polarized (Section~\ref{sec:phase_function}) \citep{sch90,esp95}.
According to our calculations, the Rayleigh ($n_{B} = 1$ branch) and Raman ($n_{B} = 2$ branch) scattering cross-sections for the 1032\AA\ and 1038\AA\ photons are 
$\sigma_{\text{Rayleigh}}(1032) = 34.40\sigma_{T}$, 
$\sigma_{\text{Rayleigh}}(1038) = 6.90\sigma_{T}$, 
$\sigma_{\text{Raman}}(1032) = 7.61\sigma_{T}$, 
$\sigma_{\text{Raman}}(1038) = 2.51\sigma_{T}$.
The Rayleigh and Raman cross-sections for the 1032\AA\ line are greater than that of the 1038\AA\ line, which explains the larger line flux ratio of the Raman-scattered optical lines $F(6825)/F(7082)$ than the original $F(1032)/F(1038)$ line flux ratio \citep{sch89}.

UV emission lines at $\lambda < \lambda_{\text{Ly}\beta}$ can be Raman scattered into optical lines via both the $s$-branch and $d$-branch.
One of such emission lines is \ion{He}{ii} ($2 - 8$) at $972.11$\AA\ \citep[e.g.,][]{peq97,lee06}, which locates at  blue wing of the Lyman-$\delta$ (972.55\AA) resonance and is Raman-scattered into blue wings of H$\beta$ via $2s$-branch and Pa$\alpha$ via $3s$ and $3d$-branches.
The central wavelengths of the Raman scattered emission lines are 4851.82\AA\ and 18594.86\AA, and the line broadening factors are $\lambda'/\lambda = 4.99$ and $19.1$, respectively.
The corresponding Raman cross-sections are 
$\sigma_{2s} = 141.61\sigma_{T}$,
$\sigma_{3s} =  43.77\sigma_{T}$, and
$\sigma_{3d} =   4.49\sigma_{T}$, 
thus about 10\% of the total flux of the 18594.86\AA\ feature could originate from the $3d$-branch.
The non-negligible flux contribution from the $d$-branch means that the 18594.86\AA\ feature tends to be slightly depolarized compared to the 4851.82\AA.

Just like the case of the broad hydrogen features due to the Raman scattering of the UV continuum as discussed in Section~\ref{sec:broad_hydrogen_feature}, measurements of the line flux ratios, line widths, and linear polarization degree for previously-unidentified optical/IR emission features in UV-bright, neutral material-rich astronomical objects will be helpful to establish their Raman scattering nature \citep[e.g.,][]{zag21}.

\section{Summary and conclusions}
\label{sec:conclusion}

We have presented explicit expressions for Rayleigh and Raman scattering cross-sections and phase matrices of ground $1s$ state hydrogen atom, based on the Kramers-Heisenberg-Waller dispersion formula.
The Rayleigh ($1s \rightarrow 1s$) scattering and Raman $s$branch ($1s \rightarrow ns$; $n \geq 2$) scattering have the angular distribution of the outgoing photon identical to the classical Rayleigh phase function.
We have shown that the phase function of the Raman $d$-branch ($1s \rightarrow nd$; $n \geq 3$) scattering is close to that of the isotropic scattering, which is because, unlike the Rayleigh case, the atomic dipole moments not only in the orthogonal directions but also the parallel direction of the propagation of the incident radiation contribute to the scattering process.
The Raman $d$-branch cross-section is non-zero at $\lambda < \lambda_{\text{Ly}\beta}$, and the Raman-scattered photons via the Raman $d$-branch tend to be much less polarized compared to the Raman-scattered photons via the $s$-branch.

The Raman scattering of the UV continuum photons by a thick neutral hydrogen gas of $N_{\text{H}} > 10^{21}~\text{cm}^{-2}$ can produce the broad optical/IR emission features with the velocity width of $\gtrsim 1,000~\text{km}~\text{s}^{-1}$ at the wavelengths of the hydrogen transitions, whose emission strength is dependent on the functional form of the cross-sections, scattering optical depth ($\propto N_{\text{H}}$), and spectral shape of the incident UV continuum.
Such broad hydrogen emission features can mimic Doppler-broadened hydrogen emission lines, which means detecting broad hydrogen emission lines alone cannot be interpreted as a signature of a compact object with extremely high kinetic energy, such as an AGN, supernova, or fast stellar wind.
We have shown that each of the broad optical/IR emission features due to the Raman scattering has different line widths and linear polarization degrees, such that emission features at longer wavelengths exhibit larger line widths, and Paschen, Brackett, and higher-level features are depolarized compared to the Balmer features due to the greater flux contribution from the less-polarized Raman $d$-branch.
We have argued that detailed observations of the line widths, line flux ratios, and linear polarization of multiple optical/IR hydrogen lines are crucial to discriminate between the broad emission features due to the Raman scattering and Doppler-broadened emission lines.

The Rayleigh and Raman cross-sections of the ground-state hydrogen atom presented in this work have a wide variety of applications in astrophysical phenomena where the hydrogen Rayleigh and Raman scattering takes place, such as symbiotic stars, planetary nebulae, star-forming regions, Active Galactic Nuclei, and Damped Lyman-$\alpha$/Lyman limit absorbers.
With the Rayleigh and Raman scattering cross-sections and the phase matrices, we can perform Monte-Carlo radiative transfer simulations to evaluate the emergent spectrum and its linear polarization of photons escaped from scattering regions of arbitrarily complex geometry surrounding some UV sources, which we will leave for the forthcoming paper.

\section*{Acknowledgements}
We thank the anonymous referee for useful comments and suggestions.
M.~K. acknowledges NAOJ for support.

\section*{Data Availability}
The analysis products of this work will be shared on a reasonable request to the corresponding author.



\bibliographystyle{mnras}
\bibliography{raman} 




%
%

\appendix

\section{Wigner-Eckart theorem}
\label{sec:wigner-eckart}

The matrix elements of the position operators can be evaluated by the use of the Wigner-Eckart theorem \citep[e.g.,][]{sak17}:
\begin{eqnarray}
\langle n, l, m | r_{q} | n', l', m' \rangle =  \langle n, l ||r|| n', l' \rangle\frac{1}{\sqrt{2l+1}} \langle 1, l'; q, m' | 1, l'; l, m \rangle,
\end{eqnarray}
where $\langle 1, l'; q, m' | 1, l'; l, m \rangle$ is the Clebsch–Gordan coefficient between an uncoupled tensor product basis of two angular momentum eigenstates $|1, l'; q, m' \rangle \equiv |1, q \rangle \otimes |l', m'\rangle$ and total angular momentum eigenstate $|1, l'; l, m \rangle \equiv |l, m \rangle$, and $\langle n, l ||r|| n', l' \rangle$ is the reduced matrix element independent of $m$ and $m'$.
$r_{q}$ ($q=-1, 0, 1$) is the vector component of the spherical position operator defined as 
\begin{eqnarray}
r_{+1}&=&-\frac{x+iy}{\sqrt{2}}\\
r_{-1}&=&\frac{x-iy}{\sqrt{2}}\\
r_{0}&=&z,
\end{eqnarray}
which is related to the spherical harmonics with $l=1$ as 
\begin{eqnarray}
r_{q} = r\sqrt{\frac{4\pi}{3}} Y_{1\:q}(\theta, \phi),
\end{eqnarray}
where $(r, \theta, \phi)$ is the spherical coordinate, and the spherical harmonics with $l=1$ are
\begin{eqnarray}
Y_{1\:+1} &=& -\sqrt{\frac{3}{8\pi}}e^{i\phi}\sin\theta, \quad Y_{1\:-1} = \sqrt{\frac{3}{8\pi}}e^{-i\phi}\sin\theta, \quad Y_{1\:0} =\sqrt{\frac{3}{4\pi}}\cos\theta.
\end{eqnarray}

The Wigner-Eckart theorem can be written in an explicit form as:
\begin{eqnarray}
\langle n,l,m|r_{q}|n',l',m' \rangle &=& \left( \int_{0}^{\infty} r^3 R_{nl}^{*}R_{n'l'}dr \right) \left(\sqrt{\frac{4\pi}{3}}\int_{\Omega}^{} d\Omega Y_{lm}^{*}Y_{1q}Y_{l'm'} \right)\nonumber \\ 
&=& R_{nl}^{n'l'} \sqrt{\frac{(2l'+1)}{(2l+1)}} \langle 1, l'; 0, 0 | 1, l'; l, 0 \rangle \langle 1, l'; q, m' | 1, l'; l, m \rangle,
\end{eqnarray}
and by using the explicit form of the Clebsch–Gordan coefficient, non-vanishing components of the matrix elements satisfying the selection rules for electric-dipole transitions ($l'=l \pm 1$ and $m'=m-q$) are given as
\begin{eqnarray}
\langle n,l,m|r_{q}|n',l',m' \rangle &=& R_{nl}^{n'l \pm 1} \left\{
\begin{array}{cl}
 (-1)^{q}\sqrt{\frac{(l+m-q+1)!(l-m+q+1)!}{(2l+3)(2l+1)(1-q)!(1+q)!(l-m)!(l+m)!}} & (l' = l + 1) \\
 \sqrt{\frac{(l-m)!(l+m)!}{(2l-1)(2l+1)(1-q)!(1+q)!(l+m-q-1)!(l-m+q-1)!}}  & (l' = l - 1)
\end{array}
\right.
\label{eqn:explicit_WignerEckart}
\end{eqnarray}

Equation~\ref{eqn:explicit_WignerEckart} for the $1s \rightarrow n_{I}p$ transition matrix reduces to
\begin{eqnarray}
\langle n_{I},1,q|r_{q}|1,0,0 \rangle &=& \frac{1}{\sqrt{3}} R_{n_{I}1}^{10}.
\label{eqn:wigner_eckert_s0branch}
\end{eqnarray}
Also, we obtain
\begin{eqnarray}
\langle n_{B},0,0|r_{q}|n_{I},1,-q \rangle &=& \frac{(-1)^{q}}{\sqrt{3}} R^{n_{I}1}_{n_{B}0}
\label{eqn:wigner_eckert_sbranch}
\end{eqnarray}
when the final state is the $s$ orbital ($l_{B}=0$, $m_{B}=0$), and
\begin{eqnarray}
\langle n_{B},2,m_{B}|r_{q}|n_{I},1,m_{B}-q \rangle &=& R_{n_{B}2}^{n_{I}1} \sqrt{\frac{(2-m_{B})!(2+m_{B})!}{15(1-q)!(1+q)!(1+m_{B}-q)!(1-m_{B}+q)!}},
\label{eqn:wigner_eckert_dbranch}
\end{eqnarray}
when the final state is the $d$ orbital ($l_{B}=2$, $m_{B}=\{-2,-1,0,1,2\})$ which is non-zero only when $-1 \leq m_{B}-q \leq 1$.

The matrix elements of the dipole operator can be explicitly expressed as:
\begin{eqnarray}
(\vec{x} \cdot \vec{\epsilon}^{(\alpha')})_{BI} &=& \sum_{q=\{-1,0,1\}} \langle n_{B},l_{B},m_{B}|r_{q}| n_{I},1,m_{I} \rangle e^{(\alpha')}_{q}\\ 
(\vec{x} \cdot \vec{\epsilon}^{(\alpha )})_{IA} &=& \sum_{q=\{-1,0,1\}} \langle n_{I},1,m_{I}|r_{q}| 1,0,0 \rangle e^{(\alpha)}_{q}
\end{eqnarray}
where $e_{q}^{(\beta)}$ ($\beta = \alpha, \alpha'$) is defined as:
\begin{eqnarray}
e^{(\beta)}_{+1} &=& \frac{-\epsilon^{(\beta)}_{x}+i\epsilon^{(\beta)}_{y}}{\sqrt{2}}, \quad e^{(\beta)}_{-1} = \frac{\epsilon^{(\beta)}_{x}+i\epsilon^{(\beta)}_{y}}{\sqrt{2}}, \quad e^{(\beta)}_{0} = \epsilon^{(\beta)}_{z}.
\end{eqnarray}
The summation of the product of the matrix elements over $m_{I}$ in Equation~\ref{eqn:kramersheisenberg_nonoriented} can be expressed as
\begin{eqnarray}
&&\sum_{m_{I}=-1,0,1}^{}(\vec{x} \cdot \vec{\epsilon}^{(\alpha')})_{BI}(\vec{x} \cdot \vec{\epsilon}^{(\alpha)})_{IA} \nonumber\\
&=& \frac{1}{\sqrt{3}}R_{n_{I}1}^{10} \sum_{q=-1,0,1}^{} \sum_{p=-1,0,1}^{} e^{(\alpha')}_{q} e^{(\alpha)}_{p} \langle n_{B}, l_{B}, m_{B}|r_{q}|n_{I}, 1, p \rangle,
\label{eqn:sum_of_matrix}
\end{eqnarray}
where Equation~\ref{eqn:wigner_eckert_s0branch} is used.
We note that Equation~\ref{eqn:sum_of_matrix} is invariant under the interchange of $\alpha$ and $\alpha'$.
Equation~\ref{eqn:sum_of_matrix} is used to obtain Equations~\ref{eqn:differential_crosssection_nB1} and \ref{eqn:differential_crosssection_nB2}.

\section{Overlap integrals of the radial wavefunctions}
\label{sec:overlap_integrals}

The overlap integrals of the radial wavefunctions of the hydrogen atom, or reduced matrix elements of the electric dipole moment operator of rank 1, can be evaluated by using the well-known \cite{gor29}’s integral formulae \citep[e.g.,][]{hat46,bet57,kar61,gol68}.
For completeness, here we summarize the definitions of the bound-state and free-state radial wavefunctions of the hydrogen atom and Gordon's integral formulae for the bound-bound and bound-free transitions using a self-consistent normalization.

The radial eigenfunctions for a bound-state hydrogen atom ($R_{nl}^{b}$) are given in many textbooks as the solution to the Schr\"{o}dinger equation \citep[e.g.,][]{bet57} \footnote{There is a minor 
misprint in Equations~3.17 of \citealt{bet57}; $(2Zr/n)$ in their Equations~3.17 should read as $(2Zr/n)^{l}$.}:
\begin{eqnarray}
R_{nl}^{b} &=& \frac{1}{(2l+1)!}\sqrt{\frac{(n+l)!}{(n-l-1)!2n}}\left(\frac{2}{n}\right)^{3/2}\left(\frac{2r}{n}\right)^{l}e^{-r/n} F(-n+l+1; 2l+2; 2r/n),
\end{eqnarray}
where $F$ is the confluent hypergeometric function of the first kind.
$R_{nl}^{b}$ is normalized such that $\int_{0}^{\infty}r^{2}(R_{nl}^{b})^{*}R_{n'l'}^{b}dr = \int_{0}^{\infty}r^{2}R_{nl}^{b}R_{n'l'}^{b}dr = \delta_{nn'}\delta_{ll'}$.
The energy levels of the bound-states are $E_{n} = -1/2n^2 < 0$ in Hartree atomic units.

The radial eigenfunctions for a free-state (continuum-state) hydrogen atom with energy $E = k^2/2 > 0$ (where $k$ is the wave number) can be expressed in the same way as the bound-state except that now $n = 1/\sqrt{-2E} =-i/k$ is purely imaginary and is continuous.
Below we use a new quantum number $n'$ defined as $n' \equiv in = 1/k$, with which $E = 1/2n'^{2}$.
The free-state radial eigenfunctions normalized in the energy-scale ($R_{El}^{c}$) and $n'$-scale ($R_{n'l}^{c}$) are respectively given as \citep[e.g.,][]{bet57,lan91}:
\begin{eqnarray}
R_{El}^{c} &=& \frac{2}{\sqrt{1-e^{-2 \pi n'}}} \left(\prod_{s=1}^{l}\sqrt{s^2+n'^2}\right)\frac{(2kr)^{l}}{(2l+1)!}e^{-ikr}F(in'+l+1; 2l+2; 2ikr)\\
R_{n'l}^{c} &=& \sqrt{\left|\frac{dE}{dn'}\right|} R_{El}^{c} = (n')^{-3/2}R_{El}^{c}.
\end{eqnarray}
The energy-scale and $n'$-scale radial eigenfunctions are defined such that $\int_{0}^{\infty}r^{2}R_{El}^{c}R_{E'l'}^{c}dr = \delta(E-E')\delta_{ll'}$ and $\int_{0}^{\infty}r^{2}R_{nl}^{c}R_{n'l'}^{c}dr = \delta(n-n')\delta_{ll'}$.
The radial eigenfunctions are real functions ($(R_{n'l}^{c})^{*}=R_{n'l}^{c}$), and the bound-state and $n'$-scale continuum-state eigenfunctions have the same functional form of $\sqrt{2}n^{-3/2}r^{-1/2}J_{2l+1}(\sqrt{8r})$ as $n, n' \rightarrow \infty$, where $J_{2l+1}$ is the Bessel function of order $2l+1$ \citep{sas69}\footnote{$\lim\limits_{s \to \infty}F(-s+l+1; 2l+2; 2r/s) = (2l+1)!(2r)^{-l-1/2}J_{2l+1}(\sqrt{8r})$ and $\lim\limits_{n \to \infty}(n+l)!/(n-l-1)! = n^{2l+1}$.}.
By replacing $n$ in the bound-state eigenfunctions by $-in'$ or $in'$, we can observe that the bound-state eigenfunctions are directly related to the continuum-state eigenfunctions as \citep[e.g.,][]{bur58,kar61}:
\begin{eqnarray}
R_{in'\:l}^{b}  &=& e^{-\frac{3}{4}\pi i} \sqrt{1-e^{-2\pi n'}} R_{n'l}^{c} = -(R_{-in'\: l}^{b})^{*}.
\label{eqn:analytic_continuation}
\end{eqnarray}

The overlap integrals for the bound-bound transition matrix under the electric dipole approximation, $R^{n'\:l\pm1}_{n\:l} = \int_{0}^{\infty}r^{3}R_{nl}R_{n'l\pm1}dr$, can be evaluated by using the Gordon's integral formula for the bound-bound transition.
A convenient form of Gordon's formula is given by \cite{kar61} as:
\begin{eqnarray}
R^{n'\:l-1}_{n\:l}~\text{(bound-bound)} &=& \int_{0}^{\infty} r^{3} R_{nl}^{b} R_{n'l-1}^{b} dr  \nonumber\\
&=& \frac{2^{2l}e^{\pi i(n'-l)}}{(2l-1)!} \sqrt{ \frac{(n+l)!}{(n-l-1)!} } \sqrt{\frac{\Gamma(n'+l)}{\Gamma(n'-l+1)}} (nn')^{l+1} \left(\frac{n-n'}{n+n'}\right)^{n'} (n^2-n'^2)^{-n} (n-n')^{2n-2l-2} \nonumber\\
&\times& \left[ {}_{2}F_{1}\left( l+1-n, l-n'; 2l; -\frac{4nn'}{(n-n')^{2}} \right) - \left(\frac{n-n'}{n+n'}\right)^{2} {}_{2}F_{1}\left(l-1-n, l-n'; 2l; -\frac{4nn'}{(n-n')^{2}}\right) \right]\label{eqn:R_minus_bb}\\
R^{n'\:l+1}_{n\:l}~\text{(bound-bound)} &=& \int_{0}^{\infty} r^{3} R_{nl}^{b}R_{n'l+1}^{b}dr  \nonumber\\
&=& \frac{2^{2l+2}e^{\pi i (n-l-1)}}{(2l+1)!} \sqrt{ \frac{(n+l)!}{(n-l-1)!} } \sqrt{ \frac{\Gamma(n'+l+2)}{\Gamma(n'-l-1)} } (n'n)^{l+2} \left(\frac{n'-n}{n'+n}\right)^{n'} (n'^2-n^2)^{-n} (n'-n)^{2n-2l-4} \nonumber\\
&\times& \left[ {}_{2}F_{1}\left( l+2-n', l+1-n; 2l+2; -\frac{4n'n}{(n'-n)^{2}} \right) - \left(\frac{n'-n}{n'+n}\right)^{2} {}_{2}F_{1}\left(l-n', l+1-n; 2l+2; -\frac{4n'n}{(n'-n)^{2}}\right) \right]\label{eqn:R_plus_bb},
\end{eqnarray}
where $n$ and $n'$ are positive integers ($n \neq n'$), and ${}_{2}F_{1}$ is the Gauss’ hypergeometric function.
Equation~\ref{eqn:R_plus_bb} is obtained by interchanging $n$ and $n'$ and replacing $l$ by $l+1$ in Equation~\ref{eqn:R_minus_bb}.
The power series of the hypergeometric functions in Equations~\ref{eqn:R_minus_bb} and \ref{eqn:R_plus_bb} terminate and reduce to polynomials because the first two arguments of the hypergeometric functions are nonpositive integers \citep[][]{bet57}.

The overlap integrals for the bound-free transitions can be obtained by direct computation \citep[e.g.,][]{sas69,lee07} or analytic continuation of Gordon's formula for the bound-bound transitions using the relation given in Equation~\ref{eqn:analytic_continuation} \citep[e.g.,][]{men35,bur58,kar61}, and here we take the latter approach.
By replacing $n'$ in Equations~\ref{eqn:R_minus_bb} and \ref{eqn:R_plus_bb} by $-in'$ and multiplying the factor given in Equation~\ref{eqn:analytic_continuation}, we obtain:
\begin{eqnarray}
R^{n'\:l-1}_{n\:l}~\text{(bound-free)} &=& \int_{0}^{\infty} r^{3}R_{n l}^{b}R_{n'l-1}^{c} dr = \frac{e^{-\frac{3}{4}\pi i}}{\sqrt{1-e^{-2\pi n'}}}R^{-in'\:l-1}_{n\:l}~\text{(bound-bound)} \nonumber\\
&=& \frac{e^{\pi i}2^{2l}(n')^{l+\frac{3}{2}}(n)^{l+1}}{(2l-1)!\sqrt{1-e^{-2\pi n'}}}\sqrt{\frac{(n+l)!}{(n-l-1)!}}\sqrt{\prod_{s=1}^{l-1}(s^2+n'^2)}e^{-2n'\text{arccot}(n'/n)}(n^{2}+n'^{2})^{-n} (n+in')^{2n-2l-2} \nonumber\\
&\times& \left[ {}_{2}F_{1}\left(l+1-n, l+in'; 2l; \frac{4inn'}{\left(n+in'\right)^{2}}\right) - \left(\frac{n+in'}{n-in'}\right)^{2}{}_{2}F_{1}\left(l-1-n, l+in'; 2l; \frac{4inn'}{\left(n+in'\right)^{2}}\right) \right]\label{eqn:R_minus_bf} \\
R^{n'\:l+1}_{n\:l}~\text{(bound-free)} &=& \int_{0}^{\infty} r^{3}R_{n l}^{b}R_{n'l+1}^{c} dr = \frac{e^{-\frac{3}{4}\pi i}}{\sqrt{1-e^{-2\pi n'}}}R^{-in'\:l+1}_{n\:l}~\text{(bound-bound)}\nonumber\\
&=& \frac{e^{\frac{1}{2}\pi i}2^{2l+2}(n')^{l+\frac{5}{2}}(n)^{l+2}}{(2l+1)!\sqrt{1-e^{-2\pi n'}}}\sqrt{\frac{(n+l)!}{(n-l-1)!}}\sqrt{\prod_{s=1}^{l+1}(s^2+n'^2)}e^{-2n'\text{arccot}(n'/n)}(n^{2}+n'^{2})^{-n} (n+in')^{2n-2l-4} \nonumber\\
&\times& \left[ {}_{2}F_{1}\left(l+2+in', l+1-n; 2l+2; \frac{4inn'}{\left(n+in'\right)^{2}}\right) - \left(\frac{n+in'}{n-in'}\right)^{2}{}_{2}F_{1}\left(l+in', l+1-n; 2l+2; \frac{4inn'}{\left(n+in'\right)^{2}}\right) \right],
\label{eqn:R_plus_bf}
\end{eqnarray}
where $n'$ is a positive real number and $n$ is a positive integer.
It can be shown that Equations~\ref{eqn:R_minus_bf} and \ref{eqn:R_plus_bf} are real numbers for any $n'>0$ \citep[e.g.,][]{kar61,sas69}.
Similar to the bound-bound case, the power series of the hypergeometric functions in Equations~\ref{eqn:R_minus_bf} and \ref{eqn:R_plus_bf} can be expressed by polynomials of finite degrees because either of the first two arguments of the hypergeometric functions are nonpositive integers $l+1-n$ or $l-1-n$.

In principle we can directly use Equations~\ref{eqn:R_minus_bb}--\ref{eqn:R_plus_bf} to numerically evaluate the overlap integrals of the radial wavefunctions for any $l \rightarrow l\pm1$ transitions.
Instead, it can be shown by the direct substitutions of $n$ and $l$ into Equations~\ref{eqn:R_minus_bb}--\ref{eqn:R_plus_bf} with some algebraic manipulation that these formulae can be reduced to simpler convenient expressions in terms of elementary functions, which are helpful for the calculations of the Rayleigh and Raman scattering cross-sections (see below).
The exact expressions of the bound-bound and bound-free overlap integrals necessary for the calculations of the Heisenberg-Kramers-Waller matrix elements are summarized in Tables~\ref{tbl:radial_integrals_boundbound} and \ref{tbl:radial_integrals_boundfree}, respectively, up to $n_{B}=12$.
It is easy to verify that the bound-bound and bound-free radial integrals in Tables~\ref{tbl:radial_integrals_boundbound} and ~\ref{tbl:radial_integrals_boundfree} have the same form as $n,n '\rightarrow \infty$ \citep[e.g.,][]{sas69}.

The overlap integrals for several low-excitation levels can be found in the literature.
For example, $R^{n' 1}_{1 0}~\text{(bound-free)} = 2^{4} n'^{7/2} e^{-2n'\text{arccot}(n')} (n'^2+1)^{-5/2} (1-e^{2\pi n'})^{-0.5}$ and $R^{n' 1}_{2 0}~\text{(bound-free)} = 2^{17/2} n'^{7/2} e^{-2n'\text{arccot}(n'/2)} (n'^2+1)^{1/2} (n'^2+4)^{-3} (1-e^{2\pi n'})^{-0.5}$ in Table~\ref{tbl:radial_integrals_boundfree} are in agreement with that given in \cite{sas69} \citep[see also][]{lee07,lee12}\footnote{The overlap integrals in the literature are sometimes provided with different normalization. \cite{bet57} give (in their Equation~71.4) the $1s \rightarrow n'p$ matrix element as $x_{W1} \equiv \langle n',1,0|x|1,0,0 \rangle =  3^{-1/2} \int_{0}^{\infty}r^{3}R_{Ep}^{*}R_{1s}dr$, thus $|\int_{0}^{\infty}r^{3}R_{Ep}^{*}R_{1s}dr|^2 = n'^{3}|\int_{0}^{\infty}r^{3}R_{n'p}^{*}R_{1s}dr|^{2} = 2^{8}n'^{10} e^{-4n'\text{arccot}(n')} (n'^2+1)^{-5} (1-e^{-2\pi n'})^{-1}$, obtained by a direct calculation of the integral using the residue theorem. The same equation is given in Appendix of \cite{sas69} in the form of $\langle E_{n'z}|z|a \rangle \equiv \langle n',1,0|z|1,0,0 \rangle = 3^{-1/2}\int_{0}^{\infty}r^{3}R_{Ep}^{*}R_{1s}dr$. Note that \cite{lee12} provide a list of the overlap integrals for both of the bound-bound and bound-free transitions up to $n_{B}=4$, but there are errors in their Equations~5--6 and Table~4 for the bound-free overlap integrals.}.

In this paper, we directly evaluate the sum-over-state ($\SumInt$ in Equation~\ref{eqn:kramersheisenberg}) to calculate the scattering cross-sections. 
We note that there is a method to replace the sum-over-state with an integral calculation \citep[related to the so-called Dalgarno-Lewis method;][]{dal55}, which allows the computation of cross-sections without explicitly using radial wavefunctions and their overlap integrals \citep[e.g.,][]{sch59b,gav67,sad92}.
Presenting the integral formulae for the sum-over-states is outside the scope of the current paper, but may be a worthwhile pursuit in future studies.


\section{Scattering cross sections in terms of the oscillator strengths}
\label{sec:oscillator_strengths}

In some literature \citep[e.g.,][]{pen69,ste80,roh22} we can find the Rayleigh and Raman scattering cross sections expressed in terms of the oscillator strengths, and here we examine the relationship between the oscillator strength formalism and the overlap integral formalism given in this work.

The average oscillator strengths for the bound-bound and bound-free $n', l' \rightarrow n, l$ transition might be defined respectively as \citep{bet57,kar61,roh22}:
\begin{eqnarray}
{f}_{n\:n'}(\text{bound-bound}) &=& \frac{1}{3}\left(\frac{1}{n'^2}-\frac{1}{n^2}\right)\frac{\text{max}(l',l)}{2l'+1}(R_{n'\:l'}^{n\:l})^2 \quad \text{[$n$: positive integer]} \\
\frac{d{f}_{n\:n'}}{dn}(\text{bound-free}) &=& \frac{1}{3}\left(\frac{1}{n'^2}+\frac{1}{n^2}\right)\frac{\text{max}(l',l)}{2l'+1}(R_{n'\:l'}^{n\:l})^2 \quad \text{[$n$: positive real]},
\end{eqnarray}
where $n'$ is a positive integer.
For the case $n'=1$, $l'=0$, and $l=1$, we obtain
\begin{eqnarray}
{f}_{n\:1}(\text{bound-bound}) &=& 2^{8}3^{-1}n^5(n^2-1)^{-4} \left(\frac{n-1}{n+1}\right)^{2n} \quad \text{[$n$: positive integer]} \\
\frac{d{f}_{n\:1}}{dn}(\text{bound-free}) &=& 2^{8}3^{-1}n^5(n^2+1)^{-4} \frac{e^{-4n\:\text{arccot}(n)}}{1-\exp\left[-2 \pi n\right]} \quad \text{[$n$: positive real]}.
\end{eqnarray}

The Rayleigh scattering cross section of unpolarized light by a nonoriented hydrogen atom in the ground state can be expressed in terms of the energy-dependent atomic polarizability $\alpha_{\text{pol}}$ as \citep[Equation~1 of][]{roh22}:
\begin{eqnarray}
\sigma_{1s} = \frac{\pi}{6}\alpha^4 \left(\frac{\lambda}{\lambda_{\text{LyLimit}}}\right)^{-4} |\alpha_{\text{pol}}|^2 = 2^{-4}\sigma_{T} \left(\frac{\lambda}{\lambda_{\text{LyLimit}}}\right)^{-4} |\alpha_{\text{pol}}|^2
\label{eqn:crosssection_polarizability}
\end{eqnarray}
in atomic units.
$\alpha_{\text{pol}}$ is generally constituted by real and imaginary parts.
The imaginary part is only significant in a very narrow wavelength range around each resonance and in the photoionization wavelength range of $\lambda/\lambda_{\text{LyLimit}} < 1$, which is out of our interest.
Thus the $\alpha_{\text{pol}}$ is well approximated by its real part at $\lambda/\lambda_{\text{LyLimit}} > 1$ \citep[Equation~29 of][]{roh22}: $|\alpha_{\text{pol}}|\approx|\alpha_{\text{R}}^{0}|$, where $\alpha_{\text{R}}^{0}$ is the real part of the polarizability, and the superscript zero means that the fine-structure and damping effects are ignored, as assumed throughout this paper.
$\alpha_{\text{R}}^{0}$ can be expressed in terms of the oscillator strengths as (Equation~15 of \citealt{pen69}; Equation~3 of \citealt{roh22}):
\begin{eqnarray}
\alpha_{\text{R}}^{0} = 4 \left[ \sum_{n_{I} \geq 2}^{} \frac{f_{n_{I}\:1}(\text{bound-bound})}{\left(1-\frac{1}{n_{I}^2}\right)^2 - \left(\frac{\lambda_{\text{LyLimit}}}{\lambda}\right)^2} + \int_{0}^{\infty} dn_{I}' \frac{\frac{df_{n_{I}'\:1}}{dn_{I}'}(\text{bound-free})}{\left(1+\frac{1}{n_{I}'^2}\right)^2 - \left(\frac{\lambda_{\text{LyLimit}}}{\lambda}\right)^2} \right],
\end{eqnarray}
then we get the Rayleigh scattering cross section as 
\begin{eqnarray}
\sigma_{1s} &=& \sigma_{T} \nonumber\\
&\times& \left| \sum_{n_{I} \geq 2} \left( 2^{7}3^{-1}n_{I}^{5}(n_{I}^2-1)^{-4} \left( \frac{n_{I}-1}{n_{I}+1} \right)^{2n_{I}} \left[ \frac{\frac{\lambda_{\text{LyLimit}}}{\lambda}}{1-\frac{1}{n_{I}^2}-\frac{\lambda_{\text{LyLimit}}}{\lambda}} - \frac{\frac{\lambda_{\text{LyLimit}}}{\lambda}}{1-\frac{1}{n_{I}^2}+\frac{\lambda_{\text{LyLimit}}}{\lambda}} \right] \right) \right.\nonumber\\
&+&\left. \int_{0}^{\infty}dn_{I}' \left( \frac{2^{7}3^{-1}n_{I}'^{5}(n_{I}'^2+1)^{-4}  e^{-4n_{I}'\text{arccot}(n_{I}')} }{(1-\exp(-2\pi n_{I}'))} \left[ \frac{\frac{\lambda_{\text{LyLimit}}}{\lambda}}{1+\frac{1}{n_{I}'^2}-\frac{\lambda_{\text{LyLimit}}}{\lambda}} - \frac{\frac{\lambda_{\text{LyLimit}}}{\lambda}}{1+\frac{1}{n_{I}'^2}+\frac{\lambda_{\text{LyLimit}}}{\lambda}} \right] \right) \right|^2.
\label{eqn:rayleigh_oscillator}
\end{eqnarray}
Equation~\ref{eqn:rayleigh_oscillator} is the total cross section, and the differential cross section can be obtained by replacing $\sigma_{T}$ with $r_{0}^2 \left(\vec{\epsilon}^{(\alpha)} \cdot \vec{\epsilon}^{(\alpha')}\right)^2$.
We can see that this Rayleigh scattering cross section is identical to Equation~\ref{eqn:rayleigh_explicit}.

\section{$1s \rightarrow 1s$ Rayleigh scattering: comparisons with approximate formulae}
\label{sec:approximate_formulae}

\begin{figure}
\center{
\includegraphics[clip, width=6.0in]{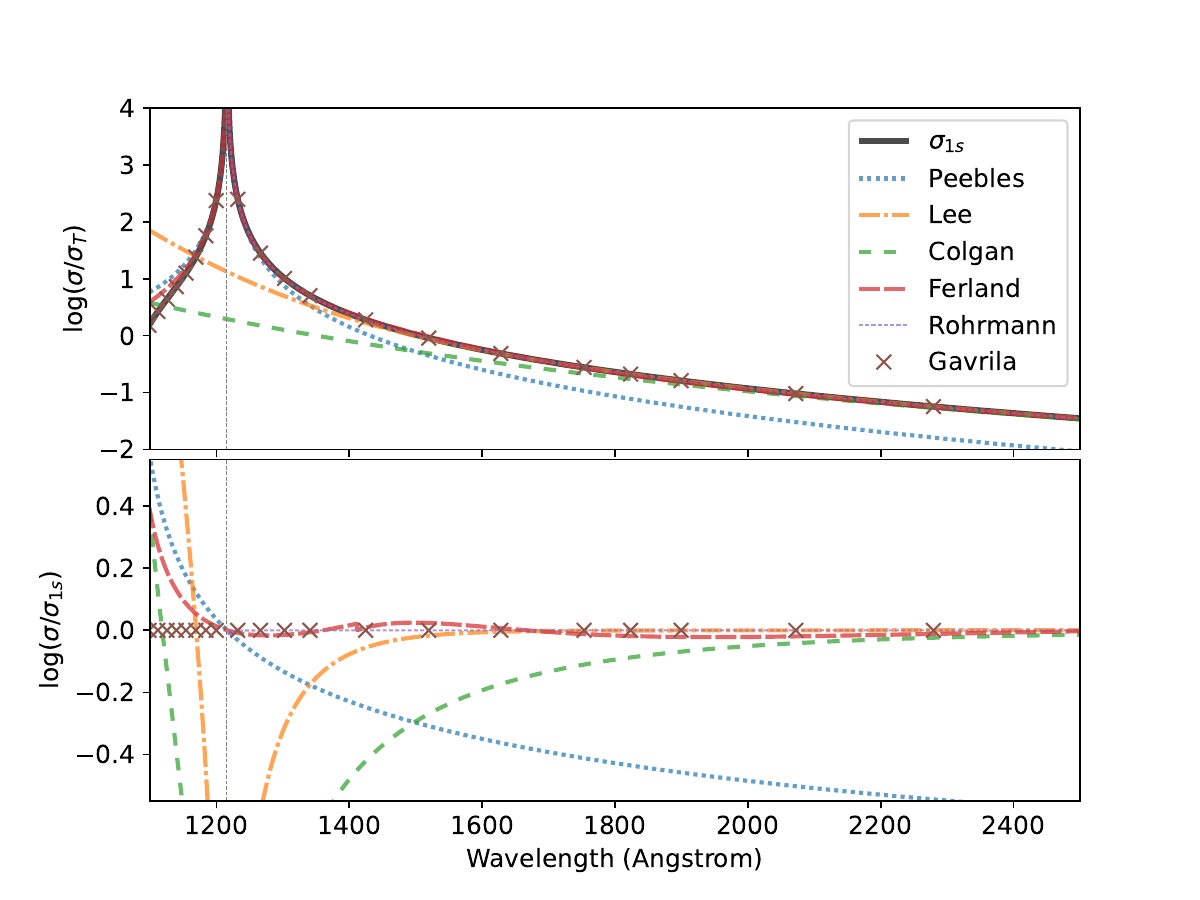}
}\vspace{0cm}
 \caption{The $1s \rightarrow 1s$ Rayleigh scattering cross-section $\sigma_{1s}$ around Lyman-$\alpha$ calculated from the Kramers-Heisenberg-Waller formula (Figure~\ref{fig:crosssections}), compared with various approximate formulae found in the literature \citep[][see Section~\ref{sec:approximate_formulae}]{pee93,lee05,col16,fer22,roh22}. The vertical line denotes the Lyman-$\alpha$ resonant wavelength. Top: the cross-sections normalized by $\sigma_{T}$. Bottom: the ratios of the approximate formulae to $\sigma_{1s}$. The cross symbol denotes the Rayleigh scattering cross-section numerically tabulated in Gavrila (1967) in the limited number of wavelengths presented there.}
 \label{fig:rayleighscattering_comparison}
\end{figure}

The Rayleigh scattering of the ground-state hydrogen atom ($1s \rightarrow 1s$) is of utmost astronomical importance because it, especially the Lyman-$\alpha$ red damping wing, can be a dominant opacity source to the UV-optical continuum in various astronomical objects.
We can find several formulae in the literature to approximate the $1s \rightarrow 1s$ Rayleigh scattering cross-section, and here we compare $\sigma_{1s}$ calculated from the Kramers-Heisenberg-Waller formula with these approximate formulae to examine their consistency and highlight the limitations of the latter.

\cite{pee93} shows that under the two-level approximation, the $1s \rightarrow 1s$ Rayleigh scattering cross-section of the ground-state atomic hydrogen can be approximately expressed as \citep[Equation~23.97 of][]{pee93}:
\begin{eqnarray}
\frac{\sigma_{\text{Peebles}}}{\sigma_{T}} = \frac{3\lambda_{\text{Ly}\alpha}^2}{8\pi \sigma_{T}}\frac{\Gamma_{1\:2}^2 \left(\omega/\omega_{\text{Ly}\alpha}\right)^{4}}{(\omega_{\text{Ly}\alpha}-\omega)^2+\frac{\Gamma_{1\:2}^2}{4}\left(\omega/\omega_{\text{Ly}\alpha}\right)^{6}},
\label{eqn:peebles}
\end{eqnarray}
where $\Gamma_{1\:2} = 6.2649\times10^{8}~\text{s}^{-1}$ is a decay rate, or damping constant, for the $2p \rightarrow 1s$ transition\footnote{The decay rate of the $np \rightarrow 1s$ transition is given as (Equations~59.11 and 60.13 of \citealt{bet57}):
\begin{eqnarray}
\Gamma_{1\:n} = \frac{4}{9} \alpha c^{-2}\left(\omega_{\text{LyLimit}} \left(1-n^{-2}\right)\right)^3 \left(R_{n\:1}^{1\:0}a_{0}\right)^2 = 2^{8}3^{-2}\alpha^{3} \omega_{\text{LyLimit}} n (n-1)^{2n-2} (n+1)^{-2n-2}.
\end{eqnarray}
For $n=2$ we get $\Gamma_{1\:2} = 6.2649\times 10^{8}~\text{s}^{-1}$.}.
As we can see in Figure~\ref{fig:rayleighscattering_comparison}, this formula provides an adequate account of the cross-section at around the resonant frequency but underestimates the low-frequency off-resonant scattering cross-section by a factor of many ($\gtrsim 40$\% at $\lambda \gtrsim 1400$\AA) since contributions from intermediate states of high excitation levels are neglected \citep[e.g.,][]{pee93,bac15}.
In the low-frequency limit, Equation~\ref{eqn:peebles} approaches $\sigma_{\text{Peebles}}/\sigma_{T} = (32\lambda_{\text{Ly}\alpha}^2\Gamma_{1\:2}^2)/(27 \pi \sigma_{T} \omega_{\text{Ly}\alpha}^2) \times (\omega/\omega_{\text{LyLimit}})^{4} = 0.1370 (\omega/\omega_{\text{LyLimit}})^{4}$, which is $9.24$ times smaller than the exact value (see below).

\cite{lee05} presents a polynomial formula of the long-wavelength Rayleigh scattering cross-section of the ground-state atomic hydrogen based on the exact low-energy expansion of the Kramers-Heisenberg-Waller formula by the use of the \cite{dal55}'s method:
\begin{eqnarray}
\frac{\sigma_{\text{Lee}}}{\sigma_{T}} = \sum_{p=0}^{p_{\text{max}}}c_{p}\left( \frac{\omega}{\omega_{\text{LyLimit}}} \right)^{2p+4},
\end{eqnarray}
where the numerical coefficients $c_{i}$ up to $p_{\text{max}}=9$ is tabulated in Table~1 of \cite{lee05}.
The Los Alamos Opacity Tables \citep{col16} use the \cite{lee05}'s formula by truncating at $p_{\text{max}}=2$; explicitly,
\begin{eqnarray}
\frac{\sigma_{\text{Colgan}}}{\sigma_{T}} = \sum_{p=0}^{2}c_{p}\left( \frac{\omega}{\omega_{\text{LyLimit}}} \right)^{2p+4} = 1.266\left( \frac{\omega}{\omega_{\text{LyLimit}}} \right)^{4} + 3.738\left( \frac{\omega}{\omega_{\text{LyLimit}}} \right)^{6} + 8.814\left( \frac{\omega}{\omega_{\text{LyLimit}}} \right)^{8}.
\label{eqn:colgan}
\end{eqnarray}
As shown in Figure~\ref{fig:rayleighscattering_comparison}, the \cite{lee05}'s formula consisting of 10 terms is accurate at $\lambda \gtrsim 1400$\AA, while the \cite{col16}'s formula consisting of 3 terms is only valid at $\lambda \gtrsim 1900$\AA.
As expected, this exact low-energy expansion asymptotically approaches $\sigma_{1s}$ at the long-wavelength regime.

\cite{fer94} present an approximate formula for the Rayleigh scattering cross-section of the ground-state atomic hydrogen, which is a sum of a polynomial fitting function to the exact calculation of the Kramers-Heisenberg-Waller cross-section at $\lambda > 1410$\AA\ \citep{gav67} and Lorentzian functions approximating radiative broadening of the Lyman lines at $\lambda_{\text{LyLimit}} < \lambda < 1410$\AA\ \citep[Equations~4-33 and 4-35 of][]{mih78}.
The \cite{fer94}'s formula is implemented in the photoionization code {\it Cloudy} \citep[][Chapter~4]{fer22} to evaluate the Rayleigh scattering cross-section for Lyman-$\alpha$ as\footnote{Last 
accessed on October 19, 2023, C23.01: \href{https://gitlab.nublado.org/cloudy/cloudy/-/blob/master/source/opacity_createall.cpp}{https://gitlab.nublado.org/cloudy/cloudy/-/blob/master/source/opacity\_createall.cpp}}:
\begin{eqnarray}
\frac{\sigma_{\text{Ferland}}}{\sigma_{T}} =
\left\{
\begin{array}{cl}
1.264 \left(\frac{\omega}{\omega_{\text{LyLimit}}}\right)^4 + 5.066\left(\frac{\omega}{\omega_{\text{LyLimit}}}\right)^6 + 708.0 \left(\frac{\omega}{\omega_{\text{LyLimit}}}\right)^{14} & (\lambda > 1410\text{\AA}) \\
\frac{3\lambda_{\text{Ly}\alpha}^2}{8\pi\sigma_{T}}\frac{\Gamma_{1\:2}^2}{ \left( \omega_{\text{Ly}\alpha} - \omega \right)^2 } & (\lambda \leq 1410\text{\AA})
\end{array}
\right.
\label{eqn:ferland}
\end{eqnarray}
Note that the $\omega^{4}$ term in Equation~\ref{eqn:ferland} (the classical Rayleigh scattering term) closely matches the lowest-order term of the exact low-energy expansion of the Kramers-Heisenberg-Waller formula given in Equation~\ref{eqn:colgan}.
As we can see in Figure~\ref{fig:rayleighscattering_comparison}, Ferland's formula is accurate to $\sim 3\%$ at $\lambda > 1200\text{\AA}$, and the asymptotic agreement between $\sigma_{\text{Ferland}}$ and $\sigma_{1s}$ at the low-frequency regime verifies the consistency between \cite{gav67}'s and our calculations.
The discrepancy with $\sigma_{1s}$ arises at $\lambda < 1200\text{\AA}$ because of the asymmetry of $\sigma_{1s}$ around the Lyman-$\alpha$ resonance compared to the symmetric Lorentzian function.
We note that {\it Cloudy} does not include the Raman scattering cross section in its opacity calculation and, by its design, cannot reproduce the Raman-scattered broad emission features discussed in the present work.

\cite{roh22} provide analytic fits to the numerically evaluated polarizability of the ground-state hydrogen atom $\alpha_{\text{pol}}$, which is related to the Rayleigh scattering cross-section (Equation~\ref{eqn:crosssection_polarizability}).
Their analytic function at the preresonance region \citep[$\lambda > \lambda_{\text{Ly}\alpha}$; Equation~30 in][]{roh22} is given as:
\begin{eqnarray}
\frac{\sigma_{\text{Rohrmann}}}{\sigma_{T}} &=& 2^{-4}\epsilon^{4} \left| \frac{1}{1-s}\left( \frac{1.46486}{0.950713-\epsilon^{2.172}}+ \frac{1.66478}{0.562516-\epsilon^{2}} \right) \right|^2,
\label{eqn:rohrmann}\\
s &=&
\left\{
\begin{array}{ll}
0.0017\sin\left( 8.2\epsilon^{1.33} \right) - 0.000093 & (\epsilon < \epsilon_{a}) \\
-0.00163\sin\left( 16.86\left|\epsilon-\epsilon_{a}\right|^{1.2} \right) & (\epsilon_{a} < \epsilon < 0.73) \\
0 & (0.73 < \epsilon < 0.745) \\
-10^{-4.9 + 0.205(0.7501-\epsilon)^{-0.3}} & (0.745 < \epsilon < 0.75)
\end{array}
\right.,
\end{eqnarray}
where $\epsilon=\lambda_{\text{LyLimit}}/\lambda$ and $\epsilon_{a} = 0.48083$.
We can see in Figure~\ref{fig:rayleighscattering_comparison} that the $\sigma_{\text{Rohrmann}}$ is in good agreement with $\sigma_{1s}$ from our formula, confirming the usefulness of their analytic fit.

For the sake of completeness, in Figure~\ref{fig:rayleighscattering_comparison} we compare $\sigma_{1s}$ with the exact numerical values of the Rayleigh scattering cross-section given by \cite{gav67} \citep[see Equation~11 and Table~I of][]{gav67}.
We confirmed that our calculation is in very good agreement with \cite{gav67}, with relative errors less 0.02~\% over the wavelength range calculated by \cite{gav67}.

\begin{table*}
\centering
\caption{The squared overlap integrals of the radial wavefunctions for the bound-bound transitions between $n\:p$ and $m\:s$ or $m\:d$ eigenstates ($m \leq 12$), in atomic units: $(R^{m\:l-1}_{n\:l})^2 = (R^{n\:l}_{m\:l-1})^2 = (\int_{0}^{\infty}r^3R^{b}_{n\:l}R^{b}_{m\:l-1}dr)^2$, where $n$ and $m$ are positive integers, and $n \neq m$. $R_{nl}^{b}\propto a_{0}^{-3/2}$ and $R^{m\:l-1}_{n\:l} \propto a_{0}$ in SI units, where $a_{0}$ is the Bohr radius.}
\label{tbl:radial_integrals_boundbound}
\begin{tabular}{l}
\hline \hline
  $(R^{1\:0}_{n\:1})^2 = 2^{8}n^{7} (n^2-1)^{-5} \left(\frac{n-1}{n+1}\right)^{2n} $ \\ 
  $(R^{2\:0}_{n\:1})^2 = 2^{17}n^{7}(n^2-1) (n^2-4)^{-6} \left(\frac{n-2}{n+2}\right)^{2n}$ \\
  $(R^{3\:0}_{n\:1})^2 = 2^{8}3^{7}n^{7}(n^2-1)(7n^2-27)^2 (n^2-9)^{-8} \left(\frac{n-3}{n+3}\right)^{2n}$ \\
  $(R^{4\:0}_{n\:1})^2 = 2^{26}3^{-2}n^{7}(n^2-1)(23n^4-288n^2+768)^2 (n^2-16)^{-10} \left(\frac{n-4}{n+4}\right)^{2n}$ \\ 
  $(R^{5\:0}_{n\:1})^2 = 2^{8}3^{-2}5^{9}n^{7}(n^2-1)(-91n^6+2545n^{4}-20625n^2+46875)^2 (n^2-25)^{-12} \left(\frac{n-5}{n+5}\right)^{2n}$ \\
  $(R^{6\:0}_{n\:1})^2 = 2^{17}3^{7}5^{-2}n^{7}(n^2-1)(7n^2-108)^{2}(-289n^6 + 10620n^{4}-97200n^2+233280)^2 (n^2-36)^{-14} \left(\frac{n-6}{n+6}\right)^{2n}$ \\
  $(R^{7\:0}_{n\:1})^2 = 2^{8}3^{-4}5^{-2}7^{9}n^{7}(n^2-1)(-29233n^{10}+2547265n^8-79704282n^6+1097665170n^{4}-6485401125n^2+12711386205)^2(n^2-49)^{-16} \left(\frac{n-7}{n+7}\right)^{2n}$ \\
  $(R^{8\:0}_{n\:1})^2 = 2^{35}3^{-4}5^{-2}7^{-2}n^{7}(n^2-1) $\\
  $\qquad \quad \times (1044871n^{12}-140890496n^{10}+7190401024n^8-175037743104n^6+2101597962240n^{4}-11499774935040n^2+21646635171840)^2 $\\
  $\qquad \quad \times (n^2-64)^{-18} \left(\frac{n-8}{n+8}\right)^{2n}$ \\
  $(R^{9\:0}_{n\:1})^2 = 2^{8}3^{16}5^{-2}7^{-2}n^{7}(n^2-1) (-1859129n^{14}+366767919n^{12}-28441442925n^{10}+1109810543211n^8-23169940624971n^6$\\
  $\qquad \quad \quad + 253715866938765n^{4}-1314709492319055n^2+2402063207770905)^2 (n^2-81)^{-20} \left(\frac{n-9}{n+9}\right)^{2n}$ \\
  $(R^{10\:0}_{n\:1})^2 = 2^{17}3^{-8}5^{9}7^{-2}n^{7}(n^2-1) (55491751n^{16}-15340645280n^{14}+1715399560000n^{12}-100302148800000n^{10}+3312354600000000n^8 $\\
  $\qquad \quad \quad -62279280000000000n^6 + 638215200000000000n^{4}-3175200000000000000n^2+5670000000000000000)^2 (n^2-100)^{-22} \left(\frac{n-10}{n+10}\right)^{2n} $ \\
  $(R^{11\:0}_{n\:1})^2 = 2^{8}3^{-8}5^{-4}7^{-2}11^{9}n^{7}(n^2-1)$\\
  $\qquad \quad \left. \times (-7887190691n^{18} + 2952526744959n^{16} - 457219036589964n^{14} + 38114601597190316n^{12} - 1865597406898265370n^{10} \right.$\\
  $\qquad \quad \quad \left. +54907078609311927570n^8 - 956546457716017794780n^6 + 9323276139285508115100n^4 - 44942282036313341370075n^2 \right.$\\
  $\qquad \quad \quad \left. +78811827918752381243175)^2 \right.$\\
  $\qquad \quad \times  (n^2 - 121)^{-24} \left(\frac{n-11}{n+11}\right)^{2n} $ \\
  $(R^{12\:0}_{n\:1})^2 = 2^{26}3^{7}5^{-4}7^{-2}11^{-2}n^{7}(n^2-1)$\\
  $\qquad \quad \left. \times (18697277347 n^{20} - 9217044652320 n^{18} + 1913951672476416 n^{16} - 218875150436106240 n^{14} + 15134741981452173312 n^{12} \right.$\\
  $\qquad \quad \quad \left. - 654425326411705221120 n^{10} + 17695723484365612646400 n^8 - 290855139916983167877120 n^6 + 2725854937338386527027200 n^4 \right.$\\
  $\qquad \quad \quad \left. - 12812479141426023628800000 n^2 + 22139963956384168830566400)^2 \right.$\\
  $\qquad \quad \times (n^2 - 144)^{-26} \left(\frac{n-12}{n+12}\right)^{2n} $ \\
  \hline
  $(R^{n\:1}_{3\:2})^2 = 2^{11}3^{7}5^{-1}n^{11}(n^2-1) (n^2-9)^{-8} \left(\frac{n-3}{n+3}\right)^{2n}$\\
  $(R^{n\:1}_{4\:2})^2 = 2^{28}3^{-2}5^{-1}n^{11}(n^2-1)(7n^2 - 48)^2 (n^2-16)^{-10}\left(\frac{n-4}{n+4}\right)^{2n}$  \\
  $(R^{n\:1}_{5\:2})^2 = 2^{11}3^{-2}5^{9}7^{-1}n^{11}(n^2-1)(29n^{4}-590n^2+2625)^2 (n^2-25)^{-12} \left(\frac{n-5}{n+5}\right)^{2n}$ \\ 
  $(R^{n\:1}_{6\:2})^2 = 2^{22}3^{7}5^{-1}7^{-1}n^{11}(n^2-1)(-167n^6 + 7092n^{4}-89424n^2+326592)^2 (n^2-36)^{-14} \left(\frac{n-6}{n+6}\right)^{2n}$ \\ 
  $(R^{n\:1}_{7\:2})^2 = 2^{12}3^{-5}5^{-1}7^{9}n^{11}(n^2-1)(2483n^8 - 186788n^6 + 4760154n^{4}-47799108n^2+155649627)^2 (n^2-49)^{-16} \left(\frac{n-7}{n+7}\right)^{2n} $ \\ 
  $(R^{n\:1}_{8\:2})^2 = 2^{37}3^{-5}5^{-3}7^{-1}n^{11}(n^2-1)(-363461n^{10} + 43842880n^8 - 1932468224n^6 + 38338560000n^{4}-335963750400n^2+1014686023680)^2$\\
  $\qquad \quad \times (n^2-64)^{-18} \left(\frac{n-8}{n+8}\right)^{2n}$ \\ 
  $(R^{n\:1}_{9\:2})^2 = 2^{12}3^{16}5^{-3}7^{-1}11^{-1}n^{11}(n^2-1)$\\
  $\qquad \quad \times (659701n^{12}-119186154n^{10} + 8270265159n^8 - 278827252236n^6 + 4746977158275n^{4}-37971082126890n^2+108735371545185)^2$\\
  $\qquad \quad \times (n^2-81)^{-20} \left(\frac{n-9}{n+9}\right)^{2n}$ \\ 
  $(R^{n\:1}_{10\:2})^2 = 2^{22}3^{-7}5^{9}7^{-2}11^{-1}n^{11}(n^2-1)$\\
  $\qquad \quad \times (-10015297n^{14}+2577444380n^{12}-263988466000n^{10} + 13815991800000n^8 - 394364460000000n^6 $\\
  $\qquad \quad \quad + 6049890000000000n^{4}-45322200000000000n^2+124740000000000000)^2$\\
  $\qquad \quad \times (n^2-100)^{-22} \left(\frac{n-10}{n+10}\right)^{2n}$ \\ 
  $(R^{n\:1}_{11\:2})^2 = 2^{11}3^{-7}5^{-3}7^{-2}11^{9}13^{-1}n^{11}(n^2-1) $\\
  $\qquad \quad \left. \times (1444880423 n^{16} - 509578839528 n^{14} + 73469720817108 n^{12} - 5610914005715480 n^{10} + 245883912301808154 n^8 \right.$\\
  $\qquad \quad \quad \left. - 6257331607246231512 n^6 + 88986307739304355380 n^4 - 634470831957535733160 n^2 + 1693477294121951993655)^2\right.$\\
  $\qquad \quad \times (n^2-121)^{-24} \left(\frac{n-11}{n+11}\right)^{2n}$ \\ 
  $(R^{n\:1}_{12\:2})^2 = 2^{28}3^{7}5^{-3}7^{-3}11^{-1}13^{-1}n^{11}(n^2-1) $\\
  $\qquad \quad \left. \times (-6941077103 n^{18} + 3252952763568 n^{16} - 636455021239296 n^{14} + 67773808902389760 n^{12} - 4294213709999505408 n^{10} \right.$\\
  $\qquad \quad \quad \left. + 166284350667412733952 n^8 - 3889155474881677099008 n^6 + 52204920798004830535680 n^4 - 358108792002857360424960 n^2 \right.$\\ $\qquad \quad \quad \left. + 932748481495814520176640)^2\right.$\\
  $\qquad \quad \times (n^2-144)^{-26} \left(\frac{n-12}{n+12}\right)^{2n}$ \\ 
  \hline \hline
\end{tabular}
\end{table*}

\begin{table*}
\centering
\caption{Same as Table~\ref{tbl:radial_integrals_boundbound} but for bound-free transitions: $(R^{m\:l-1}_{n'\:l})^2 = (\int_{0}^{\infty}r^3R^{c}_{n'\:l}R^{b}_{m\:l-1}dr)^2$ and $(R^{n'\:l}_{m\:l+1})^2 = (\int_{0}^{\infty}r^3R^{c}_{n'\:l}R^{b}_{m\:l+1}dr)^2$, where $n'$ is a positive real number and $m$ is a positive integer.}
\label{tbl:radial_integrals_boundfree}
\begin{tabular}{l}
\hline \hline
  $(R^{1\:0}_{n'\:1})^2 = \left[ 2^{8}n'^{7} (n'^2+1)^{-5} \exp\left[ -4n'\text{arccot}(n') \right] \right] \left[1-\exp(-2\pi n') \right]^{-1}$ \\ 
  $(R^{2\:0}_{n'\:1})^2 = \left[ 2^{17}n'^{7} (n'^2+1) (n'^2+4)^{-6} \exp\left[ -4n'\text{arccot}(n'/2) \right] \right] \left[1-\exp(-2\pi n') \right]^{-1}$ \\ 
  $(R^{3\:0}_{n'\:1})^2 = \left[ 2^{8}3^{7}n'^{7} (n'^2+1) (7n'^2+27)^{2} (n'^2+9)^{-8} \exp\left[ -4n'\text{arccot}(n'/3) \right] \right] \left[1-\exp(-2\pi n') \right]^{-1}$ \\ 
  $(R^{4\:0}_{n'\:1})^2 = \left[ 2^{26}3^{-2}n'^{7} (n'^2+1) (23n'^4+288n'^2+768)^{2} (n'^2+16)^{-10} \exp\left[ -4n'\text{arccot}(n'/4) \right] \right] \left[1-\exp(-2\pi n') \right]^{-1}$ \\
  $(R^{5\:0}_{n'\:1})^2 = \left[ 2^{8}3^{-2}5^{9}n'^{7}(n'^2+1)(91n'^6+2545n'^{4}+20625n'^2+46875)^2 (n'^2+25)^{-12} \exp\left[ -4n'\text{arccot}(n'/5) \right] \right] \left[1-\exp(-2\pi n') \right]^{-1}$ \\
  $(R^{6\:0}_{n'\:1})^2 = \left[ 2^{17}3^{7}5^{-2}n'^{7}(n'^2+1)(7n'^2+108)^{2}(289n'^6 + 10620n'^{4}+97200n'^2+233280)^2 (n'^2+36)^{-14} \exp\left[ -4n'\text{arccot}(n'/6) \right] \right] \left[1-\exp(-2\pi n') \right]^{-1}$ \\
  $(R^{7\:0}_{n'\:1})^2 = \left[ 2^{8}3^{-4}5^{-2}7^{9}n'^{7}(n'^2+1)(29233n'^{10}+2547265n'^8+79704282n'^6+1097665170n'^{4}+6485401125n'^2+12711386205)^2 \right.$\\
  $\qquad \quad \left. \times (n'^2+49)^{-16} \exp\left[ -4n'\text{arccot}(n'/7) \right] \right] \left[1-\exp(-2\pi n') \right]^{-1}$ \\
  $(R^{8\:0}_{n'\:1})^2 = \left[ 2^{35}3^{-4}5^{-2}7^{-2}n'^{7}(n'^2+1) \right.$\\
  $\qquad \quad \left. \times (1044871n'^{12}+140890496n'^{10}+7190401024n'^8+175037743104n'^6+2101597962240n'^{4}+11499774935040n'^2+21646635171840)^2 \right.$\\
  $\qquad \quad \left. \times (n'^2+64)^{-18} \exp\left[ -4n'\text{arccot}(n'/8) \right] \right] \left[1-\exp(-2\pi n') \right]^{-1}$ \\
  $(R^{9\:0}_{n'\:1})^2 = \left[ 2^{8}3^{16}5^{-2}7^{-2}n'^{7}(n'^2+1) \right.$\\
  $\qquad \quad \left. \times (1859129n'^{14}+366767919n'^{12}+28441442925n'^{10}+1109810543211n'^8+23169940624971n'^6 \right.$\\
  $\qquad \quad \quad \left. + 253715866938765n'^{4}+1314709492319055n'^2+2402063207770905)^2 \right.$\\
  $\qquad \quad \left. \times (n'^2+81)^{-20} \exp\left[ -4n'\text{arccot}(n'/9) \right] \right] \left[1-\exp(-2\pi n') \right]^{-1}$ \\
  $(R^{10\:0}_{n'\:1})^2 = \left[ 2^{17}3^{-8}5^{9}7^{-2}n'^{7}(n'^2+1) \right.$\\
  $\qquad \quad \left. \times (55491751n'^{16}+15340645280n'^{14}+1715399560000n'^{12}+100302148800000n'^{10}+3312354600000000n'^8 \right.$\\
  $\qquad \quad \quad \left. +62279280000000000n'^6 + 638215200000000000n'^{4}+3175200000000000000n'^2+5670000000000000000)^2 \right.$\\
  $\qquad \quad \left. \times (n'^2+100)^{-22} \exp\left[ -4n'\text{arccot}(n'/10) \right] \right] \left[1-\exp(-2\pi n') \right]^{-1}$ \\
  $(R^{11\:0}_{n'\:1})^2 = \left[ 2^{8}3^{-8}5^{-4}7^{-2}11^{9}n'^{7}(n'^2+1) \right.$\\
  $\qquad \quad \left. \times (7887190691n'^{18} + 2952526744959n'^{16} + 457219036589964n'^{14} + 38114601597190316n'^{12} + 1865597406898265370n'^{10} \right.$\\
  $\qquad \quad \quad \left. +54907078609311927570n'^8 + 956546457716017794780n'^6 + 9323276139285508115100n'^4 + 44942282036313341370075n'^2 \right.$\\
  $\qquad \quad \quad \left. +78811827918752381243175)^2 \right.$\\
  $\qquad \quad \left. \times  (n'^2 + 121)^{-24} \exp\left[ -4n'\text{arccot}(n'/11) \right] \right] \left[1-\exp(-2\pi n') \right]^{-1}$ \\
  $(R^{12\:0}_{n'\:1})^2 = \left[ 2^{26}3^{7}5^{-4}7^{-2}11^{-2}n'^{7}(n'^2+1) \right.$\\
  $\qquad \quad \left. \times (18697277347 n'^{20} + 9217044652320 n'^{18} + 1913951672476416 n'^{16} + 218875150436106240 n'^{14} + 15134741981452173312 n'^{12} \right.$\\
  $\qquad \quad \quad \left. + 654425326411705221120 n'^{10} + 17695723484365612646400 n'^8 + 290855139916983167877120 n'^6 + 2725854937338386527027200 n'^4 \right.$\\
  $\qquad \quad \quad \left. + 12812479141426023628800000 n'^2 + 22139963956384168830566400)^2 \right.$\\
  $\qquad \quad \left. \times (n'^2 + 144)^{-26}\exp\left[ -4n'\text{arccot}(n'/12) \right] \right] \left[1-\exp(-2\pi n') \right]^{-1}$ \\
  \hline
  $(R^{n'\:1}_{3\:2})^2 = \left[ 2^{11}3^{7}5^{-1}n'^{11} (n'^2+1) (n'^2+9)^{-8} \exp\left[ -4n'\text{arccot}(n'/3) \right] \right] \left[1-\exp(-2\pi n') \right]^{-1}$ \\ 
  $(R^{n'\:1}_{4\:2})^2 = \left[ 2^{28}3^{-2}5^{-1}n'^{11} (n'^2+1) (7n'^2+48)^{2} (n'^2+16)^{-10} \exp\left[ -4n'\text{arccot}(n'/4) \right] \right] \left[1-\exp(-2\pi n') \right]^{-1}$ \\ 
  $(R^{n'\:1}_{5\:2})^2 = \left[ 2^{11}3^{-2}5^{9}7^{-1}n'^{11}(n'^2+1)(29n'^{4}+590n'^2+2625)^2 (n'^2+25)^{-12} \exp\left[ -4n'\text{arccot}(n'/5) \right] \right] \left[1-\exp(-2\pi n') \right]^{-1}$ \\ 
  $(R^{n'1}_{6\:2})^2 = \left[ 2^{22}3^{7}5^{-1}7^{-1}n'^{11}(n'^2+1)(167n'^6 + 7092n'^{4}+89424n'^2+326592)^2 (n'^2+36)^{-14} \exp\left[ -4n'\text{arccot}(n'/6) \right] \right] \left[1-\exp(-2\pi n') \right]^{-1}$ \\ 
  $(R^{n'\:1}_{7\:2})^2 = \left[ 2^{12}3^{-5}5^{-1}7^{9}n'^{11}(n'^2+1)(2483n'^8 + 186788n'^6 + 4760154n'^{4}+47799108n'^2+155649627)^2\right.$\\
  $\qquad \quad \left. \times (n'^2+49)^{-16} \exp\left[ -4n'\text{arccot}(n'/7) \right] \right] \left[1-\exp(-2\pi n') \right]^{-1}$ \\ 
  $(R^{n'\:1}_{8\:2})^2 = \left[ 2^{37}3^{-5}5^{-3}7^{-1}n'^{11}(n'^2+1)(363461n'^{10} + 43842880n'^8 + 1932468224n'^6 + 38338560000n'^{4}+335963750400n'^2+1014686023680)^2\right.$\\
  $\qquad \quad \left. \times (n'^2+64)^{-18} \exp\left[ -4n'\text{arccot}(n'/8) \right] \right] \left[1-\exp(-2\pi n') \right]^{-1}$ \\ 
  $(R^{n'\:1}_{9\:2})^2 = \left[ 2^{12}3^{16}5^{-3}7^{-1}11^{-1}n'^{11}(n'^2+1) \right.$\\
  $\qquad \quad \left. \times (659701n'^{12}+119186154n'^{10} + 8270265159n'^8 + 278827252236n'^6 + 4746977158275n'^{4}+37971082126890n'^2+108735371545185)^2\right.$\\
  $\qquad \quad \left. \times (n'^2+81)^{-20} \exp\left[ -4n'\text{arccot}(n'/9) \right] \right] \left[1-\exp(-2\pi n') \right]^{-1}$ \\ 
  $(R^{n'\:1}_{10\:2})^2 = \left[ 2^{22}3^{-7}5^{9}7^{-2}11^{-1}n'^{11}(n'^2+1) \right.$\\
  $\qquad \quad \left. \times (10015297n'^{14}+2577444380n'^{12}+263988466000n'^{10} + 13815991800000n'^8 + 394364460000000n'^6 \right.$\\
  $\qquad \quad \quad \left. + 6049890000000000n'^{4}+45322200000000000n'^2+124740000000000000)^2\right.$\\
  $\qquad \quad \left. \times (n'^2+100)^{-22} \exp\left[ -4n'\text{arccot}(n'/10) \right] \right] \left[1-\exp(-2\pi n') \right]^{-1}$ \\ 
  $(R^{n'\:1}_{11\:2})^2 = \left[ 2^{11}3^{-7}5^{-3}7^{-2}11^{9}13^{-1}n'^{11}(n'^2+1) \right.$\\
  $\qquad \quad \left. \times (1444880423 n'^{16} + 509578839528 n'^{14} + 73469720817108 n'^{12} + 5610914005715480 n'^{10} + 245883912301808154 n'^8 \right.$\\
  $\qquad \quad \quad \left. + 6257331607246231512 n'^6 + 88986307739304355380 n'^4 + 634470831957535733160 n'^2 + 1693477294121951993655)^2\right.$\\
  $\qquad \quad \left. \times (n'^2+121)^{-24} \exp\left[ -4n'\text{arccot}(n'/11) \right] \right] \left[1-\exp(-2\pi n') \right]^{-1}$ \\ 
  $(R^{n'\:1}_{12\:2})^2 = \left[ 2^{28}3^{7}5^{-3}7^{-3}11^{-1}13^{-1}n'^{11}(n'^2+1) \right.$\\
  $\qquad \quad \left. \times (6941077103 n'^{18} + 3252952763568 n'^{16} + 636455021239296 n'^{14} + 67773808902389760 n'^{12} + 4294213709999505408 n'^{10} \right.$\\
  $\qquad \quad \quad \left. + 166284350667412733952 n'^8 + 3889155474881677099008 n'^6 + 52204920798004830535680 n'^4 + 358108792002857360424960 n'^2 \right.$\\ $\qquad \quad \quad \left. + 932748481495814520176640)^2\right.$\\
  $\qquad \quad \left. \times (n'^2+144)^{-26} \exp\left[ -4n'\text{arccot}(n'/12) \right] \right] \left[1-\exp(-2\pi n') \right]^{-1}$ \\ 
  \hline \hline
\end{tabular}
\end{table*}


\bsp	
\label{lastpage}
\end{document}